\documentclass[runningheads,natbib]{svjour2}
\usepackage{graphicx}

\smartqed  

\journalname{Astronomy and Astrophysics Review}

\begin{document}

\title{Formation and evolution of planetary systems: the impact of high angular resolution optical techniques
}


\titlerunning{Formation and evolution of planetary systems at high angular resolution}

\author{Olivier Absil \and Dimitri Mawet}


\institute{
O. Absil \at
  D\'epartement d'Astrophysique, G\'eophysique et Oc\'eanographie, Universit\'e de Li\`ege \\
  17 All\'ee du Six Ao\^ut, 4000 Sart Tilman, Belgium \\
  \email{absil@astro.ulg.ac.be}
\and
D. Mawet \at
  Jet Propulsion Laboratory, California Institute of Technology, \\
  4800 Oak Grove Drive, 91109 Pasadena CA, USA
}

\date{Received: date}

\maketitle


\begin{abstract}
The direct images of giant extrasolar planets recently obtained around several main sequence stars represent a major step in the study of planetary systems. These high-dynamic range images are among the most striking results obtained by the current generation of high angular resolution instruments, which will be superseded by a new generation of instruments in the coming years. It is therefore an appropriate time to review the contributions of high angular resolution visible/infrared techniques to the rapidly growing field of extrasolar planetary science. During the last 20~years, the advent of the Hubble Space Telescope, of adaptive optics on 4- to 10-m class ground-based telescopes, and of long-baseline infrared stellar interferometry has opened a new viewpoint on the formation and evolution of planetary systems. By spatially resolving the optically thick circumstellar discs of gas and dust where planets are forming, these instruments have considerably improved our models of early circumstellar environments and have thereby provided new constraints on planet formation theories. High angular resolution techniques are also directly tracing the mechanisms governing the early evolution of planetary embryos and the dispersal of optically thick material around young stars. Finally, mature planetary systems are being studied with an unprecedented accuracy thanks to single-pupil imaging and interferometry, precisely locating dust populations and putting into light a whole new family of long-period giant extrasolar planets.

\end{abstract}


\section{Introduction} \label{sec:intro}

The formation and evolution of (extrasolar) planetary systems is one of the astrophysical fields that has seen the most dramatic changes in its understanding during the past 25~years, fuelled by a wealth of breakthrough observations such as the first discovery of infrared excesses around main sequence stars due to circumstellar dust \citep{Aumann84,Aumann85}, the first resolved observation of a circumstellar disk \citep{Smith84} or the first discovery of an extrasolar planet around a main sequence star \citep{Mayor95}. Besides high resolution spectroscopy, tremendous progress in high angular resolution optical\footnote{We refer to as ``optical'' any observing technique in which the light is collected and focused by means of standard optical devices. More specifically, our review is restricted to operating wavelengths ranging from the visible to the mid-infrared regime, where the high angular resolution studies of planetary systems have been the most successful.} imaging during that same period has been one of the key drivers of the renewal of planetary science. With this paper, we aim to review and illustrate how high angular resolution imaging in the visible and infrared regimes has impacted the field of extrasolar planetary studies, from the earliest stages of planet formation in protoplanetary discs up to the late evolution of mature planetary systems. After a brief introduction setting the scene on planet formation and high angular resolution studies, the review is divided into three main sections dealing with the study of extrasolar planetary systems at various ages: it starts with the earliest stages of their formation (Section 2), describes the gradual evolution from young to fully formed systems (Section 3), and ends up with the study of mature planetary systems (Section 4). Finally, Section 5 summarises the impact of high angular resolution optical observations on the field of extrasolar planetary studies and gives some perspectives for the years to come.

    \subsection{The standard model for planetary formation}

The first step in understanding planetary formation consists in characterising the environment where these bodies form, i.e., the surroundings of young stellar objects (YSO). We will focus here on low- and intermediate-mass stars, because high-mass stars ($M_{\ast}>8M_{\odot}$) spend their whole pre-main sequence time as obscured objects, and are therefore not suited to the observing techniques used here. Moreover, massive stars are evolving so quickly into supernovae that planets are not supposed to have enough time to form or evolve in meaningful ways. Nowadays, the most widely accepted paradigm for star formation includes the following steps \citep[e.g.,][]{Adams87,McKee07}:
\begin{itemize}
\item
Star formation starts with the gravitational collapse of a dense molecular cloud. In this first phase, ``Class~0'' objects are generally not detected in the visible and infrared regimes, because light is blocked by a thick spheroidal infalling envelope. Such objects have a significant luminosity in the submillimetre to radio regime, produced by their cold envelope with typical temperatures around 15--30\,K.
\item
As gas accretes onto the star, an accretion-driven wind develops which begins to clear the envelope gas away from the rotational poles of the system. This is the embedded YSO phase with a ``Class~I'' spectral energy distribution (SED) rising beyond 2\,$\mu$m. At this stage, almost all the light from the star is still absorbed and reradiated by the circumstellar envelope of dust and gas at long wavelengths. The embedded phase is thought to last for a few times $10^5$\,yr, during which the envelope partially settles into a circumstellar disc.
\item
As the infalling envelope disperses, the central YSO and its circumstellar disc become progressively detectable in the optical regime while accretion continues. During this ``Class~II'' epoch, the near- to far-infrared SED still shows significant emission in excess to the photosphere but with a flat or falling spectrum longward of $2\,\mu$m. Circumstellar discs are generally thought to be composed of 99\% gas and 1\% dust (either rocky or icy), similarly to the molecular clouds from which they originate.
\item
Draining of the circumstellar material through accretion and other processes (such as planet formation) eventually leads to an optically thin disc mostly devoid of gas. This is the ``Class~III'' phase, where signs of accretion are scarce or absent, and the SED is dominated by the stellar photosphere up to the mid-infrared regime. In this phase, the star has finally reached the main sequence and planets (or at least planetary embryos) are supposed to be mostly formed.
\end{itemize}

In this review, we will mostly focus on the phases where planet formation and evolution are supposed to be mostly at play, i.e., for Class~I, II and III objects. Planet formation requires growth through at least 12 orders of magnitude in scale, from sub-micron-sized interstellar dust particles to bodies with radii of thousands of kilometres. The details for these complex processes are not yet fully understood, and standard models are still largely based on seminal works from the early 80's, when the study of planet formation was almost exclusively based on observations of our solar system \citep{Safronov69,Wetherill80,Pollack84}. The first step of planet formation, thought to take place in the circumstellar discs around Class I/II objects, consists in the agglomeration of small interstellar dust grains into larger particles. Theoretical models suggest that the growth to centimetre size is very efficient \citep{Weidenschilling84}, a stage at which dust particles decouple from the gas and start migrating towards the disc midplane (dust settling) and toward the central star. A rapid growth of particles beyond the ``metre barrier'' is then required to counteract the expected rapid radial inner drift of particles in this size range, which would quickly drive them to the inner evaporation zone. In this size range, fragmentation is thought to become dominant, and the mechanisms leading to the rapid growth of agglomerates beyond the metre size despite the effect of fragmentation are not yet fully understood \citep[e.g.,][]{Johansen07,Cuzzi08}.

Once the metre-size barrier is crossed and kilometre-size planetesimals are formed, runaway growth is expected to take over and produce protoplanets of Mercury- to Mars-size by gravitational accumulation within about $10^5$\,yr. These oligarchs are then expected to have strong dynamical interactions and eventually merge to produce a few planetary cores. The fate of planetary cores then depends on whether they are located inside or outside the ``snow line'', the stellocentric radius at which the temperature becomes low enough for ices to condense. Interior to the snow line, they will constitute the pool of terrestrial planets, which are thought to be fully formed in a few tens of Myr, while outside the snow line, they are subject to ``core accretion'' where large planetary cores (typically $10\,M_{\oplus}$) capture large amounts of volatile species and thereby produce giant planets with massive atmospheres. The estimated timescale for this process \citep[about 10\,Myr,][]{Pollack96} is, however, uncomfortably close to the estimated timescale for gas dispersal in protoplanetary discs \citep[about 5\,Myr, e.g.][]{Haisch01}. Several processes have been proposed to speed up the accretion of massive atmospheres, most notably planetary migration \citep{Alibert04}, which allows planetary cores to form at more ``suitable'' distances from the star and to gradually extend their ``feeding'' region. A completely different formation mechanism for gas giant, based on direct gravitational collapse of young protoplaneraty discs, has also been proposed \citep{Boss97}. Observational evidences, such as the correlation between stellar metallicity and extrasolar planet abundance, have nevertheless mostly favoured the core accretion model so far.

    \subsection{High angular resolution observing tools}

Within the multiple observing techniques that have been used to constrain planetary formation and evolution, high angular resolution techniques in the visible and near-infrared regime hold a very important place because they are among the few which have the potential of spatially resolving the various bodies and physical phenomena at play in planetary systems. In nearby star forming regions, located at distances of a few hundred parsecs, they provide a direct view at the AU-scale of the protoplanetary discs where planets are being formed. And for nearby main sequence stars, they are capable of isolating the faint signal of circumstellar material (dust, planets) from the blinding stellar light.

The near-infrared regime has been one of the most important spectral range in this context. On the one hand, it is sensitive both to the thermal emission from dust close to its sublimation temperature ($\sim 1500$\,K) and to the stellar light scattered by its surrounding material. On the other hand, it provides both the sensitivity and the technical prerequisites (optical elements, detectors, etc) for an optimum implementation on the ground. Two main types of observing techniques can be distinguished: single-pupil observations with large telescopes working close to their diffraction limit, and interferometric observations coherently combining the light from a few individual telescopes with ground separations up to a few hundred metres. In the first case, typical angular resolutions of 50\,mas can be reached (i.e., 5\,AU at 100\,pc), while in the latter case, angular resolutions down to $\sim 1$\,mas are achievable.

        \subsubsection{Single pupil imaging} \label{sub:singlepupilimaging}

Modern observatories have opened the access to unprecedented spatial scales, thereby revealing the complexity of planetary system formation. Two major technological advances have revolutionised observations at high angular resolution: space telescopes and adaptive optics systems. Both advances are used to mitigate the prime threat in the formation of good astronomical images: atmospheric turbulence. Theoretically, a diffraction-limited telescope provides images with a ``finesse'' close to $\lambda/D$, with $\lambda$ the observing wavelength and $D$ the diameter of the telescope. Unfortunately, after propagating through the mere 20~kilometres of our atmosphere, the resolution is degraded to $\lambda/r_0$, where $r_0$ is the Fried parameter indicative of the seeing cell size---typically about 10\,cm in the visible but strongly dependent on wavelength and on the observing site. Two types of solutions to this problem have emerged in the early 90's, either with a complete by-pass of the Earth atmosphere by going into space (e.g., Hubble Space Telescope), or with the design of adaptive optics systems correcting the optical effect of the turbulence in real time and thereby restoring the full resolution of the telescope.

Both solutions, each benefiting from the revolutionary advances in electronic detection techniques (CCDs and infrared focal plane arrays), have allowed tremendous progress in astronomy (see Table~\ref{tab:singlepupil} for a non-exhaustive list of single-pupil high angular resolution imaging facilities). Besides achieving the full potential of large telescopes, high angular resolution techniques also help when the scene to image contains multiple objects with orders of magnitude of difference in luminosity. The dynamic range of the best detectors is physically limited by their full-well capacity, but also in practice by the number of bits of the analog-to-digital converter (65,536:1 for 16 bits). However, diffraction from the brightest object, i.e., the central star, usually overwhelms the faint signal scattered by circumstellar material or putative low-mass companions. To mitigate this effect, coronagraphic imaging is nowadays routinely used, by placing an opaque mask at the focus of the telescope on top of the central core of the diffraction pattern of the star, and a well-dimensioned diaphragm in the relayed pupil plane to block diffraction residuals \citep{Ferrari07}. The optical system improves the dynamic range of instruments by several orders of magnitudes, allowing faint objects to be isolated in high contrast scenes, such as circumstellar discs around stars.

\begin{table}[t]
\caption{Single-pupil telescopes and instruments benefiting from high-contrast imaging / high angular resolution capabilities, used in the study of planetary formation and evolution. For ground-based near-infrared instruments, the first acronym is usually for the adaptive optics system, while the second one is for the camera. The type of coronagraph (either Lyot mask or Four Quadrant Phase Mask) is given in the last column, when available. Note that the contrast performance of these instruments varies a lot, depending on the instrument design itself and on the observing strategy. Instruments marked with a \dag\ are no longer available.}
\centering \label{tab:singlepupil}
\begin{tabular}{cccccc}
\hline\noalign{\smallskip}
Instrument & Telescope  & Wavelength   & Ang.\ res. & Coronagraph \\
              &            & ($\mu$m)  &   (mas)    &    \\
\tableheadseprule\noalign{\smallskip}
WFPC2\dag     & HST        & 0.12--1.1 & 10--100  & ... \\
NICMOS\dag    & HST        & 0.8--2.4  & 60--200  & Lyot\\
ACS           & HST        & 0.2--1.1  & 20--100  & Lyot\\
STIS          & HST        & 0.2--0.8  & 20--60   & Lyot\\
NAOS-CONICA   & VLT        & 1.1--3.5  & 30--90   & Lyot/FQPM\\
VISIR         & VLT        & 8.5--20   & 200--500 & ... \\
COME-ON+\dag  & ESO 3.6-m  & 1--5      & 60--280  & Lyot \\
PUEO          & CFHT       & 0.7--2.5  & 4--140   & Lyot \\
CIAO          & Subaru     & 1.1--2.5  & 30--70   & Lyot\\
AO-NIRC2      & Keck       & 0.9--5.0  & 20--100  & Lyot\\
LWS           & Keck       & 3.5--25   & 70--500  & ...\\
MIRLIN\dag    & Keck       & 8.0--20   & 160--400 & ...\\
ALTAIR-NIRI   & Gemini N.  & 1.1--2.5  & 30--70   & Lyot\\
NICI          & Gemini S.  & 1.1--2.5  & 30--70   & Lyot\\
T-ReCS        & Gemini S.  & 1.1--2.5  & 30--70   & ...\\
Lyot project\dag  & AEOS   & 0.8--2.5  & 60--140  & Lyot/FQPM\\
PALAO-PHARO   & Hale 200'' & 1.1--2.5  & 60--140  & Lyot/FQPM\\
WCS-PHARO     & Hale 200'' & 1.1--2.5  & 60--300  & Lyot/FQPM\\
AO-IRCAL      & Shane 120''& 1.1--2.5  & 100--150 & ... \\
\noalign{\smallskip}\hline
\end{tabular}
\end{table}

Scattered light imaging of circumstellar discs in nearby star forming regions, using ground-based telescopes equipped with adaptive optics systems in the near-infrared or space-based telescopes such as the HST in the optical, reveals physical phenomena at the 10~AU scale down to the very first tens of AUs from the central star. And in the direct solar neighbourhood (10--20\,pc), the 50--100\,mas angular resolution allows features to be distinguished at the AU scale in the disc images, some of them being possible sign-posts of dynamical interactions with planetary bodies (e.g., rings, clumps, warps, asymmetries, etc). Besides the identification of planet-related features, single-pupil imaging is most useful in studying the structure and composition of circumstellar discs, as extensively discussed in Sect.~\ref{sub:largescaledisc}.

        \subsubsection{Stellar interferometry}

By coherently combining the light collected from a few individual telescopes, stellar interferometry achieves an angular resolution equal to $\lambda/2B$ with $B$ the separation between the telescopes, referred to as ``baseline'' \citep[see e.g.,][]{Lawson00,Quirrenbach01,Monnier03}. With baselines up to a few hundred metres in the world leading facilities, interferometry currently reaches a resolving power equivalent to that of single-pupil telescopes much larger than even the most ambitious Extremely Large Telescope projects considered so far. By principle, interferometry samples the target's brightness distribution in its Fourier plane, by measuring the amplitude and phase of interference fringes, and hence images are not directly accessible. Until the late 90's, fitting simple models to the measured fringe amplitude (``visibility'') was almost the only way to derive physical information from interferometric observations, based on two-telescope arrays. Since then, multi-aperture interferometers have given access to partial information on the fringe phase through closure phase measurements \citep{Monnier03}, thereby allowing more complex morphologies to be investigated and even first images to be reconstructed from interferometric measurements.

With its exquisite angular resolution of typically 1\,mas in the near-infrared on hectometric baselines, stellar interferometry is the tool of choice to investigate the innermost parts of circumstellar discs in nearby star forming regions. In particular, interferometry is currently the only suitable method to directly characterise the most important region of protoplanetary discs where dusty grains sublimate and where accretion/ejection processes originate. It also provides a sharp view of the region where terrestrial planets are supposed to be formed. For more evolved planetary systems, interferometry can be used to constrain the presence of circumstellar material in the inner few AUs, including large amounts of warm dust, (sub-)stellar companions, or even hot planetary-mass companions. The main interferometric instruments and facilities used in the study of planetary systems are listed in Table~\ref{tab:interfero} together with their main characteristics.

\begin{table}[t]
\caption{Interferometric instruments used for the study of planetary formation and evolution. Note that the number of available telescopes (or telescope stations in case of mobile telescopes as for IOTA and VLTI) may exceed the number of individual apertures combined by a given instrument. Instruments marked with a \dag\ are no longer available.}
\centering
\label{tab:interfero}
\begin{tabular}{ccccccc}
\hline\noalign{\smallskip}
Instrument & Interf. & Baseline & Bands & Ang.\ res. & Spec.\ res. & Apert. \\
        &        &   (m)   &       &   (mas) &             &  (\#) \\
\tableheadseprule\noalign{\smallskip}
AMBER   &  VLTI  & 16--200 & J,H,K & 0.6--14 & 35--15000 & 3 \\
VINCI\dag & VLTI & 16--200 & H,K   & 0.9--14 & none      & 2 \\
MIDI    &  VLTI  & 16--200 & N     & 4--80   & 20--220   & 2 \\
V$^2$   & Keck-I & 85      & H,K,L & 2--5    & 25--1800  & 2 \\
Nuller  & Keck-I & 85      & N     & 10--16  & 40        & 2 \\
Mask    &  Keck  & 1--10   & J to L & 13--400 & none    & 15--21 \\
Classic & CHARA  & 34--330 & H,K   & 0.5--7  & none      & 2 \\
FLUOR   & CHARA  & 34--330 & K     & 0.7--7  & none      & 2 \\
MIRC    & CHARA  & 34--330 & J,H   & 0.4--5  & 40--400   & 4 \\
IONIC3\dag & IOTA& 5--38   & H     & 4--33   & 38        & 3 \\
V$^2$\dag & IOTA & 5--38   & H,K   & 4--45   & none      & 2 \\
V$^2$\dag & PTI  & 85--110 & K     & 2--3    & 22        & 2 \\
BLINC   & MMT    & 4       & N     & 250     & none      & 2 \\
\noalign{\smallskip}\hline
\end{tabular}
\end{table}


\section{Protoplanetary discs} \label{sec:proto}

As already mentioned in the introduction, high angular resolution optical imaging is precluded for the study of Class~0 objects: the optical depth of molecular envelopes around newly born systems is too high to allow visible or infrared radiation to leak through the opaque veil. For this reason, though high angular resolution mapping has been made available for Class~0 objects by radio and submillimetre interferometry, we only deal here with evolutionary phases for which most of the surrounding matter has settled into a disc, allowing us to see the star in the optical domain. It is precisely during these phases (Class I and II) that planetary formation is supposed to start, in the optically thick discs surrounding young stars.

Rigourous classification of young systems into an evolutionary sequence (e.g., between Classes I and II) is generally difficult and somewhat empirical. Indeed, observations have shown a great variety of characteristics, which are difficult to link with the system age, itself determined only with large error bars from independent and sometimes inconsistent photometric and spectroscopic measurements (lithium abundance, H$_{\alpha}$ equivalent width, X-ray flux, evolutionary track). Moreover, classifications are generally not an exact representation of an evolutionary sequence, because the disc mass distribution and the angle from which it is observed affect the way radiation is emitted towards us. Consequently, stars in different observational subclasses may have similar ages.

\begin{figure}[t]
\centering
\includegraphics[width=\textwidth]{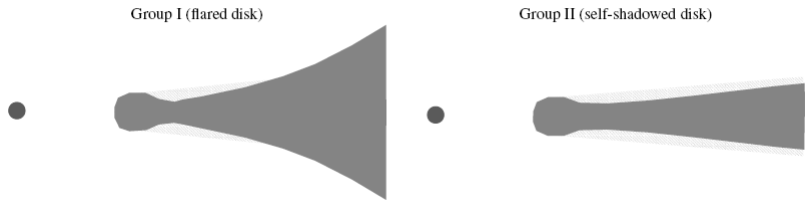}
\caption{Pictographic representation of the geometrical models associated with the SED classification of \citet{Meeus01}. Credit: \citet{Dullemond04}, reproduced with permission \copyright ESO.} \label{fig:meeus}
\end{figure}

A less ambiguous distinction can be made between T~Tauri stars, with stellar masses below $2 M_{\odot}$, and Herbig~Ae/Be stars, with stellar masses between 2 and $8 M_{\odot}$. Both of them have SEDs spanning Class I--II, showing a flat or falling SED with strong excess above the photosphere from the near-infrared to the submillimetre regime, dominated by the emission of an optically thick accretion disc:
\begin{itemize}
\item
T~Tauri stars (TTS) are low-mass pre-main sequence stars named after T~Tauri, their prototype. Their spectral types range from F to M, their age from about 1 to 10~Myr, and they are usually found in the molecular clouds that have engendered them. TTS are very active, showing evidence of starspot coverage and powerful stellar winds, which are thought to account for their X-ray and radio variable emissions. In this section we focus on classical TTS (CTTS), which have significant signs of on-going accretion unlike weak-line TTS (WTTS). The latter lack $H_\alpha$ emission in excess of 10\,$\AA$ in equivalent width, are supposedly in a more advanced evolutionary state, and are known to show a lower frequency of detected circumstellar discs.
\item
Herbig Ae/Be (HAeBe) stars, named after George Herbig (1920--), are intermediate-mass pre-main sequence stars of spectral type A or B, showing strong emission lines (especially H$\alpha$ and calcium lines). HAeBe stars have been classified into various groups depending on the shape of their infrared SED. Most notable is the classification of \citet{Meeus01}, where group I objects have flat double-peaked SEDs in the 2--100\,$\mu$m region while group II objects have weak and declining 2--100$\mu$m SEDs. This classification has been interpreted as the spectrophotometric signature of two distinct disc geometries: flared discs where the scale height increases with radius for group I, and flat self-shadowed discs for group II (see Fig.~\ref{fig:meeus}). This geometrical interpretation was later backed up by advanced theoretical models and by resolved images.
\end{itemize}
In the following, any young object that cannot be classified in either of these two categories, e.g., when considering edge-on discs where the central star is occulted by the disc mid-plane (see Sect.~\ref{sub:largescaledisc}), will be referred to by the generic term young stellar object (YSO), which includes both HAeBe and TTS but also higher mass stars.

Determining the structure of protoplanetary discs is a mandatory first step in understanding the physical processes leading to the formation of large planetary bodies. Studies based on the sole SED of such objects result in ambiguous results, which can only be disentangled by resolved images. This has been the main goal of first resolved observations of YSO discs. Besides investigating the large-scale structure of protoplanetary discs, resolved images can be used to investigate more detailed morphological aspects (e.g., persistence of spherical envelopes, wind-created cavities, accretion and outflow phenomena, location of planetary building blocks, non-symmetric features, etc), as well as the timescale for planet formation. In this section, we will first discuss the large-scale structure of protoplanetary discs seen in scattered light by high contrast single-pupil imaging facilities. Then, the second part of the section will explore the inner part of the discs as seen by near-infrared interferometers.


    \subsection{The large-scale structure of proto-planetary discs} \label{sub:largescaledisc}

Before the advent of adaptive optics on large ground-based telescopes and even the full power of the Hubble Space Telescope, the detection of infrared excess emission around young solar-type stars in star-forming regions already pointed towards the presence of circumstellar dust \citep[e.g.,][]{Strom89}. Ground-based observations soon revealed that many Class~I YSOs were nebulous in the optical and near-infrared, in particular those associated with outflows \citep{Tamura91}. The cometary morphology of the surrounding nebulae could be fitted by models of flattened envelopes with polar outflow cavities. Many of these objects were also found to be highly polarised in the optical and near-infrared due to the dominance of scattered light at short wavelengths. The strong linear polarisation in the bipolar lobes was found to be caused by single scattering on dust grains, revealing the location of the illuminating source. However, the final proof that protoplanetary discs do exist came from imaging with the Hubble Space Telescope \citep[e.g.,][]{Burrows96,McCaughrean96,Padgett99}. Since then, single-pupil imaging has been most useful in studying the structure and composition of protoplanetary discs.

\paragraph{Disc structure from scattered light.} Deriving physical parameters from images is a difficult task as the two-dimensional projection of a three-dimensional object compresses a lot of information. The inclination of the disc with respect to the line of sight is one key parameter in determining the number of physical parameters that can be constrained from the observation. Three major orientations are generally distinguished (edge-on, face-on, and intermediate):
\begin{itemize}
\item
The majority of observed discs so far are viewed edge-on (see e.g.\ HH\,30 in Sect.~\ref{sub:edgeondiscs}): this is the most favourable orientation for high contrast imaging of the circumstellar material since the central source is occulted by the equatorial plane of the optically thick disc. On each side of the mid-plane are observed two nebulae of light scattered by dusty particles away from the equatorial plane of the disc. Scattered light images from edge-on discs provide limited but essential information on the disc properties: inclination, relative opacities at different wavelengths, scale height in the outer part of the disc, and degree of forward scattering.
\item
Face-on discs are also a very interesting case in that any departure from radial symmetry in the observed image is dominated by inclination rather than vertical structure, which means that any departure from constant ellipticity implies non-axisymmetric illumination or disk structure. For example, the disc around TW~Hya (see Sect.~\ref{sub:ttauridiscs}) is elliptical between 70 and 140\,AU, and essentially circular further out, which suggests that the disc is warped. The most important structural information that can be extracted from images of such discs is the dependence of the surface brightness on radius (from which temperature and density profile can be calculated) and ellipticity (warped discs).
\item
The third kind of orientation is the intermediate-inclination regime where the star is directly visible but both nebulae on each side of the disk mid-plane are still present, although one may be too faint to be detected. Scattered light images of those discs are more difficult to interpret than those of edge-on discs because the mass-opacity product and the inclination are degenerate, so that ancillary data (e.g., SED or submillimetre images) are needed to break the inclination degeneracy. The outer radius of the discs is the only parameter that can be cleanly extracted.
\end{itemize}

\paragraph{Composition from multi-wavelength imaging.} When visible or near-infrared light interacts with dust, it can be either scattered or absorbed. The fate of an individual photon depends on its wavelength and on the dust scattering properties (composition, size, shape). When scattered, photons are deviated by an angle whose probability distribution is called the scattering phase function, characterised by the asymmetry factor $-1<g<1$. Forward scattering ($g\approx 1$) is found for larger grains whereas small grains preferentially scatter isotropically ($g\approx 0$). For optically thick protoplanetary discs, the scattered stellar photons seen by an outside observer have interacted mostly with the uppermost layers of the disc (optical depth $\tau<1$), so that scattered light imaging of YSOs only probes the surface layer of disc. This inevitably causes degeneracies in the parameters, which can only be alleviated by multi-wavelength analysis.

In the following, we review the characteristics of three families of protoplanetary discs based on the geometry and mass classifications introduced here above. We deliberately chose to present a few prototypical objects, which are well known from the observational standpoint, and theoretically dissected. In the present section, we focus on the large-scale structure of circumstellar discs without discussing direct signatures of planetary formation (grain growth, evidence for planetary-mass companions); this will be the topic of Section~\ref{sec:transition}.

	\subsubsection{Large scale structure of prototypes edge-on YSOs} \label{sub:edgeondiscs}

As already mentioned, any object that cannot be unambiguously classified as TTS or HAeBe star will be referred to by the generic term YSO. This is usually the case when considering edge-on discs, where the central star is occulted by the disc mid-plane thereby preventing the identification of specific spectral features. Edge-on YSOs such as HH~30 \citep{Burrows96}, CB-26 \citep{Stecklum04} or Haro 6-5B \citep{Padgett99}, and other prototypes, share many common features, though the detailed morphologies of the individual objects are unique. When seen edge-on, optically thick protoplanetary discs naturally occult their central stars and offer their vertical structures to direct view. All of the observed circumstellar reflection nebulae are crossed by dark lanes from 450 to 900\,AU in extent and from $<50$ to 350 AU in apparent thickness. The absorption lanes extend perpendicularly to known optical and millimetre outflows in these sources. The dark lanes are optically thick circumstellar discs seen in silhouette against bright reflection nebulosity. The bipolar reflection nebulae extending on both sides of the dust lanes appear to be produced by scattering from the upper and lower surfaces of the discs and from dusty material within or on the walls of the outflow cavities.

\begin{figure}[t]
\centering
\includegraphics[width=0.9\textwidth]{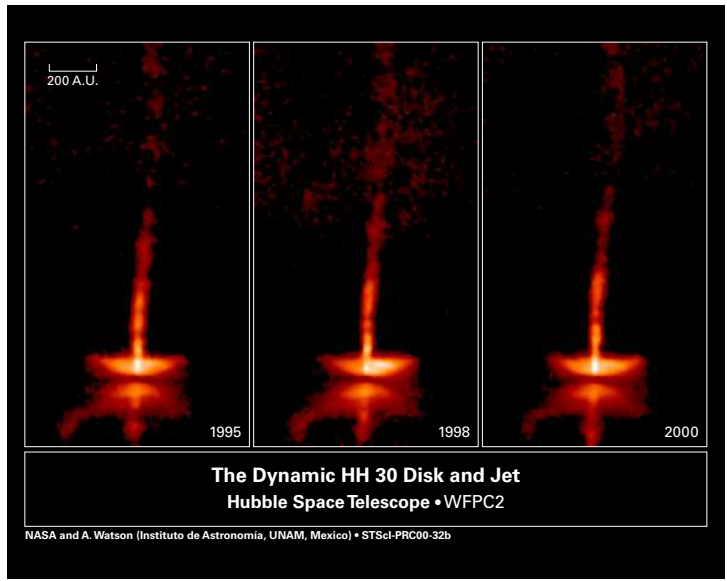}
\caption{Hubble Space Telescope image of HH~30, a prototype edge-on YSO. The images of this edge-on seen disc show an evolving compact bipolar reflection nebula bisected by a dark lane. Courtesy of K.~Stapelfeldt, with acknowledgments to NASA and STScI.} \label{fig:hh30}
\end{figure}

If we were to pick a prototypical edge-on YSO, we would certainly choose HH~30 \citep{Burrows96}. This well-studied object is located in the L1551 molecular cloud at a distance of 140\,pc. HST images taken in 1995 already revealed a complex system, showing a compact bipolar reflection nebula bisected by a dark lane (Fig.~\ref{fig:hh30}). HH~30 is actually an edge-on, flared optically thick circumstellar disc, about 450 AU in diameter. Noticeably, and reinforcing its status of prototype, HH~30 shows highly collimated bipolar emission-line jets. The analysis of the jet wiggling has recently led to the conclusion that the HH~30 system is a binary system \citep{Anglada07}. The energy source for these jets is believed to be the gravitational energy of the matter accreted by the central protostars, and their collimation is believed to be caused by a helical magnetic field that is coupled either to the inner part of a circumstellar disc or to the central binary. It seems almost certain that both rotation and magnetic fields are involved in the origin of the bipolar outflows.

YSOs and their protoplanetary discs are dynamical systems. We see them evolving on time-scales of a few years \citep[Fig.~\ref{fig:hh30}]{Watson07}. Variability is thought to originate from the accreting star (or binary system) itself, which would explain uniform photometric variability, and from something close to the star that would create non-uniform asymmetric morphological variability.

Very few exhaustive surveys of YSOs actually exist, which can be mainly explained by the observational difficulties. \citet{Connelley07} have, nevertheless, presented a unique sample of nearby Class~I YSOs. Their 106-YSO atlas, compiled with the modest 2.2-metre telescope of the University of Hawaii, is not biased towards well known star forming regions or famous individual objects. The general trend from this imaging survey is that the near-infrared luminosities and physical sizes of the nebulae increase with the bolometric luminosity of the illuminating sources. The great variety of nebulae makes it difficult to classify them based on morphology. However, as expected, a strong correlation between the spectral index $\alpha$ (slope of the linear regression through the infrared SED data points) and the flux ratio between the nebulae and the central star is noted.

The subtle observational variety in edge-on YSOs (characteristic scales of the dark lane, relative brightness of the bipolar reflection nebulae, etc.) appears to be entirely due to variations in disc mass \citep{Padgett99}. It seems plausible that once the envelope has dispersed, the disc begins to diminish in mass owing to accretion onto the forming star and planetary system, as well as associated outflow events.

	\subsubsection{Large scale structure of T Tauri prototypes} \label{sub:ttauridiscs}

The majority of classical TTS (CTTS) have been known for quite a long time to possess infrared excess-producing circumstellar material \citep{Strom89}. For weak-line TTS (WTTS), considered to be the missing link between YSOs and more evolved debris discs, the fraction of objects presenting such excesses is much smaller, especially for those objects that are located outside major star forming clouds \citep{Padgett06}\footnote{The infrared excess frequency increases by a factor between 3 and 6 for WTTS located in major star forming clouds, but remains significantly lower than for CTTS \citep{Cieza07}.}. About a dozen TTS discs have been imaged at high angular resolution in the optical regime (see Table~\ref{tab:ttauri}). The most interesting results in terms of morphological characterisation have been obtained by the Hubble Space Telescope, chronologically using better performing instruments: WFC2, NICMOS, STIS and ACS. The typical resolution of HST for this kind of object is of the order of 10\,AU, a scale at which the large-scale morphology of the circumstellar material becomes observable.

\begin{table}[t]
\caption{Discs around TTS that have been imaged in scattered light (see http://www.circumstellardisks.org/ for more details and an up-to-date list).}
\centering \label{tab:ttauri}
\begin{tabular}{cccccc}
\hline\noalign{\smallskip}
Name & Type & Dist.\ & Diam.\ & Incl.\  &Ref.\\
     &      & (pc)   & (AU)   & (deg)   &  \\
\tableheadseprule\noalign{\smallskip}
PDS 70	          	&K5	   &140	&350	 	&62	&\citet{Riaud06}\\
HH 30	        		&K7	   &140	&420	 	&83	&\citet{Burrows96}\\
GG Tau	         	&K7	   &140	&800 	&37	&\citet{Guilloteau99}\\
TW Hya	          	&K7	   &56	&448	 	&0	&\citet{Krist00}\\
GM Aur	        		&K7	   &140 	&700 	&54	&\citet{Schneider03}\\
CoKu Tau 1		&M0	   &140	&896		 & ... &\citet{Padgett99}\\
LkHa 263C		&M0	   &275	&302	 	 &87	&\citet{Jayawardhana02}\\
HV Tau C		 	&M0	   &140      &168	&84	&\citet{Stapelfeldt03}\\
IM Lup 	        		&M0	   &150	&210	 	&50  &\citet{Pinte08b}\\
HK Tau B	         	&M1	   &140	&210	 	&84	&\citet{Stapelfeldt98}\\
DG Tau B	         	& ... 	   &140	&550		&75	&\citet{Padgett99}\\
IRAS 04325+2402 	& ... 	   &140	&60	         &83	&\citet{Hartmann99}\\
ASR 41	         	& ... 	   &316 	&6320	&80	&\citet{Hodapp04}\\
\noalign{\smallskip}\hline
\end{tabular}
\end{table}

Perhaps one of the most significant discs imaged around a directly visible TTS is that of TW~Hydrae, a K8V star approximately 55\,pc away in the constellation of Hydra. This star is the closest TTS to the solar system, which makes it one of the most studied targets in its class. TW Hydrae is similar in mass to the Sun, but is only about 5--10\,Myr old. The star appears to be accreting from a face-on protoplanetary disc of dust and gas, which has been resolved in images from the Hubble Space Telescope \citep{Krist00}. TW~Hydrae is accompanied by about twenty other low-mass stars with similar ages and space motions, which form the TW~Hydrae association or TWA, one of the closest regions of recent ``fossil'' star-formation.

\begin{figure}[t]
\centering
\includegraphics[width=1.0\textwidth]{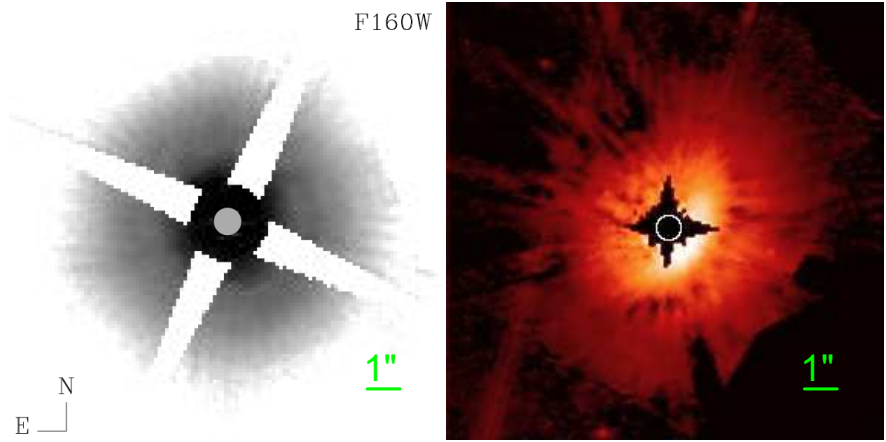}
\caption{Left: PSF-subtracted, roll-combined HST/NICMOS image of the face-on disc around TW~Hya in the F160W filter \citep[credit:][reproduced by permission of the AAS]{Weinberger02}. The disc becomes visible just outside the coronagraph (0.3") and continues out to a radius of 4" ($\sim $230 AU). The noise is dominated by subtraction residuals in the disc. Right: The disc around IM Lup, as seen by NICMOS at F160W, after PSF subtraction \citep[credit:][reproduced with permission \copyright ESO]{Pinte08b}. Both images are shown on a logarithmic stretch.} \label{fig:ttauri}
\end{figure}

HST observations of this isolated TTS revealed the presence of a compact circumstellar nebula (Fig.~\ref{fig:ttauri}). After subtraction of a reference point-spread function (PSF)---a widely used technique, with or without a coronagraph---a smooth, symmetrical, circular halo can be seen in both R- and I-band WFPC2 images \citep{Krist00}. Its intensity at those wavelengths declines with radius until reaching an outer sensitivity limit at 3.5" (200 AU). The radial brightness profile of the face-on disc is complex and can be described with multiple, contiguous zones with individual power-law intensity relations. The halo appears slightly blue relative to the star, especially in the outer zones. The zonal structure found in the disc could arise from radial variations in the dust properties that determine the local equilibrium temperature or perhaps via dynamical effects of unseen companions.

Stellar light scattered from the optically thick dust disc is also seen from 20 to 230\,AU in the J and H bands, from the ground \citep{Trilling01}, and from space \citep{Weinberger02}. The surface brightness seen with HST/NICMOS declines following the power law $r^{-2.6\pm0.1}$ between 45 and 150 AU. The scattering profile indicates that the disc is flared, not geometrically flat, showing a sharp break in slope at a radius of 150 AU. The disc also presents a strong polarisation signature between 0.1'' and 1.4'' (i.e., 5 to 77\,AU) from the star \citep{Apai04}. The polarised intensity is consistent with shorter wavelength surface brightness observations. Polarimetry is an excellent imaging technique to be used from the ground. Simultaneous imaging in orthogonal polarisation channels allows unpolarised atmospheric turbulence effects to be removed, while gaining further knowledge of the disc geometry dust grain properties, since polarisation is caused by scattering and its amplitude depends on the scatterer geometry with respect to the emitter.

\citet{Roberge05} also show evidence that the disc around TW~Hya is elliptical between 65 and 140\,AU and circular beyond this region, consistent with the previously detected break in the surface brightness distribution at 150\,AU from the central star. This observation suggests that the disc is warped, the deformation being presumably caused by perturbing bodies.

The case of IM~Lup is also particularly interesting. This star has recently been the subject of the most exhaustive study so far, using a novel method that couples multi-wavelength spatially-resolved and spectral data into a single data set, fits the whole data set with a single general radiative transfer model (MCFOST), and uses a Bayesian approach to constrain the large parameter space \citep{Pinte08b}. This study certainly represents the future of circumstellar disc analysis, but necessitates a wealth of data which is not always available. The results on the IM Lup case presented in \citet{Pinte08b} are impressive, with a single model that can reproduce satisfactorily all the observations of the disc. This analysis illustrates the importance of combining a wide range of observations in order to fully constrain the disc model, with each observation providing a strong constraint only on some aspects of the disc structure and dust content. PSF subtracted images reveal, at all wavelengths, a compact nebula adjacent to the SW of the star (Fig.~\ref{fig:ttauri}). Morphologically, the nebula is a broad, symmetrical arc nearly 4" in size. The nebulosity is brightest to the SW, but is clearly seen circumscribing the star. A strong front/back asymmetry is detected in F606W, F814W, and F160W indicating that it is dominated by scattered light from the star. To the SW of the elliptically shaped circumstellar nebulosity, an arc-like dark band (presumably the higher opacity disc midplane centred at 1.5" along the disc morphological minor axis), bifurcates into an isophotally concentric lower surface brightness scattering region (presumably the ``lower'' scattering surface of the back side of the disc) extending to 2.5" on the morphological minor axis of the disc. This feature is most obvious in the F814W and F160W images. The dark lane between the upper and lower scattered light nebulae is best seen 1.4" southwest of the star. This coincides with the probable forward scattering direction given the purported disc inclination in the range 40 -- $60^\circ$.

\begin{figure}[t]
\centering
\includegraphics[width=0.9\textwidth]{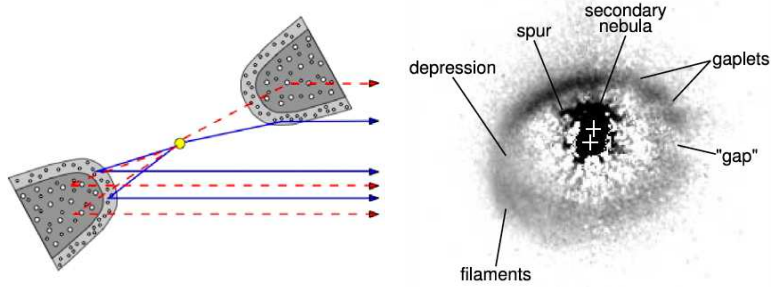}
\caption{\textit{Left} Sketch of the scattering geometry of the circumbinary disc around GG Tau \citep[reprinted from][with permission from Elsevier \copyright]{Duchene08}. \textit{Right} HST/ACS image of the same disc \citep[credit:][reproduced by permission of the AAS]{Krist05b}. In this representative scattering example, with the observer on the right, the scattering anisotropy between the top/front part and the bottom/back one is well illustrated.} \label{fig:ggTau}
\end{figure}

GG~Tau is another particular example of TTS disc. Its unique feature is its confirmed circumbinarity. \citet{Krist05b} used HST/ACS to take the most detailed images of this system to date (Fig.~\ref{fig:ggTau}). They confirmed features in the disc including a gap apparently caused by shadowing from circumstellar material, an asymmetric distribution of light on the near edge of the disc, enhanced brightness along the near edge of the disc due to forward scattering, and a compact reflection nebula near the secondary star. New features are seen in the ACS images: two short filaments along the disc, localised but strong variations in the disc intensity (``gaplets''), and a ``spur'' or filament extending from the reflection nebulosity near the secondary (Fig.~\ref{fig:ggTau}). The brightness asymmetries along the disc suggest that it is asymmetrically illuminated by the stars due to extinction by nonuniform circumstellar material or that the illuminated surface of the disc is warped by tidal effects (or perhaps both). Localised, time-dependent brightness variations in the disc are also seen.

Generally, the large-scale structure of discs seen around TTS in scattered light is dominated by inclination effects (edge-on vs face-on, or intermediate inclination regimes). The variety is, therefore, mainly due to the observer viewpoint with respect to the disc orientation. Most of the observed disc structures, when deconvolved of projection effects, have in common an optically thick flared geometry, consistent with hydrostatic equilibrium. When interpreting images of such objects, as already noted, one must keep in mind that protoplanetary discs are optically thick. Subsequently, the scattered photons have only interacted with the uppermost layers of the disc.

		\subsubsection{Large scale structure of Herbig Ae/Be disc prototypes} \label{sub:haebediscs}

\begin{table}[t]
\caption{Discs around HAeBe stars that have been imaged in scattered light (see http://www.circumstellardisks.org/ for more details and an up-to-date listing). }
\centering \label{tab:Haebe}
\begin{tabular}{cccccc}
\hline\noalign{\smallskip}
Name & Type & Dist.\ & Diam.\ & Incl.\  &Ref.\\
     &      & (pc)   & (AU)   & (deg)   &  \\
\tableheadseprule\noalign{\smallskip}
AB Aur		&A0e	&144		&2592	&21.5 &\citet{Grady99}\\
HD 163296	&A1e	&122		&902		&60	 &\citet{Grady00}\\
HD 150193A	&A2IV	&150		&390		&38	&\citet{Fukagawa03}\\
PDS 144N	&A2IV	&1000	&800		&83	&\citet{Perrin2006}\\
HD 169142	&B9		&145		&435		&0	 &\citet{Kuhn01}\\
HD 100546	&B9		&103		&721		&51	&\citet{Grady01}\\
HD 142527	&F6		&140		&980		&30	 &\citet{Fukagawa06}\\
\noalign{\smallskip}\hline
\end{tabular}
\end{table}

HAeBe stars are excellent targets for disc imaging since they are bright and massive, and susceptible of possessing large amounts of circumstellar material. Table~\ref{tab:Haebe} lists the protoplanetary discs around HAeBe stars that have been resolved through their near-infrared scattered light. Thermal emission from such discs has also been resolved in the mid-infrared regime using ground-based telescopes in a few cases. One of the most striking indications for a flared disc structure around a group~I HAe star actually comes from this type of observations, when \citet{Lagage06} have used the VISIR instrument at the VLT to study the protoplanetary disc around HD~97048. By choosing a filter centred around a prominent Polycyclic Aromatic Hydrocarbon (PAH) emission band at 8.6\,$\mu$m, they have been able to probe the disc surface layer up to about 600\,AU from the central star. A strong east-west asymmetry is detected: the brightness isophotal contours are elliptical in shape, and the ellipse centres are offset from the peak of emission with the offset increasing at lower fluxes. Such features are characteristic of a flaring disc (i.e., with a scale height $H(r)=H_0(r/r_0)^{\beta}$), vertically optically thick and inclined to the line of sight. Modelling the isophotal contours provides a flaring power-law exponent $\beta = 1.26 \pm 0.05$, consistent with hydrostatic, radiative equilibrium models of passive flared discs. This study supports the classification of \citet{Meeus01}, where group~I HAeBe objects are associated with flared discs. Other mid-infrared images of HAe discs \citep{Doucet06} seem to further confirm this classification, with strong disc flaring found only around group~I objects.

\begin{figure}[t]
\centering
\includegraphics[width=1.0\textwidth]{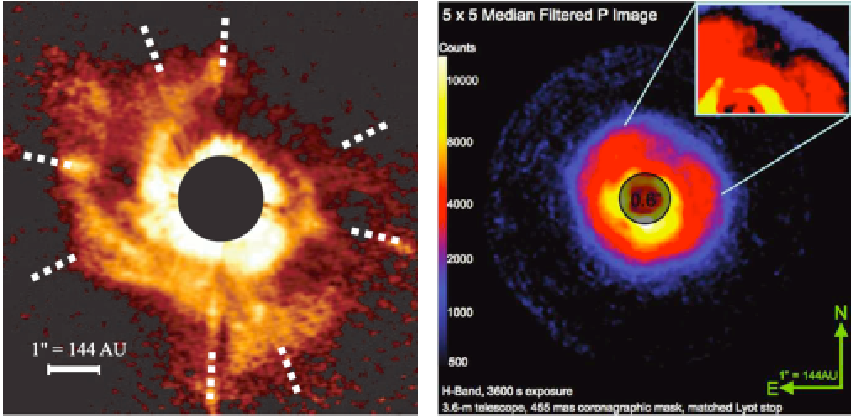}
\caption{\textit{Left}: Subaru H-band image of the circumstellar disc around AB~Aur \citep[credit:][reproduced by permission of the AAS]{Fukagawa04}. Directions of the spider patterns are indicated by dashed lines. \textit{Right}: polarimetric differential imaging of the central part of the same system seen through the Lyot project coronagraph \citep[credit:][reproduced by permission of the AAS]{Oppenheimer08}.} \label{fig:HAeBe}
\end{figure}

Because it is bright ($V=7$) and close (144\,pc), the prototype protoplanetary disc around the young star AB~Aur is presumably one of the best studied HAeBe stars. This star belongs to the Taurus-Auriga star forming region, about 1--3\,Myr old. Like many other HAeBe stars, AB~Aur shows a variety of phenomena and a wealth of structures in its disc. It is surrounded by a large reflection nebula extending thousands of AUs from the star \citep{Grady99}. At millimetre wavelengths, an annular region 100--750 AU radius indicates the presence of molecular gas of CO, HCO, H$_2$. The gas melange seems to be depleted closer-in in favour of dust, which is a direct consequence of the strong radiation pressure and wind from the massive central A0Ve star. According to \citet{Fukagawa04}, who took images at near-infrared wavelengths using adaptive optics coronagraphy on the Subaru telescope, the inner dust disc extends within 130\,AU, spiraling inwards down to a few AUs (Fig.~\ref{fig:HAeBe}). Recently, in the framework of the Lyot project on the AEOS telescope at Maui (the first on-sky extreme adaptive optics systems), \citet{Oppenheimer08} imaged the dust in an annulus between 43 and 300\,AU in polarised light. Remarkably, they detected a region of lower polarised intensity towards the position angle ${\rm PA} \approx 330^\circ$, which was interpreted as an azimuthal physical gap at a radius of 102\,AU, presumably suggesting the presence of an unseen low-mass companion at this orbital distance (Fig.~\ref{fig:HAeBe}). However, very recent polarisation images taken with NICMOS reveal the same pattern of polarised intensity, but also provide a measurement of total intensity (M.~Perrin et al., in preparation). The ratio of polarised intensity and total intensity represents the polarisation fraction, which is a more meaningful physical quantity. Based on this measurement, the darker area is now interpreted as a region of lower polarisation, and not a physical gap. The region of lower polarisation is consistent with the back-scattered light polarisation fraction as predicted by Mie scattering theory. The gap interpretation is subsequently discarded. As noted by M.~Perrin et al., this cautionary tale indicates the need for measuring the polarisation fraction in order to draw firm conclusions about the nature of disk geometries or scattering bodies.

Another typical object belonging to the HAeBe family is PDS~144. This system consists of a $V\sim13$ primary (south) and a fainter companion (north), with the spectra of both stars showing evidence for circumstellar discs and accretion. In Lick adaptive optics polarimetric imaging studies of this system, \citet{Perrin2004,Perrin2006} resolved the extended polarised light scattered from dust around the northern star. Follow-up Keck adaptive optics and mid-infrared observations showed that this star is entirely hidden by an optically thick disc at all wavelengths from 1.2 to 11.7 microns. The disc major axis subtends $\sim0.8$" on the sky, corresponding to $\sim 800$ AU at a distance of 1000 pc. Bright ``wings'' extend 0.3" above and below the disc ansae, most likely due to scattering from the edges of an outflow cavity in a circumstellar envelope.

HD\,100546, yet another HAeBe star, possesses a disc showing two bright apparent spiral arms \citep{Grady01}, and some weaker ones only recently observed with HST/ACS \citep{Ardila07}. \citet{Quillen06a} first proposed that the spiral structures could be caused by the illumination of a warped outer disc outside 200\,AU, which could be the result of precession induced by a Jupiter-mass body embedded in the disc. However, \citet{Ardila07} argue that only the geometrically thick, optically thin envelope of the disc is seen, while the optically thick disc responsible for the far-IR emission is undetected. The observed spiral arms are then structures on this envelope. The disc colours, indicating that the extended nebulosity is not a remnant of the infalling envelope but reprocessed disc material, are similar to those derived for Kuiper Belt objects, suggesting that the same processes responsible for their colours may be at work here.

The general trend for discs around intermediate-mass HAeBe stars follows that of TTS except for the greater variety of disc substructures. The larger mass of those systems certainly triggers dynamical effects that can explain the wealth of features (asymmetries, spirals, etc.). But again, because young discs are optically thick, caution must always be applied to the interpretation of features. It may be tempting to conclude that they trace spiral density waves generated by a protoplanet formation in the disc, but optical effects, due to opacity and the propagation geometry, can generally explain the same phenomenon \citep{Duchene08}. Moreover, enhanced density does not necessarily mean increased scattered light flux in an optically thick medium.


    \subsection{The interferometric view of the first ten AUs} \label{sub:ysointerf}

Long-baseline infrared interferometry provides a suitable resolving power to explore the innermost part of proto-planetary discs: assuming a $B=100$\,m baseline, the angular resolution $\lambda/2B$ ranges between about 2\,mas in the near-infrared and 10\,mas in the mid-infrared, which corresponds to a linear resolution between 0.25\,AU and 1.5\,AU at the distance of the nearest star forming region (Taurus-Aurigae, 140\,pc). Interferometry is therefore one of the most appropriate tools to study the physical phenomena at play in this most important region where star-disc interactions are taking place.

The most straightforward information brought by interferometric observations is an estimation of
the size of the emitting region at the observing wavelength, which can be directly derived from
simple visibility measurements assuming a given model for the object brightness distribution. Since a large part of the infrared flux in YSOs is produced by the circumstellar disc, interferometry can determine the size of the main emitting region within the disc, and thereby constrain theoretical models.

The first successful observation of a YSO with stellar interferometry was reported by \citet{Malbet98}, who resolved the young outbursting star FU~Ori on a 103-m long baseline in the $K$ band with the Palomar Testbed Interferometer (PTI). Although the single visibility measurement presented in this paper only provides limited constraints on the size and morphology of the circumstellar disc, an accretion disc model of the type proposed by \citet{Hartmann85} was shown to be in good agreement with both the interferometric observations and the SED of FU~Ori. The success of standard accretion disc models to reproduce interferometric observations of YSOs was however short-lived, as \citet{MillanGabet99} soon obtained resolved interferometric observations of the Herbig~Ae star AB~Aur with the Infrared Optical Telescope Array (IOTA). Fitting simple geometric models to these observations indicates a characteristic size of 5\,mas (0.7\,AU) for the $K$-band emission, significantly larger than predicted by classical accretion disc models such as those proposed by \citet{Hillenbrand92}. This early result, suggesting the presence of a large inner hole (or of an optically thin cavity), was then confirmed on a larger sample of 15~HAeBe stars by \citet{MillanGabet01}, and extended to two TTS by \citet{Akeson00}.

        \subsubsection{Inner cavities in HAeBe dust discs}

\begin{figure}[t]
\centering
\includegraphics[width=0.9\textwidth]{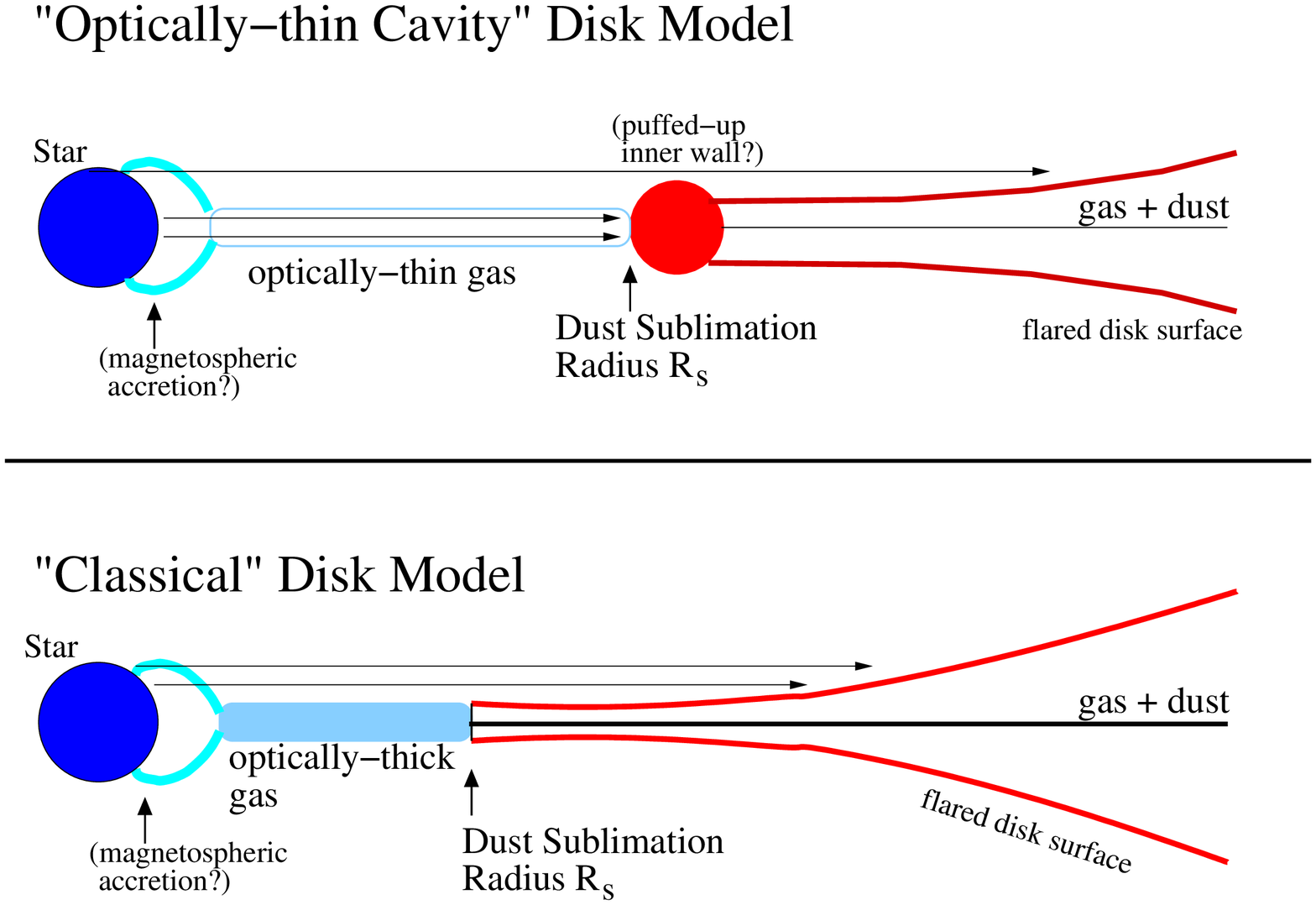}
\caption{\textit{Top}: Cross section in the inner disc region for the ``inner cavity'' disc model proposed by \citet{Monnier05} to explain interferometric observations of HAeBe stars. The presence of a puffed-up inner rim is predicted by theoretical models \citep{Natta01,Dullemond01}. \textit{Bottom}: Classical accretion disc model. The presence of optically thick gas in the midplane partially shields the innermost dust from stellar radiation, causing the dust sublimation radius to shrink for the same sublimation temperature. Credit: \citet{Monnier05}, reproduced by permission of the AAS.} \label{fig:innerdisc}
\end{figure}

Direct evidence for the presence of inner holes in the circumstellar discs around HAeBe stars was provided by interferometric images of LkH$\alpha$~101 \citep{Tuthill01}, one of the brightest Herbig~Be stars in the near-infrared. These images were reconstructed from visibilities and closure phases obtained by placing an annulus-shaped aperture mask close to the secondary of the Keck~I telescope, thereby providing diffraction-limited imaging in the $H$ and $K$ bands \citep{Tuthill00}. Based on these observations, the authors propose that the position of the disc inner edge is related to the sublimation of dust grains by direct stellar radiation, rather than to disc-reprocessing or viscous heating processes. By modelling state-of-the-art SEDs of four Herbig~Ae stars, \citet{Natta01} independently proposed a similar scenario, where dust evaporation clears up the inner disc and produces a puffed-up inner wall of optically thick dust at the sublimation radius. This scenario, illustrated in Fig.~\ref{fig:innerdisc}, was then further developed in a seminal paper by \citet{Dullemond01}, where the description of the disc structure is formalised, based on the flaring disc models of \citet{Chiang97}.

\begin{figure}[t]
\centering
\includegraphics[width=1.0\textwidth]{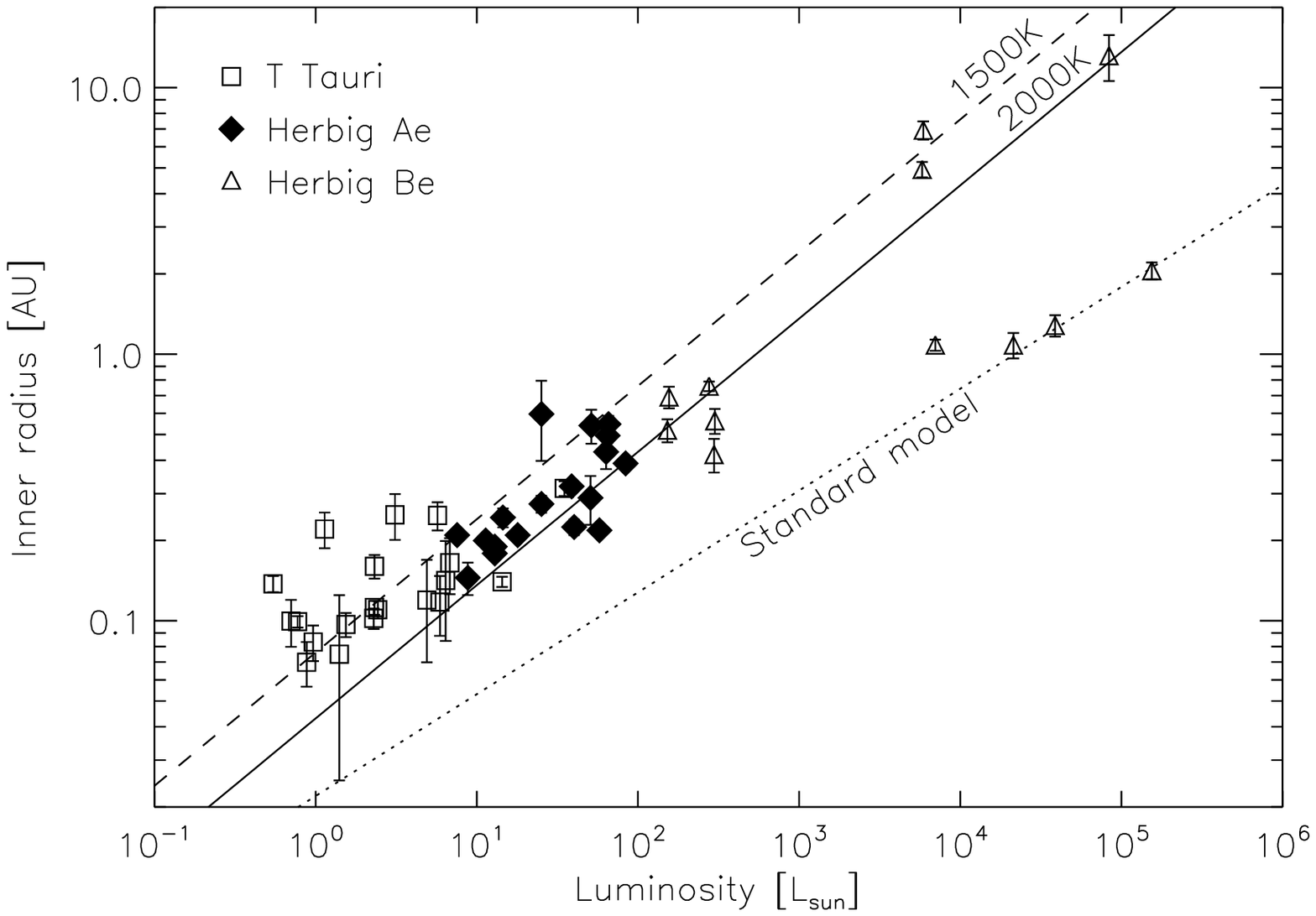}
\caption{Measured sizes of HAeBe and T~Tauri objects from the literature plotted as a function of the central object's luminosity (stellar + accretion). Observations are generally inconsistent with a standard accretion disc model \citep{Hillenbrand92}, and are more plausibly reproduced by a puffed-up inner rim model with an evacuated central cavity \citep[][here plotted for two different dust sublimation temperatures, following]{Dullemond01}. A significant deviation from the expected inner dust radius of puffed-up models is however noted at large luminosities, where observations are more consistent with standard accretion disc models. The inner radii of TTS discs also seem to disagree with puffed-up rim model predictions at low masses.} \label{fig:sizelum}
\end{figure}

This new circumstellar disc model has been tested and refined in many ways during the past years, and interferometric observations have played a leading role in this context. One of the first tests was to investigate the colour-dependence of the measured inner disc radii. \citet{MillanGabet01} found that the sizes were similar in the $H$ and $K$ bands, as predicted by a spatially localised emission region. This result was also confirmed in the mid-infrared regime by \citet{Tuthill02} in the case of the bright Herbig Be star LkH$\alpha$~101. Another important test was performed by \citet{Monnier02}, who examined the correlation between near-infrared disc size and stellar luminosity for HAeBe stars. Even though this first version of the now famous size-luminosity diagram showed a large scatter in its data points, the authors clearly showed that the dust inner radius increases with stellar luminosity following a power-law close to the expected $R_{\rm sub} \propto L_{\ast}^{1/2}$ relation for the dust sublimation radius, and that the measured near-infrared sizes are consistent with the dust sublimation radii of relatively large dust grains ($\ge 1$\,$\mu$m) with sublimation temperatures around 1500\,K. The size-luminosity diagram, displayed in Fig.~\ref{fig:sizelum} for an enlarged sample, was however already showing some discrepancies with the inner wall model in the case of high-luminosity Herbig Be stars, for which the dust sublimation radius generally overestimates the disc inner radius measured by interferometry. This trend was confirmed (with much less scatter in the diagram) by subsequent surveys of HAeBe stars, suggesting that the inner wall model should be replaced by standard accretion models in the case of high-luminosity ($> 10^3 L_{\odot}$) Herbig Be stars
\citep{Eisner03,Eisner04,Monnier05}. This distinction is assumed to be due to enhanced inner gas
disc, which may shield the inner dust disc and thereby lead to reduced dust sublimation distance
\citep{Monnier02}, or even extend down to the stellar surface due to increased accretion rate and/or reduced stellar magnetic fields \citep{Eisner04}.

Further constraints on the inner wall model have been obtained with the first closure phase survey of HAeBe stars \citep{Monnier06}, which has shown only small departures from centrosymmetry in the brightness distribution of the inner disc. This result is consistent with the puffed-up inner wall model, provided that the inner rim is smoothly curved as predicted by \citet{Isella05}. Vertical
inner rims are incompatible with the data as they would produce large skewness in the brightness
distribution, even for moderate inclinations. The curvature of the inner rim is a natural
consequence of the dependence of grain evaporation temperature on density, due to the large
vertical density gradient. \citet{Isella06} further show that archival visibility measurements of
Herbig~Ae and late Herbig Be stars are consistent with this model. However, it must be noted that
the puffed-up inner rim wall is still a controversial topic, and that it is not the only model to simultaneously reproduce the interferometric and SED observations: a compact ($\le 10$\,AU) optically thin dusty halo around the disc inner region is another possible explanation \citep{Vinkovic06}. Such a halo could potentially be produced by a dusty outflow above the disc \citep{Vinkovic07}.

        \subsubsection{Larger cavities for T Tauri discs?}

First observations of a few bright T~Tauri objects with small-aperture interferometers have shown that, as in the case of HAeBe stars, the inner disc radii are significantly larger than those inferred from classical accretion disc models, and are rather consistent with the estimated dust sublimation radii predicted by puffed-up inner wall models \citep{Akeson00,Akeson02,Akeson05a}. However, the analysis of a larger sample of TTS with more typical luminosities using the Keck Interferometer (KI) soon revealed that the inner radii inferred from simple geometric models (e.g., uniform disc or ring) are often even larger than the expected dust sublimation radii by a factor up to 2--3 \citep{Colavita03,Eisner05,Akeson05b}.

Even though part of the discrepancy can be explained by taking into account the accretion luminosity in addition to the intrinsic stellar luminosity \citep{Muzerolle03}, a large dispersion remains in the inner radii of T~Tauri discs. This spread could directly result from different dust properties around the observed stars. In particular, smaller grains are expected to lead to larger dust sublimation distances \citep[e.g.][]{Monnier02}. Alternatively, the spread could be due to the evolutionary status of the objects. This hypothesis is backed up by the fact that the ratio of measured to predicted inner radii increases for smaller accretion rates \citep{Akeson05b}, so that the inner cavity might be progressively cleared up while the accretion rate decreases, assuming that objects with lower accretion rates are older \citep{Hartmann98}. \citet{Eisner07b} further proposes that the dust disc around TTS with low accretion luminosity may actually be truncated by the stellar magnetic field, as lower accretion rates allow the magnetospheric radius to extend further away from the star.

Very recently, a re-analysis of all interferometric measurements of TTS by \citet{Pinte08a} has shown that the spread in the inner disc sizes estimated with simple geometric models may be entirely due to the effect of scattered light, which is negligible for HAeBe stars but not for cooler TTS. The low interferometric visibilities recorded on several T~Tauri objects may therefore be due to the contribution of extended scattered emission. In that case, the inner disc radii estimated from dust sublimation temperature are entirely compatible with the interferometric observations, so that including additional physical phenomena to reproduce them is not required any more.

Particular among T~Tauri objects are the young outbursting FU~Ori stars. The first observations of FU~Ori itself by \citet{Malbet98}, which supported classical accretion disc models, was later confirmed by \citet{Malbet05} and \citet{Quanz06}, who have collected additional visibility measurements of FU~Ori with various interferometers and at various wavelengths. Although accretion discs alone could not reproduce the observations of three other FU~Ori type objects by \citet{MillanGabet06a}, who suggested that additional extended dusty structures must be present \citep[as already proposed by][]{Kenyon91}, the structure of outbursting YSO discs seems significantly different from quiet YSO discs. Further observations of such objects are required to disentangle the contributions of discs, envelopes and putative close companions, and thereby perform detailed test of accretion models.

        \subsubsection{Middle disc structure and composition}

While near-infrared interferometry is most sensitive to the inner rim of circumstellar discs, mid-infrared observations on 50--100\,m baselines have the potential to unveil the structure of the disc in the 1--10\,AU region thanks to a lower spatial resolution and to a better sensitivity to colder dust (at a few hundred Kelvins). First interferometric observations in the mid-infrared have however been strongly limited in baseline length, as they were either based on single-aperture experiments such as the BLINC nulling interferometer \citep{Hinz01,Liu03}, or performed with the $\sim$10-m baseline Infrared Spatial Interferometer \citep[ISI,][]{Tuthill02}. They have nonetheless resolved a few of the brightest HAeBe stars, and shown that simple flared models might not always be appropriate to describe the disc geometry.

The advent of long-baseline mid-infrared instruments such as VLTI/MIDI or KI/Nuller has provided a new view of circumstellar disc structure and of the dust chemical composition in the disc ``atmosphere''. In particular, the low spectral resolution offered by MIDI across the $N$ band provides a way to disentangle the contribution of these two effects. The first observations with MIDI, performed by \citet{Leinert04} on a sample of seven HAeBe objects, have shown that the characteristic mid-infrared size of circumstellar discs around intermediate-mass stars is in the range 1--10\,AU. Moreover, these observations have shown that the reddest objects in the sample are also those with the largest size, a result consistent with the classification of \citet{Meeus01}, in the sense that group~II objects, which are assumed to be self-shadowed, have a smaller mid-infrared excess and a smaller size as they are dominated by the bright inner rim, while the flared group~I objects have larger mid-infrared excesses and also larger mid-infrared sizes as stellar light directly reaches the disc (see illustration in Fig.~\ref{fig:meeus}). Additional observations of the Herbig~Ae star HR~5999 are in broad agreement with the size/colour correlation, although they show a rather complex structure with a potentially truncated disc \citep{Preibisch06}. It is expected that further mid-infrared observations of Herbig~Ae stars will definitively answer the question of the presence of puffed-up inner walls and confirm the hypothesis that group~I and group~II sources correspond to flaring and self-shadowed discs \citep{vanBoekel05}.

The size/colour correlation might also extend to lower mass stars, as shown in the case of an eruptive TTS by \citet{Abraham06}. However, \citet{Schegerer08} show that the presence of a puffed-up wall is not mandatory to reproduce the mid-infrared observations of another TTS, RY~Tau. Using self-consistent radiative transfer modelling, the authors fit both the SED and the interferometric data with an active accretion disc, possibly surrounded by an optically thin dusty envelope as suggested by \citet{Vinkovic06}. MIDI observations of FU~Ori \citep{Quanz06} also show that a flat accretion disc is appropriate to reproduce both interferometric and photometric observations without any additional component (either a puffed-up wall or an extended dusty envelope), as already suggested by \citet{Malbet98,Malbet05}.

Besides unveiling the middle disc structure, mid-infrared interferometric observations have provided new constraints on the dust composition and its gradient within the disc. By comparing spectrally dispersed visibility measurements at different baselines with an unresolved spectrum of the whole disc, and analysing the shape of the silicate features in the spectra, \citet{vanBoekel04} have shown that the dust grains within the first two AU of three Herbig~Ae stars are significantly more crystalline and significantly larger than in the outer (2--20\,AU) disc region. Combined with the very young ages of the observed objects, these observations imply that crystallisation of almost the entire inner disc takes place very early in the disc history, most probably well before the onset of planet formation. Moreover, the spectral shape of the inner disc spectra shows remarkable similarity with the spectra of outer solar system comets, suggesting that the building blocks of planetesimals are formed early in the inner part of circumstellar discs, and are at some point transported outward by radial mixing as suggested by theoretical models \citep{Gail04}. The gradient of dust composition (more olivine and less pyroxene in the inner disc) observed towards one of the three objects is also in qualitative agreement with such models.

More recently, spectrally dispersed mid-infrared observations have also been obtained for TW~Hya and RY~Tau, two classical TTS \citep{Ratzka07,Schegerer08}. They suggest that crystalline grains are mostly located in the innermost part of the circumstellar discs as in the case of Herbig~Ae stars. On the other hand, in the peculiar case of FU~Ori, the MIDI observations obtained by \citet{Quanz06} do not show any evidence of crystalline grains, which could be related to the very early evolutionary stage of this object. This hypothesis is however contradicted by the fact that some level of dust processing (in particular, grain growth) seems to have already taken place.

        \subsubsection{Gas: accretion and winds}

A number of the aforementioned studies have suggested that the presence of gas and related accretion processes in the inner disc of YSOs may play a significant role in the interpretation of interferometric observations. Detecting and locating circumstellar gas in the inner region has therefore become a major subject of study during the past couple of years. The advent of multi-colour and/or spectrally dispersed interferometric observations has played a preponderant role in this context.

A first type of constraint on the presence of an inner gas accretion disc comes from enhanced studies of the dust sublimation region in the near-infrared. \citet{Eisner07a} obtained resolved visibility measurements of 11~HAeBe stars with the PTI across five spectral channels in the $K$ band, and showed a global increase of the infrared emission region size as a function of wavelength. This result is inconsistent with a single-temperature ring at a fixed radius, as expected from inner wall models, and rather suggests a two-ring model (which also nicely fits the SEDs). The best-fit temperatures of the two rings support a model where the inner ring emission traces hot gas, while the outer ring emission arises from dust. The low spectral resolution used in this study did not provide evidence for spectral line emission of CO or H$_2$O. Similar results have subsequently been obtained on a number of HAeBe stars with various instruments. \citet{Kraus08a} combined VLTI/AMBER, PTI and VLTI/MIDI data to show that the characteristic emission size of the Herbig~Be star MWC~147 increases strongly towards longer wavelength and is 2--4 times smaller than the size predicted by dust sublimation. The addition of optically thick gas emission from an active gaseous disc inside the dust sublimation zone is proposed to fit both the interferometric data and the SED, so that MWC~147 may be dominated by accretion luminosity emerging from an inner gaseous disc. \citet{Isella08} and \citet{Tatulli08} came to similar conclusions respectively for a HAe and a late HBe star, as their observations show significant emission within the expected dust sublimation radius. Using significantly longer baselines ($\sim$300\,m) at the CHARA Array, \citet{Tannirkulam08a,Tannirkulam08b} show that a single ring model is incompatible with the broadband observation of two HAe stars, because it would produce oscillations in the visibility curve which are not seen in the data. They also suggest that the inner gas disc proposed to explain the observations may partially shield the dust from direct stellar emission, and thereby reduce the dust sublimation distance as proposed by the model of \citet{Muzerolle04}. Very recently, spectro-interferometric observations at the KI \citep{Eisner09} further confirmed the two-component model on an enlarged sample of HAeBe and TTS, with hot gas within the dust sublimation ring needed to explain the increase of visibility as a function of wavelength. Among the possible sources for gaseous continuum emission, \citet{Eisner09} argue that free-free transitions of hydrogen is the most viable mechanism, but no firm conclusion can be given yet. All these observations seem to provide a smoother transition between the accretion dominated high-luminosity HBe stars and the lower luminosity HAeBe stars, whose near-infrared emission is now understood to arise from both dust and gas. Observations of a larger sample will however be needed to understand the trends in inner disc properties with respect to stellar type, stellar mass, accretion rate, etc.

A second type of constraint comes from the variation of complex visibility within the spectral lines associated to circumstellar gas. This requires a spectral resolution of at least a few hundred in the near-infrared, which has recently been enabled by VLTI/AMBER and by the new grism mode of the KI. The first AMBER observations of a YSO in medium spectral resolution (MR, $R \simeq 1500$) performed by \citet{Malbet07} on an early-type Herbig Be star (MWC~297) have shown that the visibility amplitude across the Br$\gamma$ line, tracing atomic hydrogen, is significantly smaller than in the continuum, which implies a larger size for the Br$\gamma$ emission region than for the dusty disc. This result was interpreted as the signature of an outflowing stellar radial wind, although the exact nature of the wind could not be constrained. The Br$\gamma$ emission line has subsequently been observed around a number of HAeBe stars with spectro-interferometry, showing a large diversity in physical size for the line emission region, and therefore most probably a large diversity in the emission mechanisms. In practice, the size of the Br$\gamma$ emitting region seems to be generally smaller than (or similar to) the size of the dusty disc emission region \citep{Tatulli07,Kraus08b,Eisner09}. Depending on its actual size, the Br$\gamma$ emission could originate either from a compact disc wind or stellar wind when larger than the corotation radius, or from magnetospheric accretion in the most compact cases. The latter hypothesis is supported by an apparent correlation between the H$\alpha$ line profile shape and the size of the Br$\gamma$ emission region. The unique behaviour of MWC~297, showing an increased size for the Br$\gamma$ emission region, could be due to the fact that the continuum emission is dominated by optically thick inner gas instead of pure dust emission. These observations imply that, for some HAeBe stars (at least), Br$\gamma$ is not a primary tracer of accretion, which makes the empirical correlation between the Br$\gamma$ luminosity and the accretion luminosity quite remarkable, and suggests a tight connection between accretion and ejection processes in YSOs \citep{Kraus08b}.

Besides atomic hydrogen, two molecular species have recently been resolved around HAeBe stars by infrared interferometry: water vapour around MWC~480 \citep[band at 2.0\,$\mu$m,][]{Eisner07} and carbon monoxide around 51~Oph \citep[band at 2.3\,$\mu$m,][]{Tatulli08}. CO emission is also tentatively observed towards several objects in the survey of \citet{Eisner09}. In all cases, the line emission seems more compact than the adjacent continuum emission, suggesting that it originates in a dust-free hot gaseous disc. The replenishment of molecular gas is supposedly due to the infall of icy bodies in the case of H$_2$O, or to chemical reactions in the case of CO. Self-shielding may also be invoked in the case of CO molecules, provided that the column density is high enough. Note however that follow-up high-resolution $K$-band spectroscopy of MWC\,480 has not confirmed the presence of water vapour in the inner disc \citep{Najita09}, and strongly questions the interpretation of spectrally dispersed visibilities by \citet{Eisner07}.


\section{Transition from protoplanetary to debris discs} \label{sec:transition}

The so-called ``transition'' discs have been identified for about twenty years \citep{Strom89,Skrutskie90}, but have only recently become well-defined: one of the first thorough investigation of a transition object was actually done by \citet{Calvet02} on TW~Hya, suggesting the presence of an inner hole that was later on confirmed by millimetre imaging \citep{Hughes07}. Transition objects are characterised by a low dust content in their inner disc regions ($<1-10$\,AU), based on the absence of significant near-infrared excess combined with strong excess emission at longer wavelengths. When compared to standard Class~II objects, this is suggestive of an evolved nature, where the inner disc is depleted while the outer disc is still optically thick.

The evolutionary processes transforming massive, gas-rich circumstellar discs into tenuous, gas-poor debris discs are still not well understood. Several mechanisms for disc dispersal have been proposed to explain the apparent inside-out dispersal of protoplanetary discs:
\begin{itemize}
\item
Grain growth, possibly followed by the formation of rocky planetary cores \citep[e.g.,][]{Dullemond05};
\item
Dynamical clearing of a large gap by a Jovian mass planet \citep[e.g.,][]{Lin79,Artymowicz94}, possibly followed by the isolation and viscous draining of the inner disc in the case of high-mass planets \citep{Lubow99};
\item
Disc photoevaporation by ultraviolet starlight \citep{Clarke01,Alexander06};
\item
Magneto-rotational instability \citep{Chiang07}.
\end{itemize}
Because this class of objects only makes up a small fraction of the total disc population, the transition phase from primordial to debris discs is suspected to be short-lived. There is however no evidence for transition objects to be older than typical members of the cluster or group they belong to, suggesting a large diversity of disc properties. It must also be noted that some transition discs still exhibit signatures of accretion, implying that gas can accrete through the dust free inner regions.

Transition from primordial to debris disc seems to happen first in the inner disc. Infrared interferometry is therefore a privileged tool to directly witness the processes at play during this important epoch. A direct view of these regions can bring crucial constraints on the various dust dispersal scenarios, which cannot be distinguished based solely on the SED \citep{Henning08}. Furthermore, infrared interferometry can help determine the composition and properties of the dust in the upper disc layer, where optical and UV photons are absorbed, thereby determining the local disc temperature and hence its vertical scale-height.

Angular resolution from 1 to 100\,mas is typically needed to resolve the inner regions where dust depletion is thought to start, allowing single pupils to only partially resolve the closest individuals. Nevertheless, single-pupil imaging can also be used to constrain planet formation, for instance by investigating the grain size at larger stellocentric distances ($>10$\,AU) in the upper layer of optically thick discs or in the whole vertical profile of bright optically thin discs. Single-pupil imaging can also be used to reveal the formation of larger bodies in the inner disc, which can lead to the appearance of structures such as density waves or gaps at several tens of AUs from the central star.


    \subsection{Evidence for grain growth and evolution} \label{sub:growth}

The most common way to probe the growth and processing of circumstellar dust grains has long been based on the interpretation of SED shapes in the (sub-)millimetre region \citep[e.g.,][]{Beckwith90}, and more recently on the shape and strength of silicate features in the mid-infrared region \citep[e.g.,][]{Bouwman01,KesslerSilacci06}. These studies have shown that dust grains around young stars are generally significantly larger and more crystalline than in the interstellar medium, where grains are mostly sub-micronic and amorphous. Dust settling has also been suggested by SED modelling in several cases \citep[e.g.,][]{SiciliaAguilar07}. In this context, high-angular resolution observations provide a more direct means to confirm and further refine models for grain growth and processing. Grain growth can be interpreted as the first sign of planet formation in circumstellar discs, although it is not clear whether this process actually starts in the prestellar phase (dense molecular cloud) or in the disc phase.

        \subsubsection{Single-pupil observations}

The first evidence for grain growth from resolved single-pupil imaging was obtained by probing the opacity law of dust grains in young circumstellar discs seen in silhouette against background nebular light. \citet{Cotera01} and \citet{Throop01} obtained similar results on two silhouette discs located in Taurus (HH~30) and Orion (114--426): the measured extinction law is flatter than the typical reddening produced by interstellar dust grains. This is almost unambiguously interpreted as the evidence for significant grain growth, as composition and shape of dust grains have little influence on opacities. These results were confirmed at different wavelengths by \citet{Watson04} and \citet{Shuping03}, suggesting however that grains are not much larger than a few microns in the first case (HH~30), and that visible and near-infrared opacities are still dominated by small grains in the second case (114--426).

Single-pupil observations of light scattered off the upper layers of protoplanetary discs can also probe the grain size distribution in various ways. However, because young discs are optically thick, this type of diagnostic brings information only on the uppermost layers of the disc. Examining the colour of scattered light can be used to some extent to assess grain sizes: observations of TW\,Hya \citep{Weinberger02,Roberge05} have shown scattered light to have the same colour as the star in the visible and near-infrared, at all radii except in the innermost regions. This indicates that the grains are larger than about 1\,$\mu$m all the way out to large radii, since smaller grains would produce blue colours. The colour structure of the outer nebula of silhouette discs has also been investigated \citet{Stapelfeldt03}, giving results consistent with the size of interstellar dust grains. This discrepancy with previous opacity-based studies might be due to the fact that different regions are probed by these two techniques, suggesting that larger grains are more abundant in the disc mid-plane and/or in the inner regions.

\begin{figure}[t]
\centering
\includegraphics[width=0.9\textwidth]{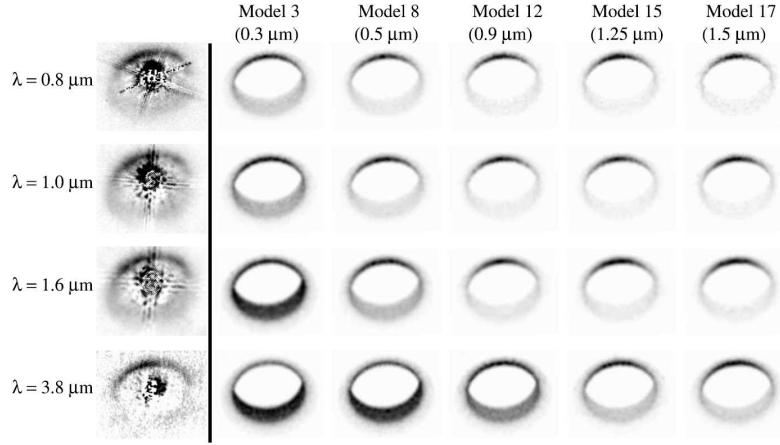}
\caption{Evidence for grain growth from scattered images of GG~Tau at multiple wavelengths: the apparent morphology of the circumstellar disc is best reproduced with large grain populations ($>1\mu m$) thanks to their forward-scattering properties. Credit: \citep{Duchene04}, reproduced by permission of the AAS.} \label{fig:duchene}
\end{figure}

Another successful way to exploit the properties of scattered light is based on the scattering phase function: large grains tend to scatter preferentially forward, while small grains scatter isotropically. This type of study can however be entirely successful only when the disc geometry and orientation are known from other observations, as the appearance of young discs in scattered light is a complex combination of dust properties and disc geometry. Azimuthal intensity measurements have the decisive advantage over colour measurements that they do not depend on grain models. Reconstructing the scattering phase function from azimuthal intensity variations in discs with known geometry has allowed grain size to be estimated in the upper layers of GG\,Tau \citep{McCabe02,Duchene04}. The phase function was found to be strongly forward-scattering and only weakly dependent on the wavelength, which are both direct evidence for the presence of large ($\ge1\,\mu$m) grains. By using scattered light observations at several wavelengths up to 3.8\,$\mu$m, \citet{Duchene04} were furthermore capable of probing different depths in the disc upper layers, and proposed a stratified structure in which the uppermost layers contain grains very similar to interstellar grains while deeper layers contain significantly larger grains (Fig.~\ref{fig:duchene}). This result suggests that vertical settling of dust grains is on-going in this disc. Further evidence for grain growth and stratification in GG\,Tau comes from the first detection of mid-infrared scattered light from its disc \citep{McCabe03}: in the mid-infrared, only grains a few microns in size or larger can scatter efficiently, so that the detection of significant scattered light implies copious amounts of grains larger than 1\,$\mu$m. And because mid-infrared imaging penetrates deeper in the disc, these large grains are thought to be located deeper in the disc than the small grains seen in the short-wavelength ($\lambda\sim1\,\mu$m) images. More recently, hydrodynamical simulations of GG\,Tau have shown that the expected amount of vertical settling is fully consistent with those observations \citep{Pinte07}.

Polarisation measurements of scattered light images can also be used to probe grain size, as the degree of linear polarisation decreases for larger dust grains. Such studies have been performed at 1\,$\mu$m for the GG\,Tau disc \citep{Silber00} and in the $JHK$ bands for the TW\,Hya disc \citep{Hales06}, showing high polarisation degrees for both objects (up to $\sim50$\%). This has been interpreted as the signature of small dust grains. This result may seem in contradiction with the aforementioned studies, unless grain stratification is invoked. In the small polarimetric imaging survey of \citet{Hales06}, extended polarised signals have only be found for objects that are suspected to harbour flared discs based on previous modelling. The fact that high polarisation degrees are observed towards such objects may also be consistent with on-going settling of large grains.

Finally, let us note that the most comprehensive theoretical study of a TTS disc so far, performed by simultaneously fitting photometry, spectroscopy, millimetre interferometry, visible and near-infrared imaging with a 3D continuum radiative transfer code \citep{Pinte08b}, has brought to light several signs of early planetary formation, including grain growth up to a few millimetre in size, vertical settling, and possibly the presence of fluffy aggregates and/or ice mantles around dust grains.

        \subsubsection{Interferometric observations}

Grain growth has also been suggested by infrared interferometry observations of the innermost part of circumstellar discs. Near-infrared observations of six intermediate mass pre-main sequence stars obtained with PTI and IOTA, already discussed in Section \ref{sub:ysointerf}, have been reanalysed by \citet{Isella06} with a more detailed physical model, computing self-consistently the shape and emission of the disc inner edge based on a puffed-up inner rim model \citep{Dullemond01}. By fitting three free parameters (grain size, disc inclination and position angle) to the measured visibilities, they show that grains larger than $\sim1$\,$\mu$m are either required or consistent with the observations in four cases. This suggests that grains in the disc inner rim are often much larger than interstellar grains, confirming that grain growth also takes place in the innermost disc regions. A similar result was subsequently obtained in the case of MWC\,758 with VLTI/AMBER \citep{Isella08}. Besides these observational results, radiative transfer and hydrostatic equilibrium calculations by \citet{Tannirkulam07} show that dust growth and settling can significantly affect the shape of the inner dust rim, which will be within reach of imaging interferometers in the coming years.

\begin{figure}[t]
\centering
\includegraphics[width=0.7\textwidth]{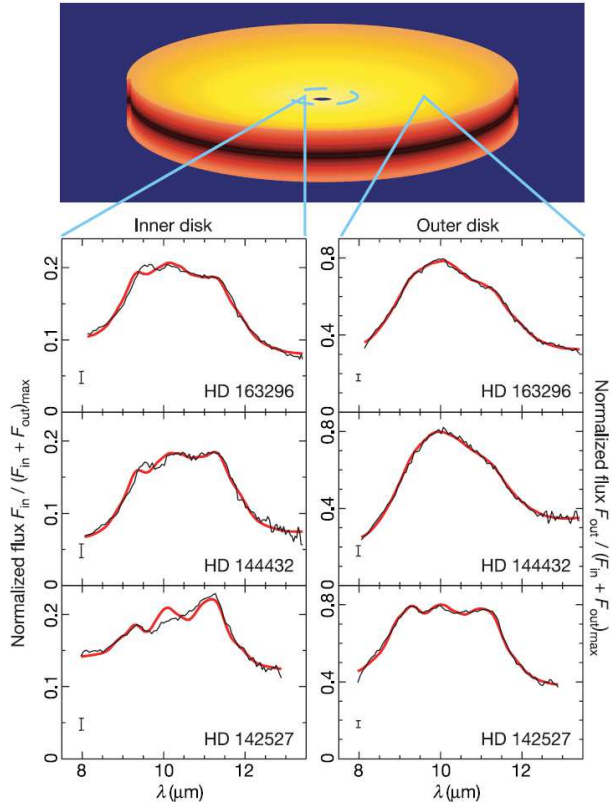}
\caption{Mid-infrared spectra of the inner (1--2\,AU) and outer (2--20\,AU) disc regions of three HAeBe stars, as seen with VLTI/MIDI \citep[reprinted by permission from Macmillan Publishers Ltd: Nature 432:479-482, \copyright 2004]{vanBoekel04}. The differences in shape between the inner and outer disc spectra indicate a difference in dust mineralogy, with more crystalline grains located in the inner part. Reprinted by permission from Macmillan Publishers Ltd: Nature (432: 479--482), copyright (2004).} \label{fig:vanboekel}
\end{figure}

Many of the interferometric constraints on dust size and composition actually come from the mid-infrared regime, where silicates have strong emission features. \citet{vanBoekel04} were the first to perform a compositional analysis of dust grains with mid-infrared interferometry. Three protoplanetary discs were spectrally and spatially resolved with VLTI/MIDI, showing the grain population to be highly crystalline in the innermost two astronomical units of these discs and dominated by large olivine grain ($\ge 1\,\mu$m). Thanks to spectrally dispersed visibilities obtained at different baseline lengths (i.e., probing different regions of the disc), a significant compositional difference was observed with the outer disc ($2-20$\,AU), where grains are less crystalline and smaller, and where pyroxene grains are proportionally more abundant (see Fig.~\ref{fig:vanboekel}). The measured chemical gradient is in qualitative agreement with theoretical models accounting for chemical equilibrium and radial mixing of disc material. These observations also indicate that silicate dust in the inner regions is already highly crystalline before planet formation occurs. It is however not clear whether standard crystallisation processes (gas-phase condensation and thermal annealing) can account for the high degree of crystallinity of the inner discs. The similarity of the observed spectra with those of solar system comets also suggests that the building blocks of larger bodies are present early on in protoplanetary discs.

Similar results have subsequently been obtained with VLTI/MIDI for two TTS: TW\,Hya \citep{Ratzka07} and RY\,Tau \citep{Schegerer08}. In both cases, grains are found to be more crystalline in the innermost disc region. In the case of TW~Hya, the observations furthermore show that dust settling is efficient in this disc, as the presence of large grains has been inferred from previous millimetre observations while only small grains ($<1\,\mu$m) are seen in the uppermost disc layers probed by VLTI/MIDI. Together with the relative segregation of crystalline grains, this suggests that mixing (both radial and vertical) is not very efficient in this case. Another kind of mid-infrared interferometric observation of young discs was obtained by controlling the tilt and piston of individual Keck mirror segments to produce four independent and non redundant interference patterns, using six segments each. This method was used by \citet{Tannirkulam08b} to observe MWC~275 and AB~Aur. The size and flux of the MWC~275 outer disc were found to be significantly smaller than in the case of AB~Aur, and could only be reproduced if the disc is depleted in small grains beyond $\sim7$\,AU. This result indicates that dust grains have undergone significant growth and settling in the outer region of this disc.


    \subsection{Planet formation and disc dissipation} \label{sub:formation}

When dust grains have grown to sufficiently large aggregates and have settled in the mid-plane of the disc, the next step is the building of large bodies, i.e., planets. Indeed, once planetesimals have grown beyond the km-size through physical processes that are still poorly understood \citep[e.g.,][]{Henning08}, runaway growth is thought to drive such protoplanets to planetary size. Those which are far enough from the central star are then expected to accrete the surrounding gas located in their ``feeding zone''. In order to produce giant gas planets, this process must however happen on a timescale smaller than the disc dissipation timescale, i.e., the timescale of the process(es) through which the disc gas and dust content is progressively depleted. Disc dissipation is likely to occur through a set of different and interrelated mechanisms which eventually lead to the formation of the evolved planetary system. The interplay between disc dissipation and planet formation can be probed by high angular resolution imaging, either by revealing direct signs of planetary core formation or by constraining the disc dissipation mechanisms and timescale.

One of the most striking pieces of evidence for disc dispersal and/or planet formation is the observation of ubiquitous inner holes in young debris discs. This will be discussed in Section~\ref{sub:youngdebris}, while we focus here on earlier stages where discs are still optically thick. The presence of inner holes in young, optically thick discs has been more elusive than in young debris discs, especially with single-pupil imaging for which only large inner holes ($>10$\,AU) can be detected. Large inner holes have nevertheless been detected with mid-infrared imaging on 8-m class telescopes around one Herbig Ae star \citep[HD\,142527,][]{Fujiwara06} and one T~Tauri star \citep[IRS\,48,][]{Geers07}. Similar results have also been recently obtained around a few other young stars with submillimetre interferometry \citep{Pietu06,Hughes07,Brown08,Dutrey08}. In the case of IRS\,48, the inner hole has a radius of 30\,AU and is not completely empty: it is filled with PAH emission, suggesting that very small grains are still present in this part of the disc. This behaviour could either be a pure effect of grain growth, although the presence of a sharp inner edge would be difficult to explain, or the evidence for a (possibly low-mass) companion to the central star. In the latter case, large grains would be filtered at the outer edge of the hole due to gravitational interactions with the companion, while smaller grains (e.g., PAHs) would continue to accrete inwards along with the gas.

In the case of HD\,142527, evidence for copious amounts of warm dust in the innermost part of the disc argue for a gap-like structure rather than a hole, which could have been opened by a secondary companion within the disc. This can however not be considered as direct evidence for planetary formation, as the gap could also be formed by a (sub-)stellar companion. This actually turned out to be the case for CoKu\,Tau/4, which was until recently considered as a bona-fide transition disc. Aperture masking observations on the Keck~II telescope by \citet{Ireland08} demonstrated the presence of a near-equal binary companion at an angular distance of 53\,mas ($\sim8$\,AU). The authors argue that more transition discs could actually be circumbinary discs around compact binary systems, although first results around other transition objects show no evidence of companions in the 20--160\,mas projected separation range (M.~J.~Ireland et al., in preparation). Closer companions could eventually be detected with infrared interferometry, provided that the sensitivity of current instruments can be improved by one or two magnitudes (e.g., up to $K\sim9$).

Besides inner holes, single-pupil imaging is sensitive to gaps at large orbital distance formed through the dynamical clearing by a bound low-mass companion. This has recently been illustrated by the detection of a low-density cavity within the optically thick disc of AB~Aur, as already discussed in Section~\ref{sub:haebediscs} (see also Fig.~\ref{fig:HAeBe}). This could be the first evidence for dynamical interactions between a forming extrasolar planet and its parent disc, although this interpretation must be taken with care (see Sect.~\ref{sub:haebediscs}).

Inner holes have also been investigated with interferometry, down to much smaller stellocentric radii. TW\,Hya in particular has been the focus of two studies which aimed at confirming the presence of a 4\,AU inner hole suggested from SED modelling by \citet{Calvet02}. Near-infrared observations with the Keck interferometer first showed that optically thin dust must be present within the hole, down to a radius of 0.06\,AU \citep{Eisner06}. The second study, performed in the mid-infrared at the VLTI, further refined the disc geometry by showing the inner hole to extend only up to 0.5\,AU, followed by a region with a progressive transition from zero to full disc height between 0.5 and 0.8\,AU \citep{Ratzka07}. Combined with the near-infrared interferometric observations of \citet{Eisner06}, these observations further confirm that optically thin material is present within the inner hole. These two studies strikingly illustrate the need for resolved images in the interpretation of young star SEDs. The physical processes producing this peculiar disc structure are however not constrained by the observations. Planet formation, locally enhanced particle growth and inside-out disc dispersal are all possible explanations. Recent spectroastrometric imaging of TW\,Hya and two other transition discs with VLT/CRIRES \citep{Pontoppidan08} also indicates the presence of gas within the inner hole and favours planetary formation (either through grain growth or the influence of a planetary companion) as the mechanism for inner disc clearing.

Another type of interferometric study was performed by recombining two parts of the 6.5-m MMT pupil in the mid-infrared with the BLINC nulling interferometer. By efficiently dimming the stellar radiation thanks to destructive on-axis interference, \citet{Liu03} have been able to study the disc emission of HD~100546 at several wavelength from 10 to 25\,$\mu$m. The fact that the disc size (measured by the null depth) appears similar at all wavelengths suggests the presence of a large inner disc gap, whose size was however not possible to directly constrain. Such an inner hole, extending out to about 10\,AU, was already proposed by \citet{Bouwman03} from SED modelling. Subsequent spatially resolved visible spectroscopy by \citet{Acke06} also supports this interpretation.

Finally, direct evidence for bright hot spots in young accretion discs has recently been reported around FU\,Ori \citep{Malbet05} and AB\,Aur \citep{MillanGabet06b} using near-infrared interferometry. In the first case, fitting simple models to a large number (almost 300) of individual visibility measurements suggests the presence of a bright spot at a projected distance of 10\,AU with an estimated blackbody temperature of 4600\,K, assuming a radius of $5R_{\odot}$. An almost radial displacement of this spot is marginally detected over five years of observations. This would suggest a stellar companion to FU\,Ori on a highly eccentric orbit (or inclined with respect to the disc), although a young (accreting) planet is not excluded by the observations. In the second case, localised off-centre emission is detected at 1--4\,AU in the circumstellar environment of AB\,Aur, based on closure-phase measurements with IOTA/IONIC. Here again, the nature of the disc hot spot cannot be ascertained. The most likely explanations are viscous heating due to gravitational instabilities in the disc, or the presence of a close stellar companion or accreting substellar object. In both cases, model-independent imaging based on a larger pool of closure-phase measurements would be needed to better constrain the morphology of these hot spots.


    \subsection{Witnessing the last stage of the transition phase} \label{sub:youngdebris}

Among intermediate age systems (5--10~Myr), the variety of disc evolution states is very large. In the two previous sections, we have discussed optically thick discs showing first signs of grain evolution and/or planetary formation. In this section, we turn to somewhat more evolved individuals where most of the circumstellar gas has already cleared out, leaving optically thin discs of dust. Such discs are thought to be still partly primordial, and partly replenished by collisional grinding of larger rocky bodies. Because circumstellar gas has cleared, giant planets are supposed to be mostly formed in these systems, but are probably still subject to major dynamical changes.

Discs at this age are relatively bright compared to their older counterparts but not so dense as to be optically thick, with typical bolometric fractional luminosities around $10^{-2}$. Besides the fact that they are relatively easy to image, such objects are particularly interesting to study because they trace the early evolution of fully formed planetary systems. In particular, the influence of the early dynamical evolution of massive planets can be traced in the structure of dusty rings. In this section, we describe three striking examples of such optically thin transition discs.

        \subsubsection{HD 141569: a prototypical HAeBe transition disc}

\begin{figure}[!t]
\centering
\includegraphics[width=0.9\textwidth]{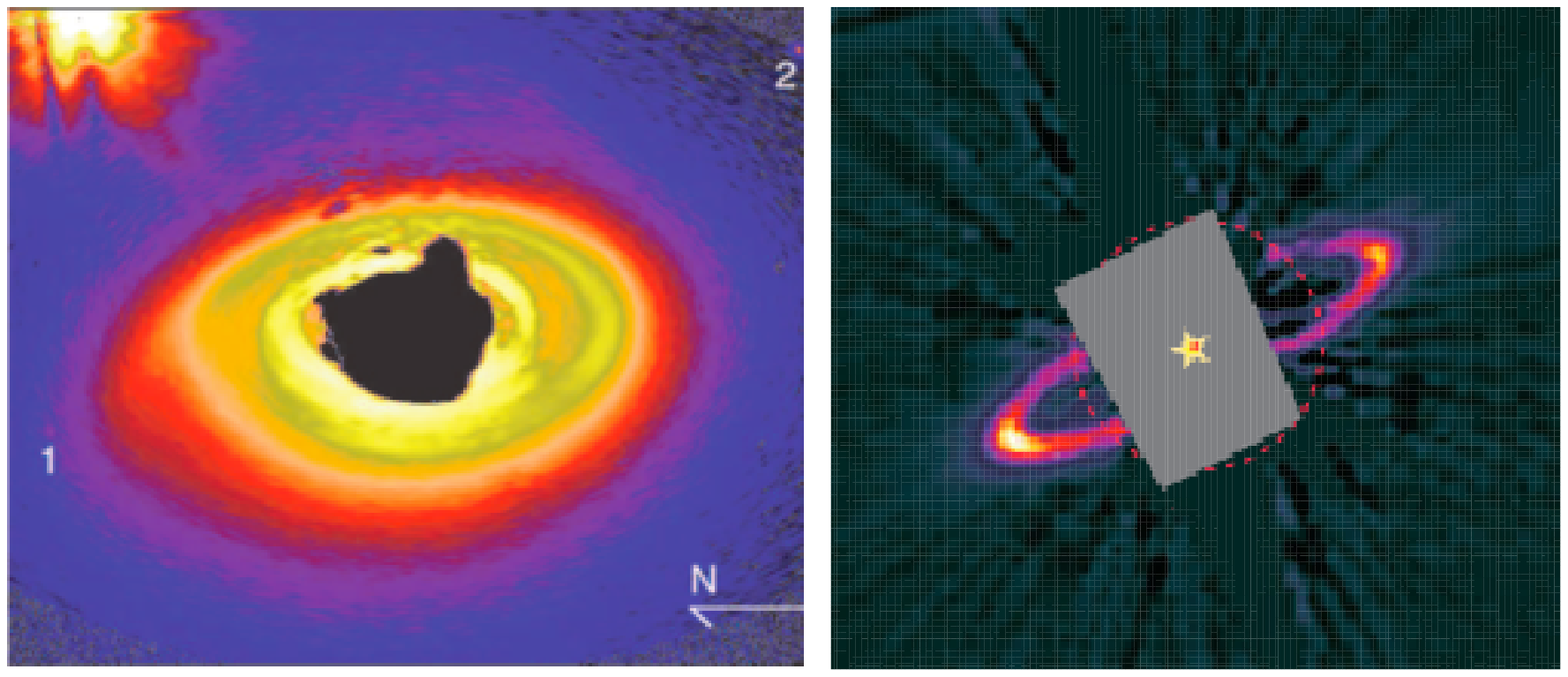}
\caption{\textit{Left} HST/ACS image of the disc around the transition star HD141569 \citep[credit:][reproduced by permission of the AAS]{Clampin03}. \textit{Right} HST/STIS image of the disc around HR 4796A \citep[credit:][reproduced by permission of the AAS]{Schneider09}.} \label{fig:transitiondisc}
\end{figure}

The disc around the B9.5Ve star HD~141569, simultaneously imaged by \citet{Augereau99b} and \citet{Weinberger99} using HST/NICMOS, is the prototype of optically thin transitional discs (Fig.~\ref{fig:transitiondisc}). Because it is bright, and given its transient state, it has been the subject of a great deal of both observational and theoretical work. It has an age of about 5--10\,Myr and a fractional infrared luminosity of 0.018 \citep{Merin04}. Starlight scattered by the dust in its disc has been resolved by HST in both optical and near-infrared coronagraphic images revealing a complex structure with rings at 200 and 325\,AU \citep{Augereau99b,Weinberger99,Mouillet01}, which are actually tightly wound spiral structures \citep{Clampin03}. Scattering from dust is seen up to 1200\,AU in a more open spiral arm structure \citep{Clampin03}. Moreover, the mid-infrared emission from the inner disc has been marginally resolved within 100~AU \citep{Fisher00,Marsh02}, showing a significant depression in optical depth for the region within 30\,AU. Grains are inferred to be mostly sub-micronic from multi-wavelength modelling, suggesting that the disc is dominated by collisional debris. Circumstellar gas has also been detected within 50\,AU of the star by mid-infrared spectroscopy \citep{Brittain02}.

The explanations for the complex structure of HD~141569 invoke either the gravitational perturbations of binary companions, which might be the two coeval M stars seen at about 1000~AU \citep{Augereau04,Quillen05}, or the secular perturbations of a moderately eccentric, relatively low mass planet orbiting within the disc, between the two observed spiral structures \citep{Wyatt05}. Following the work of \citet{Ardila05}, a recent dynamical simulation by \citet{Reche09} suggest that the spiral structure seen in the scattered images of the disc can be explained by the cumulated influence of a single fly-by of the binary companion between 5000 and 8000 years ago and of the presence of a massive planet (a few Jupiter masses) in a low-eccentricity orbit with an apastron at about 130\,AU (i.e., within the innermost spiral structure). This study illustrates how the structure of an optically thin debris disc bears the memory of dynamical events that happened in the history of the planetary system.

        \subsubsection{HR 4796A: a young A-type main sequence star}

The young ($8\pm 2$\,Myr) A0V star HR~4796A is the primary member of a binary system at a distance of 73\,pc. Resolved imaging of the disc was first performed in the mid-infrared \citep{Jayawardhana98,Koerner98}, showing an inner hole with a 55\,AU radius. \citet{Koerner98} also found excess flux at the stellar position at 12\,$\mu$m and 20\,$\mu$m, with values that imply a dust temperature of 250\,K and subsequently, a position for this inner ring at 4.5\,AU from the star. Subsequent coronagraphic observations by \citet{Schneider99} with HST/NICMOS resolved the dusty disc at 1.1 and 1.6\,$\mu$m, revealing a ring 1.05'' (76 AU) in radius with a photometric FWHM of 18.5\,AU (Fig.~\ref{fig:transitiondisc}). Based on these images, preferential forward scattering by the dust grains is studied in the disc models of \citet{Augereau99a}, showing a weak asymmetry consistent with blow-out size of grains. The NICMOS imagery also suggests that the Northeast side of the disc is brighter than the Southwest side, which is modelled by \citet{Wyatt99} as the gravitational influence of a companion body in the system on a slightly eccentric orbit. A similar brightness asymmetry is seen in mid-infrared observations by \citet{Telesco00}.

Using Keck/MIRLIN at wavelengths of 7.9, 10.3, 12.5 and 24.5\,$\mu$m, \citet{Wahhaj05} confirmed the presence of the outer ring and of an unresolved dust population at about 4\,AU from the star. They further show that a two-component outer ring is necessary to fit both Keck thermal infrared and HST scattered-light images, with a narrow ring centered at 66\,AU and a wide ring extending from about 45\,AU to 125\,AU. Dust grains in the narrow ring are about 10 times larger and have lower albedos than those in the wider ring. These large grains, which are less susceptible to radiation pressure, are considered as the direct tracers of the parent body population. The properties of the narrow and wide dust rings are consistent with a picture in which radiation pressure dominates the dispersal of an exo-Kuiper belt.

\citet{Schneider09} recently presented high spatial resolution optical images of the HR~4796A circumstellar debris dust ring using HST/STIS in coronagraphic mode. Two significant flux density asymmetries are noted: (1) preferential forward scattering by the disc grains and (2) an azimuthal surface brightness anisotropy about the morphological minor axis of the disc with corresponding differential ansal brightness. The debris ring is found to be offset from the location of the star by $\sim$14\,AU, a shift insufficient to explain the differing brightness of the northeast and southwest ansae simply by the $1/r^2$ diminution of starlight. The STIS data also better quantify the radial confinement of the debris to a characteristic region less than 14\,AU.

The inferred spatial distribution of the disc grains is consistent with the possibility of one or more unseen co-orbital planetary-mass perturbers, and the colours of the disc grains are consistent with a collisionally evolved population of debris, possibly including ices reddened by radiation exposure to the central star. The presence of organic tholin-like material in the outer parts of the disc was even proposed by \citet{Debes08}, who used HST images of the disc at several wavelengths to isolate the contribution of the outer disc from 0.5 to 2.2\,$\mu$m. Porous grains composed of silicates, carbons and ices is another plausible explanation to the observed SED of the outer disc \citep{Kohler08}.

        \subsubsection{49 Ceti: a transition disc seen in the mid-infrared}

In another proof of the utility of ground-based mid-infrared imaging, \citet{Wahhaj07} used the Keck telescope to resolve the disc thermal emission around the young ($8\pm 2$\,Myr) A1V star 49~Ceti. At 17.9\,$\mu$m, the emission is brighter and more extended toward the Northwest than the Southeast. Here again, such an asymmetry could be due to the perturbing influence of large bodies imposing a forced eccentricity on the disc, which causes one side of the disc to be hotter and brighter (an effect referred to as pericenter glow). Modelling of the mid-infrared images indicates that the bulk of the mid-infrared emission comes from very small grains ($a \sim 0.1 \, \mu$m) confined between 30 and 60\,AU from the star. This population of dust grains contributes negligibly to the large excess observed in the spectral energy distribution. Most of the non-photospheric energy is radiated at longer wavelengths by an outer disc of larger grains ($a \sim 15\, \mu$m), with an inner radius of about 60\,AU and an outer radius of about 900\,AU. The global properties of the 49~Cet disc show stronger affinity with the HR~4796A disc than with typical ``Vega-like'' debris discs, confirming its classification as an optically thin transition disc.

For all of the three transition discs discussed here, several structures such as rings, brightness asymmetries, warps, etc., seem to reveal the presence of massive Jupiter-type bodies at several tens of AUs from the central stars. Such an interpretation of the observed features would be consistent with a rapid (and ubiquitous) formation of giant gaseous planets in a few Myr, before the optically thick disk of gas has cleared out. However, it must be noted that the confinement of dusty grains could, at least in some cases, be explained without the influence of shepherding planets. \citet{Besla07} propose that if a local enhancement of dust grains (e.g., due to a catastrophic collision) occurs somewhere in the disc, photoelectric emission from the grains can heat the gas to temperatures well above that of the dust. The gas then orbits with a super-Keplerian (resp.\ sub-Keplerian) speed inward (resp.\ outward) of the associated pressure maximum, which tends to concentrate the grains into a narrow region. The rise in dust density leads to further heating and a stronger concentration of grains, and a narrow dust ring forms as a result of this instability. This mechanism does not only operate around early-type stars that have high ultraviolet fluxes, but also around stars with spectral types as late as K. This implies that this process is generic and may have occurred during the lifetime of any circumstellar disc.


\section{Mature planetary systems} \label{sec:mature}

By mature planetary systems, we refer to systems where planets are already formed, where the primordial discs have dissipated, and gas is mostly absent. These systems have ages larger than about 10\,Myr and are in a state similar to the solar system, where the dynamical evolution of large bodies and the collisional grinding of rocky bodies dominate. The presence of circumstellar dust in optically thin discs around stars with ages above about 10\,Myr is attributed to populations of planetesimals that were not used to make up planets \citep[e.g.,][]{Mann06}. These leftovers are supposed to produce dust by mutual collisions or cometary activity, and therefore replenish the disc that can persist over much of the star's lifetime. In our solar system, two main dust populations are distinguished: the zodiacal dust cloud in the inner terrestrial planet zone and the Edgeworth-Kuiper belt beyond the orbit of Neptune. Besides indicating the presence of planetesimals (i.e., asteroids and comets), dusty debris discs can also show direct signs of dynamical interactions with larger bodies under the form of clumps, asymmetries, warps, etc.

More and more of those debris discs have been detected through indirect and direct detection techniques. Indirect methods such as SED modelling based on spectro-photometric data in the infrared (mainly from IRAS, ISO and Spitzer) and submillimetre regimes allow the temperature of the emitting material to be constrained but fail to give precise and unambiguous spatial information. On the other hand, imaging directly sees the scattered or thermal emission from the disc in a given spectral band, only probing certain particles with proper sizes with respect to the wavelength (i.e., the most efficient scatterers). Gathering spatially resolved information, at multiple wavelengths, if possible, is necessary to alleviate the degeneracies of SED modelling and construct a solid physical model by locating the dust populations and improving the characterisation of dust grains, to study planet/dust interactions, and possibly to find a new population of planets and/or (hot) dust grains.

Despite the observing challenge (debris discs are faint, with a low fractional luminosity), more than a dozen debris discs have been imaged in scattered light by HST and ground-based adaptive optics instruments, often aided by coronagraphs (see Table~\ref{tab:debris}). Those coronagraphs provide a means to probe debris discs down to the outskirts of the rocky planet region (5--10\,AU), where warm dust is likely to reside. Infrared interferometry provides a complementary view by giving access to the innermost part of the discs ($<5$--10\,AU). Imaging such spatial scales is necessary to fully understand the planetesimal formation mechanisms as well as the interactions between dust populations and (otherwise invisible) planetary bodies.


    \subsection{The large-scale structure of debris discs}

In debris discs, dust is present as a result of the collisional grinding of larger planetesimals. Where protoplanetary discs are characterised by dust growth, debris discs are thus characterised by planetesimal destruction. Studying their composition and structure is essential to understand their formation history. Dust grain composition is presumably the same as the composition of their parent bodies, so that their characterisation (elemental composition, crystalline/amorphous silicate ratio) provides indirect constraints on the thermal and coagulation history of the planetesimals. Those bodies are on one hand the original building blocks of larger planets, and on the other hand potential contributors to the volatile element content of formed planets (e.g., delivered during the Late Heavy Bombardment). But most importantly, the structure of dusty discs can reveal the distribution of planetesimals and therefore constrain the architecture of the whole planetary system. In this section, we first describe the general structure of debris discs as seen by single-pupil imaging and its connection to the Edgeworth-Kuiper belt, and then discuss three interesting cases in more details.

    	\subsubsection{Structure of debris discs and connection to the Edgeworth-Kuiper belt}

Most debris disc observations are spatially unresolved and the debris discs are thus generally identified from the thermal emission excess contributed by dust in their spectral energy distributions. In a few cases, the discs are close and bright enough so that spatially resolved images can be obtained. High angular resolution images show a rich diversity of morphologies, predominantly governed by inclination:
\begin{itemize}
\item edge-on discs: HD 32297, AU~Mic, HD 139664, HD 15115, $\beta$ Pic;
\item face-on discs: HD 53143, HD 106146, HD 181327;
\item intermediate inclinations: HD 10647, Fomalhaut, HD 92945, HD 15745.
\end{itemize}

\begin{table}[t]
\caption{Debris discs around main sequence stars that have been imaged in scattered light.}
\centering \label{tab:debris}
\begin{tabular}{ccccccc}
\hline\noalign{\smallskip}
Name & Type & Dist.\ & Diam.\ & Incl.\  & Colour & Ref.\\
     &      & (pc)   & (AU)   & (deg)   &        &  \\
\tableheadseprule\noalign{\smallskip}
HR 4796A      	&A0    	&67	     	&140 	&73		&Red 	&\citet{Schneider99}\\
HD 32297	      	&A0    	&113	     	&655		&79		&Blue	&\citet{Mawet09}\\
Fomalhaut	&A3		&7.2		&259		&66		&Red	&\citet{Kalas05}\\
$\beta$ Pic	&A5    	&19.3   	&501 	&90		&Red 	&\citet{Boccaletti09}\\
HD 15745	    	&F2		&64		&67		&67 	&?		&\citet{Kalas07b}\\
HD 15115	 	&F2		&45		&872		&90 	&Blue	&\citet{Kalas07a}\\
HD 181327   	&F5		&50.6	&172		&31.7	&Red	&\citet{Schneider06}\\
HD 139664	&F5	&17.5	&96		&87		&?		&\citet{Kalas06}\\
HD 10647    	&F8 	&17.3  	&130 	&50 		&? 		&\citet{Stapelfeldt07}\\
HD 107146	&G2		&28.5	&399		&25		&Red	&\citet{Ardila04}\\
HD 61005	 	&G8		&34.6	&207		&63.5	&?		&\citet{Hines07}\\
HD 53143      	&K1		&18.4	&110		&45		&?		&\citet{Kalas06}\\
HD 92945    	&K1 	&22  		&57 		&30 		&Gray 	&\citet{Golimowski07}\\
AU Mic		&M1		&9.9	&290		&90		&Blue 	&\citet{Krist05a}\\
\noalign{\smallskip}\hline
\end{tabular}
\end{table}

Morphological features also include warps (AU~Mic, $\beta$~Pic), offsets of the disc centre with respect to the central star (Fomalhaut), brightness asymmetries (HD~32297, Fomalhaut, HD~10647, HD~15115, HD~15745), clumpy rings (AU~Mic, $\beta$~Pic, Fomalhaut, HD~139664, HD~181327) and sharp inner edges (Fomalhaut), features thought to be due to gravitational perturbations induced by massive planets (which is the case at least for $\beta$~Pic and Fomalhaut, see Section~\ref{sub:fompic}). Similar features have actually been observed in our own solar system, due to the gravitational influence either of terrestrial planets on the zodiacal dust cloud \citep[e.g.,][]{Kelsall98,Leinert07} or of the outer giant planets on the structure of the Edgeworth-Kuiper belt \citep{Morbidelli08}. Among the observed structural features, also note the very unusual case of the disc around HD~61005, exhibiting a strong asymmetry about its morphological major axis, but a mirror-symmetry about its minor axis.

Even though the origin of individual features is still under discussion and the associated models require further refinements (e.g., in the dust collisional processes and the effects of gas drag), the complexity of these features, in particular the azimuthal asymmetries, indicates that planets likely play a major role in their creation. This is of interest because such structures, in particular those created by the trapping of particles in mean motion resonances with planets, are sensitive to the presence of moderately massive planets at large distances as illustrated by direct observations of planets in discs (see Section~\ref{sub:fompic}) or by theoretical studies \citep{Wyatt03,Kuchner03,Deller05,Reche08}. In this context, high angular resolution debris disc observations over a wide wavelength range are of critical importance, and future observatories such as Herschel, JWST and ALMA will play a key role.

Besides the similarities in morphological features between resolved debris discs and our solar system, the observed populations of cold dust at several tens of AU from their parent star can generally be directly related to the Edgeworth-Kuiper belt (EKB) in our solar system. Nowadays, almost 1000 objects at least 100\,km in size have been found in this region of the solar system extending from the orbit of Neptune at 30\,AU up to 85\,AU from the Sun. Furthermore, the frequency of low-inclination comets with periods $< 200$~years indicates that there must actually be more than 100 million of such large bodies. To explain the formation of 100-km bodies on time scales less than the age of the Solar system, \citet{Stern97} suggested that the original mass of the EKB was at least 10 Earth masses. Most of this mass has since then been ejected or collisionally eroded to explain the current mass (0.1--$1M_{\oplus}$) of the EKB. To match the fractional luminosity of debris discs with an EKB-like model would require between a couple of Earth masses in cometary objects up to 100 Earth masses for the brightest individuals (e.g., $\beta$~Pic). Those are reasonable numbers, indicating the nature of debris discs may be similar to a young EKB. Besides the steady-state collisional grinding of planetesimals, which reduces the dust content as a function of age, major dynamical events such as the Late Heavy Bombardment \citep{Gomes05} may explain the difference in mass between the outer discs of solar and extrasolar systems.

In addition to morphological studies, high angular resolution imaging (combined with SED modelling) allows degeneracies to be lifted in theoretical models, and can be used to probe spatial variations in grain size and composition. Scattered-light colours are known for about half of the 14 debris discs imaged so far (see Table~\ref{tab:debris}). For discs in which the mid-infrared emission has also been resolved \citep[like HD~32297,][]{Schneider05}, the amount of scattered light compared to the mid-infrared emission from the same physical area provides a means to compute the albedo of the dust, mostly believed to be composed of silicate compounds. The canonical albedo of silicates gives us the following tendencies regarding disc colours:
\begin{itemize}
\item grains smaller than 0.1\,$\mu$m produce Rayleigh scattering in the visible and near-infrared regimes, and are thus blue;
\item grains larger than 2\,$\mu$m scatter neutrally;
\item grains around 1\,$\mu$m in size appear slightly red.
\end{itemize}
In the most general case of a grain size distribution created by a collision cascade, scattering is dominated by small grains. It must however be noted that more realistic grains may be porous aggregates mixed with ices, for which the albedo is by far more complex \citep{Lisse08}.

	   \subsubsection{AU~Mic: a well-studied edge-on disc}

\begin{figure}[t]
\centering
\includegraphics[width=0.9\textwidth]{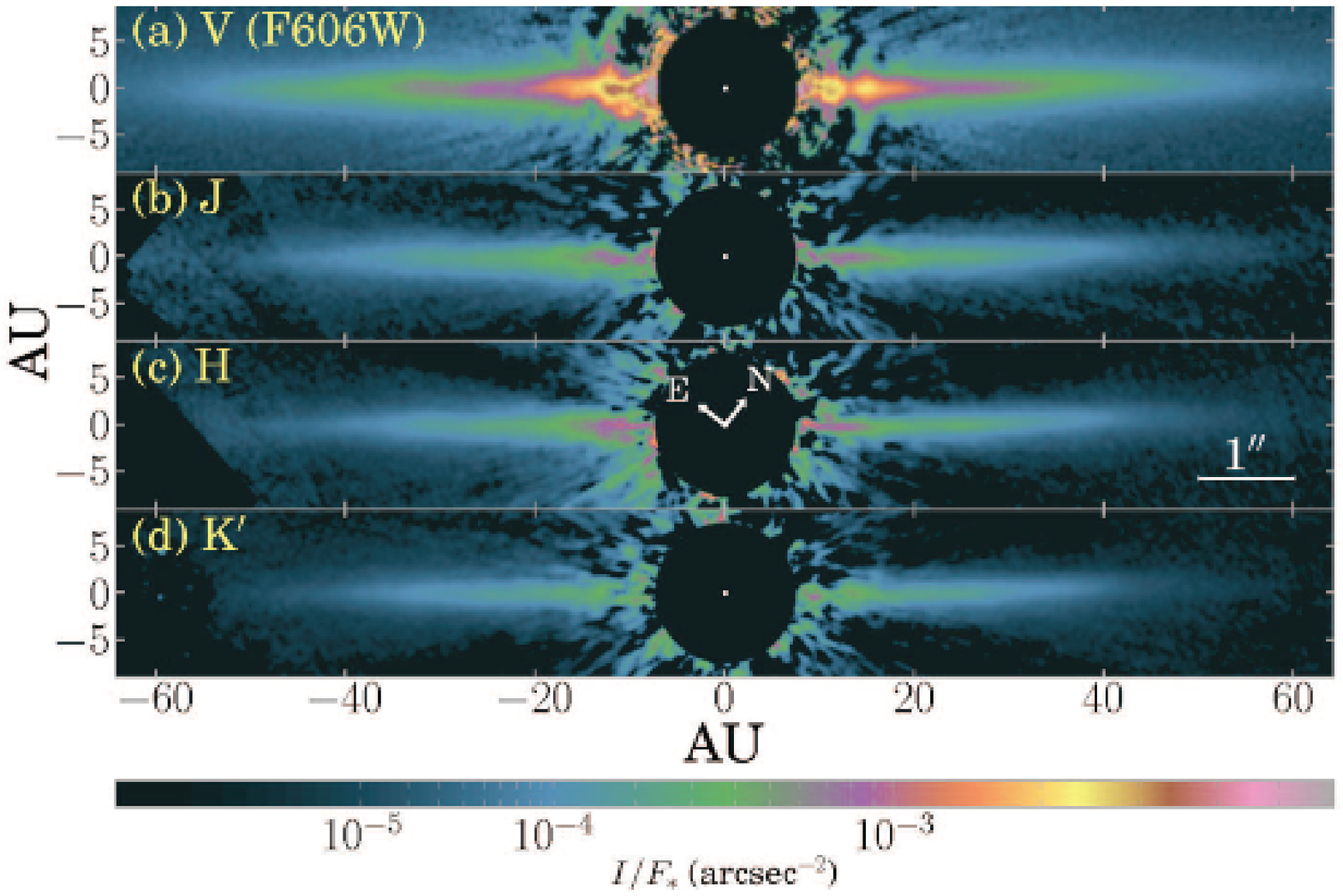}
\caption{HST/ACS F606W image \citep{Krist05a} and Keck AO near-infrared images \citep{Fitzgerald07} of the AU Mic debris disc. Credit: \citet{Fitzgerald07}, reproduced by permission of the AAS} \label{fig:dd1}
\end{figure}

AU~Mic, a young and nearby M-type star (10\,pc), possesses one of the best studied edge-on debris discs. \citet{Kalas04} used the University of Hawaii 2.2-m telescope equipped with a coronagraph to image its edge-on disc for the first time, tracing it in the red up to 210\,AU. Later on, this system was imaged with Keck adaptive optics in the near-infrared by \citet{Liu04} and \citet{Metchev05}. These higher resolution observations trace the disc from 17 to 60\,AU, pointing to a central clearing at least 20 AU wide. \citet{Krist05a} then took images in the blue using HST/ACS, adding more colour information and tracing the disc in this wavelength range from 7.5 to 150\,AU (Fig.~\ref{fig:dd1}). The central clearing initially suggested by SED modelling is also clearly confirmed, refining its size at $\sim12$\,AU. Remarkably, the AU~Mic disc appears blue, which could be due to a surplus of very small grains compared with other imaged debris discs that have more neutral or red colours. The origin of small grains in this system could be related to the low radiation pressure exerted by the late-type star.

Comprehensive modelling of the AU~Mic disc was performed independently by two teams \citep{Augereau06,Strubbe06} using complementary approaches. Both teams interpret the break in the density profile seen around 40\,AU as the evidence of a narrow ``birth ring'' at that location. The inner disc is then expected to be populated by grains larger than 1\,$\mu$m that migrate inward by corpuscular drag due to the intense stellar wind and by Poynting-Roberston drag due to the stellar photons, while the outer disc is populated by sub-micronic dust grains pushed outward by radiation and stellar wind pressure. This steady-state disc evolution picture is consistent with the colour gradient seen in the outer disc, with the disc becoming bluer at larger distance due to an increased amount of small grains.

The most recent high angular resolution study of this system, performed by \citet{Fitzgerald07} in the near-infrared JHK bands with Keck adaptive optics, confirmed a blue colour across the near-infrared bands and the presence of substructure in the inner disc. Some of the structural features exhibit wavelength-dependent positions. Recent measurements of the scattered-light polarisation \citep{Graham07} indicate the presence of porous grains. The scattering properties of these porous grains are different from previously modelled grain types, which has a significant effect on the inferred disc structure: the regions of the AU~Mic disc inside the birth ring must then be depleted of small grains.

        \subsubsection{Fomalhaut and $\beta$ Pic: two discs with embedded planets} \label{sub:fompic}

Fomalhaut and $\beta$~Pic are amongst the most famous debris discs imaged so far. Recently, two teams almost simultaneously announced the discovery of the exoplanets that had been suspected for long to be at the origin of the peculiar structure of these two discs.

The arguably most spectacular example of an extrasolar Edgeworth-Kuiper belt is found around the 200\,Myr-old A4V star Fomalhaut. \citet{Kalas05} imaged this system using HST/ACS and reported the detection of optical light reflected from the dust grains orbiting Fomalhaut. The system is inclined 24\,degrees away from edge-on, enabling the measurement of disc structure around its entire circumference, at a linear resolution of 0.5\,AU. The dust is distributed in a belt 25\,AU wide, with a very sharp inner edge at a radial distance of 133\,AU. \citet{Kalas05} measured an offset of 15\,AU between the belt's geometric centre and Fomalhaut itself. Both the sharp inner edge and the disc centre offset strongly suggest the presence of planetary-mass objects orbiting Fomalhaut \citep{Quillen06b}.

\begin{figure}[t]
\centering
\includegraphics[width=0.9\textwidth]{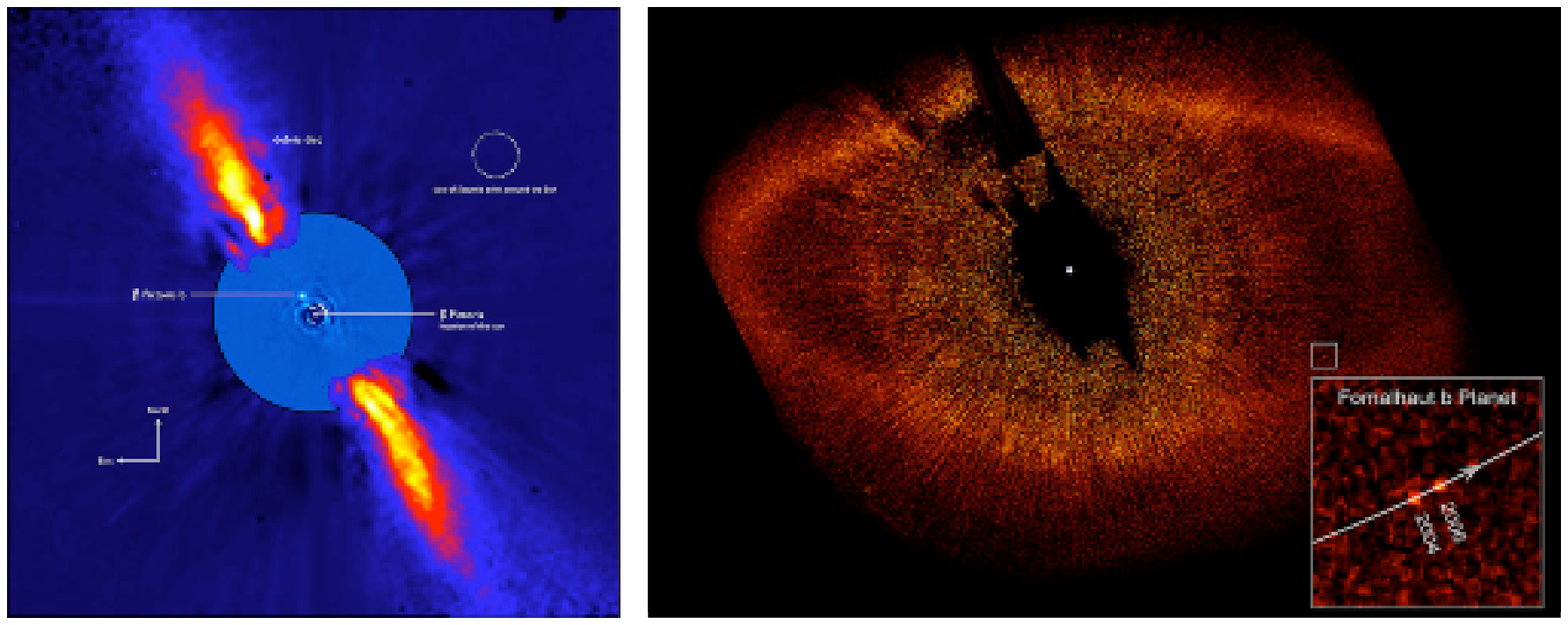}
\caption{\textit{Left} The $\beta$ Pic disc and its putative inner planet \citep[reproduced with permission, \copyright ESO]{Lagrange09}. \textit{Right} The Fomalhaut disc and its outer planet \citep[credit:][reprinted with permission from AAAS]{Kalas08}} \label{fig:betapic_fom}
\end{figure}

Very recently, using HST/ACS, \citet{Kalas08} have directly detected a common proper motion companion, Fomalhaut~b, with properties consistent with the previously predicted exoplanet (Fig.~\ref{fig:betapic_fom}). Two epochs of optical data furthermore reveal its Keplerian orbital motion. The estimated mass of Fomalhaut~b is less than three Jupiter masses: a more massive object would disrupt the dust belt and would have been detected in deep H and L' images obtained with the Keck and Gemini Observatories. Surprising features in the data set suggest contamination from other sources of optical luminosity, such as stellar light reflected from a circumplanetary dust disc.

Since the discovery of its dusty disc in 1984 by Smith \& Terrile, $\beta$~Pic has become the prototype of young early-type planetary systems. In more than 20 years of extensive studies of this system, indications of the existence of a planet around it have accumulated. Among the most striking features in that context are the $5^{\circ}$ warp seen in the inner disc up to about 80\,AU \citep{Heap00}, the detection of ``Falling Evaporating Bodies'' producing circumstellar gas \citep{Beust00}, the presence of a bright mid-infrared clump \citep{Telesco05} and the identification of possible belt-like structures of planetesimals \citep{Okamoto04}. Most of these features have been shown by dynamical models to be consistent with the presence of a Jupiter-sized planet at a distance of about 12\,AU from the central star \citep{Freistetter07}.

In this context, \citet{Lagrange09}, recently revisited an archival data set of deep adaptive-optics L'-band images of $\beta$~Pic obtained with the NACO instrument at the Very Large Telescope. They detected a faint point-like signal at a projected distance of 8 AU from the star, within the northeastern extension of the dust disc. Common proper motion has not been confirmed yet, but there is a very low probability for a background object, and atmospheric and instrumental artifacts have been cautiously ruled out. Its $L'=11.2$ apparent magnitude would indicate a temperature of $\sim 1500$\,K and a mass of $8\,M_{\rm Jup}$. If confirmed, this companion could explain the main morphological and dynamical peculiarities of the $\beta$~Pic disc. Its closeness and location inside the disc suggest a formation process by core accretion rather than binary-like formation processes.

    \subsection{Spatially resolved exozodiacal discs}

Unlike single-pupil imaging, which is most suited to image scattered light from dust grains located in the outer part of planetary systems, near- and mid-infrared interferometry has the appropriate angular resolution and wavelength range to probe the thermal emission from warm grains located in the terrestrial planet region ($<5$--10\,AU) around nearby main sequence stars. In fact, even with the advent of the Spitzer Space Telescope, the presence of warm dust populations around mature main sequence stars has remained rather elusive. Due to the overwhelming stellar radiation at these wavelengths and to the limited accuracy of both spectro-photometric measurements and photospheric models, Spitzer observations have probed warm dust discs only at least 1000 times as dense as the solar zodiacal cloud. Such exozodiacal discs are responsible for mid-infrared excesses of the order of a few percent with respect to the stellar flux.

\begin{figure}[t]
\centering
\includegraphics[width=0.8\textwidth]{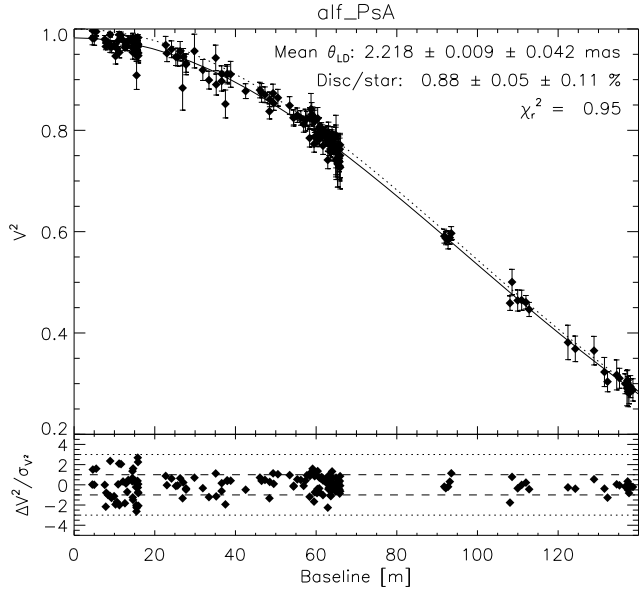}
\caption{Result of the fit of a star-disc model to squared visibilities obtained on Fomalhaut with the VLTI/VINCI interferometer \citep[credit:][reproduced by permission of the AAS]{Absil09}. The solid line represents the best fit star-disc model, while the dotted line represents the best fit with a single star for comparison. In the latter case, a systematic offset between the model and the observations is noticed at short baselines, which reveals the presence of resolved circumstellar emission around Fomalhaut. The star-disc model provides a good fit to the full data set and gives simultaneous estimations of the limb-darkened diameter of the star and of the star-disc flux ratio.} \label{fig:fomzodi}
\end{figure}

Using infrared interferometry, it has recently become possible to directly detect warm dust populations with similar sensitivity levels as Spitzer in terms of exozodiacal disc density. Two types of observations have been carried out in this context: near-infrared high-precision visibility measurements, and mid-infrared nulling observations. The principle of debris disc detection by interferometric visibility measurements is based on the fact that the stellar photosphere and its surrounding dust disc have different spatial scales. For an A-type star at a distance of 20\,pc, the angular diameter of the photosphere is typically about 1\,mas, while the circumstellar disc extends beyond the sublimation radius of dust grains, typically located around 10 to 20\,mas for black body grains sublimating at $T_{\rm sub}\simeq 1500$\,K. In the near-infrared regime, debris discs are therefore generally fully resolved at short baselines ($10-20$\,m), while photospheres are only resolved at long baselines ($\sim$200\,m). One can take advantage of this fact to isolate the contribution of circumstellar dust by performing visibility measurements at short baselines, where the stellar photosphere is almost unresolved. The presence of resolved circumstellar emission then shows up as a deficit of squared visibility with respect to the expected visibility of the sole stellar photosphere, which is proportional to the disc/star flux ratio (see Fig.~\ref{fig:fomzodi}). Due to the expected faintness of the circumstellar emission ($\le 1$\% in most cases), exozodiacal disc detection needs both a high accuracy on the measured visibilities and on the estimated visibility of the stellar photosphere at short baselines.

The first attempt to directly detect hot dust with near-infrared interferometry was performed by \citet{Ciardi01} at the PTI around Vega, the brightest prototype debris disc star. The absence of short baselines at the PTI precluded a firm conclusion at that time, although a simple model of a star surrounded by a uniform dust disc producing a 3--6\% flux ratio was proposed to explain the residuals seen in the fit of a single stellar photosphere. A second study was performed by \citet{DiFolco04} on five debris disc stars with VLTI/VINCI, but only upper limits on infrared excesses could be obtained, once again mostly due to inappropriate baseline lengths. Vega was later re-observed by \citet{Absil06} using an optimised set of short and long baselines at the CHARA array. Combined with the high accuracy of the single-mode FLUOR instrument, this strategy has allowed hot dust to be directly detected around a main sequence star for the first time. The main outcome of these observations is an accurate estimate of the $K$-band flux excess due to the dust ($1.26\% \pm 0.29\%$). Based on this sole photometric information, complemented by (much less accurate) photometric data at other infrared wavelengths, it is then possible to put useful constraints on the disc morphology, composition and dynamics by using a standard radiative transfer code to reproduce the disc spectral energy distribution. The resolved emission seems to emanate mostly from hot sub-micronic carbonaceous grains located within 1\,AU of the photosphere, close to their sublimation limit. Although the estimated dust mass is only about $10^{-7} M_{\oplus}$, the fractional luminosity $L_{\rm dust}/L_{\ast}$ of this dust population amounts to $\sim5\times10^{-4}$ due to its high temperature.

Follow-up studies with CHARA/FLUOR performed by \citet{DiFolco07}, \citet{Absil08} and \citet{Akeson09} have allowed near-infrared excesses to be detected around two more A-type stars ($\zeta$\,Aql and $\beta$\,Leo) and one G-type star ($\tau$\,Cet). The re-interpretation of archival VLTI/VINCI data also shows that Fomalhaut has a $K$-band infrared excess at the 1\% level \citep[see Fig.~\ref{fig:fomzodi}]{Absil09}, while an on-going survey at the CHARA array is currently revealing more detections of this kind. These results suggest that Vega is not an isolated case, and that the presence of large amounts of hot dust might actually be common around A-type and even later-type main sequence stars. Transposed to the solar system, the fractional luminosities of such discs would correspond to density levels of a few thousand times the zodiacal dust cloud. However, their composition and morphology seem quite different from the zodiacal cloud, with smaller dust grains located closer to the star. Such dust has very short radiation pressure and collision lifetimes, and is actually inconsistent with theoretical models for the steady state evolution of debris discs \citep{Wyatt07,Lohne08}: even massive asteroid belts would not produce such amounts of small dust so close to the central star at ages of a few hundreds Myr. Major dynamical perturbations within these planetary systems, akin to the Late Heavy Bombardment that happened early in our solar system, might be responsible for this hot dust population. Such a scenario would require the presence of (migrating) giant planets at a few tens of AUs from the target stars, and would make the solar system even more common among planetary systems.

The second type of interferometric observations targeting inner debris discs have been obtained in the mid-infrared with nulling instruments. The principle of these observations is to combine the light beams collected by two separated pupils in a destructive way, by adding a $\pi$ phase shift in one arm of the interferometer. In this way, the flux from the on-axis star is significantly reduced, while its surrounding environment is transmitted for angular distances as small as $\lambda/4B$, with $B$ the interferometer baseline. Using the BLINC instrument, which combines two sub-pupils of the 6.5-m MMT, \citet{Liu04} were the first to apply this technique to the study of debris discs, and to Vega in particular. With their very short baseline, these observations were only sensitive to dust located farther than about 1\,AU and were used to put an upper limit of 2.1\% on the excess due to the dust at 10.6\,$\mu$m. This result, combined with the near-infrared detection of \citet{Absil06}, suggests that warm dust is confined to the first AU around the star. Very recently, a refurbished version of BLINC has provided its first detection of warm dust around $\beta$\,Leo, showing spatially resolved emission at the level of about 3\% of the photospheric flux (N.~Stock et al., in prep.). Together with the near-infrared detection reported by \citet{Akeson09}, this should allow for a more precise characterisation of the dust population. The Keck Interferometer Nuller (KIN) should also soon bring more constraints on exozodiacal dust populations. Thanks to a longer interferometric baseline than MMT/BLINC, the KIN is sensitive to dust located closer to the star. Preliminary results include some constraints on the mid-infrared excess around Vega \citep{Serabyn06}, the discovery of a two-component inner debris disc around 51~Oph \citep{Stark09}, a marginal detection around Fomalhaut (B.~Mennesson et al., in prep.) and the detection of warm dust around $\eta$~Crv (G.~Bryden, personal communication).

Finally, even though interferometry provides the most suitable angular resolution to study exozodiacal discs, single-pupil observations have been able to resolve hot dust populations around main sequence stars in a few cases. In the first study of this kind, \citet{Chen01} marginally resolved the dust disc around $\zeta$~Lep, an A-type main sequence star at 21.5\,pc, using Keck/LWS imaging at 11.7 and 17.9\,$\mu$m. The dust population, shown by this study to mostly reside within 6\,AU, was recently re-observed with T-ReCS at Gemini South \citep{Moerchen07}, yielding a characteristic radius of 3\,AU for the dust disc, with an fainter dust population at 4--8\,AU. Grain size seems consistent with blow-out radius, and the dust mass corresponds to only a small fraction (0.37\%) of the total mass in the solar asteroid belt. This population can be reproduced by the steady state collisional evolution of an asteroid belt only if $\zeta$~Lep is younger than 100\,Myr, while a transient event such as a catastrophic collision would be privileged at older ages. Thermal emission from warm dust was also recently resolved around the young (12\,Myr) A-type star $\eta$~Tel with T-ReCS \citep{Smith09}, showing the presence of two dust populations at 24\,AU (resolved) and $\sim4$\,AU (unresolved). Such structure is consistent with a young analogue of our solar system. The innermost population might also be the signature of on-going terrestrial planet formation, while the 24\,AU population could trace the recent formation of Pluto-sized objects.


    \subsection{Direct observation of extrasolar planets}

Direct imaging of exoplanets is a very challenging task, not only because of the small angular separation between planets and stars (e.g., $<500$\,mas for a 5\,AU orbital radius at 10\,pc), but also and mostly because the contrast between a planet and its host star ranges from $10^{-3}$ for hot giant planets in the infrared to $10^{-10}$ for Earth-like planets in the visible. Nevertheless, direct imaging is a most promising method as it provides a straightforward means to characterise planetary atmospheres with spectro-photometric measurements.

        \subsubsection{Young low-mass objects at the planet-brown dwarf transition}

\begin{figure}[t]
\centering
\includegraphics[width=0.9\textwidth]{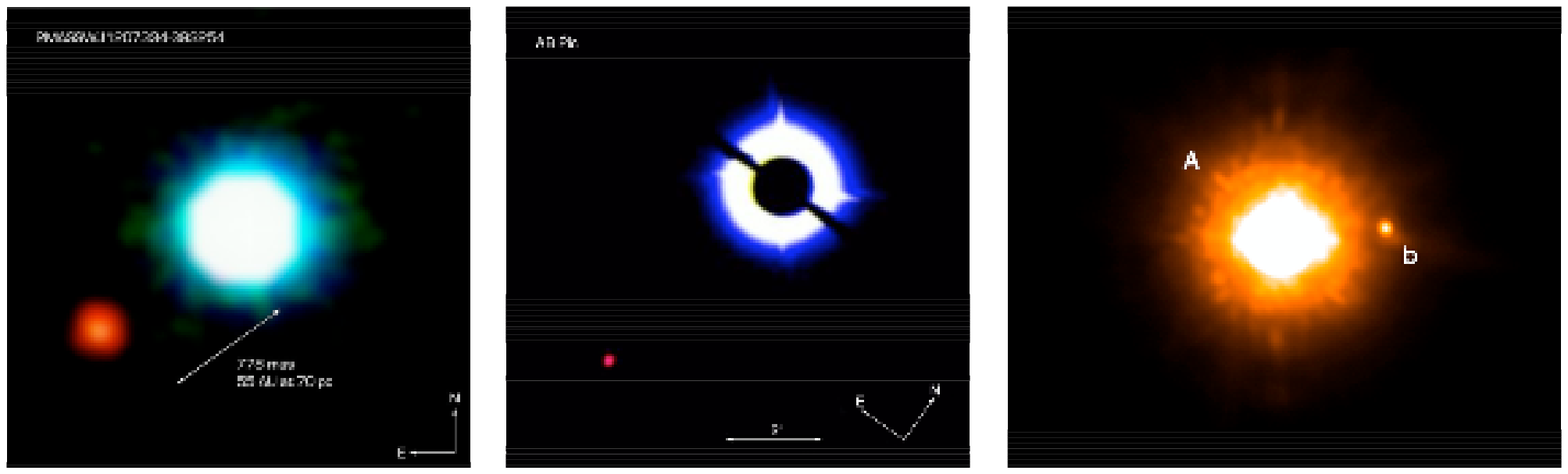}
\caption{Low-mass substellar companions imaged at the VLT. \textit{Left} 2M1207b \citep[credit:][reproduced with permission \copyright ESO]{Chauvin04}. \textit{Middle} AB~Pic~b \citep[\copyright ESO]{Chauvin05b}. \textit{Right} GQ~Lup~b \citep[\copyright ESO]{Neuhauser05}.} \label{fig:planets_2m_abp_gql}
\end{figure}

The first resolved observation of a bound planetary mass object was obtained by \citet{Chauvin04}, using deep VLT/NACO infrared images of the brown dwarf 2MASSWJ\,1207334--393254. This $25\,M_{\rm Jup}$ brown dwarf located at about 70\,pc is identified as a member of the TW~Hydrae association (age $\sim 8$\,Myr). Using adaptive optics infrared wavefront sensing to acquire sharp images of its circumstellar environment, \citet{Chauvin04} discovered a very faint and red object at an angular separation of 778\,mas (55\,AU, Fig. \ref{fig:planets_2m_abp_gql}). This discovery is considered as the first image of an exoplanet, obtained almost ten years after the detection of 51~Peg~b, which demonstrates the difficulty of the task. The observation was made relatively easy by the large angular distance of the companion and the moderate contrast in the near-infrared ($\Delta m \sim 5$). According to evolutionary models, the characteristics of the planet are a mass $M = 5\pm 2 \, M_{\rm Jup}$ and an effective temperature $T_{\rm eff}=1250 \pm 200$\,K.

A few months later, \citet{Neuhauser05} announced the detection of a potentially planetary-mass co-moving companion to the T~Tauri star GQ~Lup ($<10$\,Myr). Orbiting at a distance of 103\,AU (Fig.~\ref{fig:planets_2m_abp_gql}), GQ~Lup~b was later on classified as a brown dwarf companion, with a still rather uncertain mass comprised between $M=13 \, M_{\rm Jup}$ and $M=36 \, M_{\rm Jup}$ \citep[Neuh\"auser et al 2009, submitted to A\&A]{McElwain07,Marois07}. The same year, \citet{Chauvin05b} imaged AB~Pic~b (Fig.~\ref{fig:planets_2m_abp_gql}), a companion to the young K2V star AB~Pic~A ($\sim 30$\,Myr), which is orbiting far away from its host star (275~AU) and has a mass at the planet-brown dwarf transition ($\sim 13.5 \, M_{\rm Jup}$). Imaging these three objects was made easier because young planets and substellar companions with ages of a few Myr are generally much brighter than Gyr-old objects, because of the on-going contraction of and possibly accretion onto these young bodies.

The planetary status of the objects discovered so far is however not clear yet. Their masses are derived from spectro-photometric observations that are compared to evolutionary models \citep{Burrows97,Baraffe03}. Standard thermal evolution models of giant planets employ arbitrary initial conditions selected more for computational expediency than physical accuracy. Since the initial conditions are eventually forgotten by the evolving planet, this approach is valid for mature planets. However, the luminosities and early cooling rates of young planets are highly sensitive to their internal entropies, which depend on the formation mechanism and are highly model dependent.

Recently, \citet{Marley07} and \citet{Fortney08}, using an original implementation of the core accretion model, proposed a solution to the problem of the sensitivity to initial conditions. As a result of the accretion shock through which most of the planetary mass is processed, they find lower initial internal entropies than commonly assumed in published evolution tracks, with a surprising result: young Jupiters are smaller, cooler, and subsequently could be 1--6\,mag fainter than predicted by previous cooling tracks that include a ``hot start'' initial condition. Furthermore, the time interval during which the young Jupiters are fainter than expected depends on the mass of the planet. Jupiter mass planets (1M$_J$) align with the conventional model luminosity in as little at 20\,Myr, but 10M$_J$ planets can take up to 1\,Gyr to match commonly cited luminosities. Furthermore, early evolution tracks should be regarded as uncertain for much longer than the commonly quoted 1\,Myr.

These results have important consequences both for detection strategies and for assigning masses to young Jovian planets based on observed luminosities. If correct, this would make true Jupiter-like planets much more difficult to detect at young ages than previously thought. The modelling of substellar objects is thus not perfect, and as the mass derived from spectro-photometry is highly model dependent, more observations are then needed to break the degeneracy. Observing an enlarged sample of brown dwarfs and massive planets is a possible way to validate the evolutionary models developed for these objects, especially at young ages. A more elegant avenue consists of determining the orbital parameters of brown dwarf and planetary companions, so as to independently derive their dynamical masses as already done for six systems so far \citep[see][and references therein]{Dupuy09}.

Another reason for studying young brown dwarfs and bound brown dwarfs in general in the context of planetary formation and evolution is that they occupy the gap between regular stars and giant planets. The current paradigm for substellar objects identifies two distinct classes of stars later than M: the so-called L and T types. The L class \citep{Kirkpatrick99} includes objects with effective temperatures lower than M dwarfs ($T_{\rm eff} < 2000$\,K), featuring strong lines of neutral alkali elements, an absence of TiO and VO absorption and strong H$_2$O absorption in the visible. Later than L, the T class was originally proposed by \citet{Kirkpatrick99} to account for the methane brown dwarf Gl~229B, the famous companion to an M dwarf 6.3\,pc away, which was actually the first object to be incontrovertibly identified as a brown dwarf by \citet{Nakajima95} with an allowed mass range from 20 to $50\,M_{\rm Jup}$. These methane dwarfs are cooler than L-type dwarfs ($T_{\rm eff} < 1300$\,K) and exhibit strong H$_2$O and CH$_4$ absorption in the H and K bands. Their spectra resemble those of the solar-system objects Jupiter and Titan more than those of stars, supporting the theoretical view that the ultimate state of a cooling BD is a degenerate object much like a planet.

Even with all the observational efforts carried out so far, bound brown dwarfs are still rare, and the so-called brown dwarf desert is only slowly being populated. Up to now, the most successful method to detect close ($\le 5$\,AU) substellar companions has been the Doppler technique based on radial velocity measurements. However, this method is currently insensitive to larger separations, and is difficult to apply to young stellar systems. Thanks to the development of high contrast and high angular resolution imaging on large AO-assisted telescopes equipped with coronagraphs or differential imagers, brown dwarfs can now be rapidly probed at large separations (typically $\ge 10-100$\,AU). Since our knowledge of substellar multiplicity at wide separations derives exclusively from imaging efforts, the importance of developing new optimised techniques is vital. Nowadays, these techniques are becoming mature enough to increase our knowledge of substellar companions, and even to reach the exoplanet regime.

\begin{table}[!t]
\caption{List of bound substellar companions imaged so far at or below the planet-brown dwarf transition. Most of the values given in this table are actually affected by large error bars. See {\tt http://exoplanet.eu} for up-to-date information.}
\centering \label{tab:planets}
\begin{tabular}{cccccc}
\hline\noalign{\smallskip}
Name & Mass            & Rad.            & Sep. &  Age  & Ref. \\
     & ($M_{\rm Jup}$) & ($R_{\rm Jup}$) & (AU) & (Myr) &  \\
\tableheadseprule\noalign{\smallskip}
2M1207 b	&4	&1.5		&46	& 8 &\citet{Chauvin04}	\\
GQ Lup b	&21.5	&1.8	&103	& 1 &\citet{Neuhauser05}	\\
AB Pic b	&13.5	& ...   &275	& 30 &\citet{Chauvin05b}	\\
SCR 1845 b	&$>8.5$	& ...	&4.5	& ? &\citet{Biller06}	\\
UScoCTIO 108 b	&14	& ...	&670	& 5 &\citet{Kashyap08}	\\
1RXS J1609--2105 b	&8	&1.7		&330	& 5 &\citet{Lafreniere08}	\\
CT Cha b 	&17	&2.2		&440	& 2 &\citet{Schmidt08}	\\
Fomalhaut b	&$< 3$	 &	... &115	& 200 &\citet{Kalas08} \\
HR 8799 b	&7	&1.1		&68	& 60 &\citet{Marois08}\\
HR 8799 c	&10	&1.2		&38	& 60 &\citet{Marois08}\\
HR 8799 d	&10	&1.2		&24	& 60 &\citet{Marois08}\\
$\beta$ Pic b	&8	& ... &8	& 12 &\citet{Lagrange09}\\
\noalign{\smallskip}\hline
\end{tabular}
\end{table}

	\subsubsection{Many low-yield extrasolar planet imaging surveys}

Since the discovery of \citet{Chauvin04}, and despite the extensive observational efforts produced by several teams around the world, only 11 other bound low-mass objects have been imaged so far at or below the planet-brown dwarf transition (see Table~\ref{tab:planets}). A significant number of surveys have been and are currently performed in almost every world-leading observatory, with an unanticipated low yield. Among those surveys, let us cite the survey of \citet{McCarthy04} at Lick, Steward and Keck observatories, the Cornell survey at Palomar \citep{Carson05}, the Gemini deep planet survey \citep{Lafreniere07}, the survey of \citet{Masciadri05} with VLT/NACO, the survey of \citet{Biller07} at the VLT and the MMT, the survey by \citet{Chauvin06} using VLT and CFHT, an HST survey by \citet{Lowrance05}, and the Lyot project at the AEOS telescope in Maui \citep{Sivaramakrishnan07}. Those surveys have mainly targeted young solar-type F, G, K stars in search for substellar companions, such as brown dwarfs and massive young/hot exoplanets.

\citet{Nielsen08} examined the implications for the distribution of extrasolar planets based on the null results from two of the largest direct imaging surveys published to date. Combining the measured contrast curves from 22 of the stars observed with the VLT/NACO adaptive optics system by Masciadri and coworkers and 48 of the stars observed with the SDI\footnote{SDI, for Spectral Differential Imaging, is a method of speckle subtraction/calibration that makes use of spectral features in the observed target, most often the 1.6\,$\mu$m methane absorption feature robustly observed in substellar objects cooler than 1400\,K (T~dwarfs). Images are taken simultaneously both within and outside the absorption feature. Due to the simultaneity of the observations, the star and the coherent speckle pattern are largely identical, while any faint companion with the absorption feature is brighter in one filter than the other. Subtracting the two images allows getting rid of the speckle pattern down to chromatic instrumental aberrations and non-common path errors.} mode on VLT/NACO and MMT by Biller and coworkers (for a total of 60 unique stars), they set the following upper limit with 95\% confidence: the fraction of stars with young planets with semimajor axis between 20 and 100\,AU, and mass above $4\,M_{\rm Jup}$, is 20\% or less. Note that this result is based on the ``hot start'' COND model \citep{Baraffe03} for mass-luminosity conversion, which, as already discussed before, can highly overestimate the luminosity of young objects. It is then likely that the derived fraction of 20\% is underestimated.

The important conclusion of this study, though, is that even null results from direct imaging surveys are very powerful in constraining the distributions of young giant planets (0.5--13\,$M_{\rm Jup}$) at large separations, but also that more studies need to be done to close the gap between planets that can be detected by direct imaging, and those to which the radial velocity method is sensitive.

        \subsubsection{Planets around A-type stars: the case of HR 8799}

Since higher stellar luminosities offer less favourable planet-to-star contrasts, bright A-type stars have been mostly neglected in imaging surveys. However, because of their higher mass, A-type stars have still some advantages:
\begin{itemize}
\item they can retain heavier and more extended discs ;
\item they can therefore form massive planets at wider separations.
\end{itemize}
Radial velocity surveys of evolved stars that were A~stars when on the main sequence confirm these hypotheses by showing a trend of a higher frequency of planets at wider separations \citep{Johnson08}. The recent discoveries of the planetary companions to Fomalhaut and $\beta$ Pic (see Section~\ref{sub:fompic}), both A stars, is a testimony to the potential of these stars to sustain planet formation farther out from the central star, making their planets easier to detect.

\begin{figure}[!t]
\centering
\includegraphics[width=0.8\textwidth]{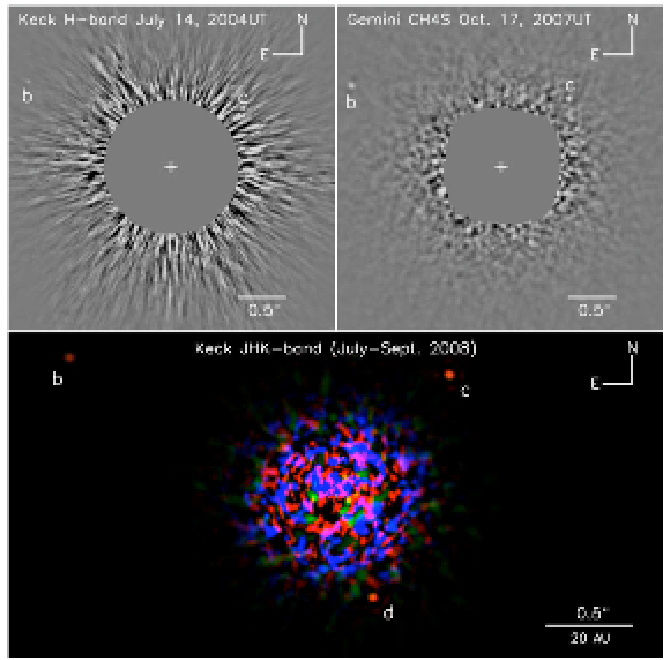}
\caption{HR~8799~bcd images \citep[credit:][reprinted with permission from AAAS]{Marois08}. \textit{Upper left} A Keck image acquired in July 2004. \textit{Upper right} Gemini discovery ADI image acquired in October 2007. Both b and c are detected at the two epochs. \textit{Bottom} A colour image of the planetary system produced by combining the J-, H-, and Ks-band images obtained at the Keck telescope in July (H) and September (J and Ks) 2008. Reprinted with permission from AAAS.} \label{fig:hr8799}
\end{figure}

High-contrast observations with the Keck and Gemini telescopes have also recently revealed three planets orbiting the young ($\sim60$\,Myr) and nearby (39.4\,pc) A-type main sequence star HR~8799, with projected separations of 24, 38, and 68~AU. This star was previously known to harbour a debris disc, with a mean temperature of about 50\,K for the dust grains derived from a black body fit to Infrared Astronomical Satellite (IRAS) and Infrared Space Observatory (ISO) photometric measurements. Such black body grains, in an optically thin disc, would reside at 75\,AU from HR~8799. Multi-epoch data show counter-clockwise orbital motion for all three imaged planets as seen from our terrestrial location. The low luminosity of the companions and the estimated age of the system imply planetary masses between 5 and 13 times that of Jupiter. The dust is presumably just outside the orbit of the most distant companion seen in the images (see Fig.~\ref{fig:hr8799}), similar to the way the Edgeworth-Kuiper belt is confined by Neptune in our Solar system. This system resembles a scaled-up version of the outer portion of our Solar System.


\section{Conclusions} \label{sec:conclusion}

    \subsection{The early stages of planet formation}

A considerable number of new constraints on planetary formation and evolution processes have been brought by high angular resolution observations in the optical regime, especially during the past decade. In particular, the early phases of planet formation have been a major application of visible/infrared imaging on large diffraction limited telescopes (either ground-based with adaptive optics, or space-borne) and of infrared interferometry. By obtaining resolved observations of the environment of nearby Herbig Ae/Be (HAeBe) and T~Tauri stars (TTS), it has become possible to test the predictions of protoplanetary disc models that were generally based solely on spectral energy distributions (SED) until the late 90's. A first important result was obviously to confirm that disc-like structures are actually ubiquitous around ``Class I'' and ``Class II'' objects. Regarding large-scale morphology, single-pupil observations of light scattered by the upper layers of these optically thick discs have confirmed that their structure comes in two main flavours: geometrically flat (and usually self-shadowed), or flared with an increasing scale height as a function of stellocentric radius. The morphology of the disc inner regions has also been investigated by infrared interferometry, showing unexpectedly large inner radii with respect to classical accretion disc models for both TTS and HAeBe stars (except for the most luminous ones). These observations have called for a significant review of inner disc models, leading to the now widely accepted ``puffed-up inner rim'' model, where the inner disc radius is set by the sublimation temperature of dust grains. The presence of circumstellar gas within or just outside the dust inner rim has also been inferred in a few cases, providing new constraints on accretion and wind launching mechanisms. All these studies have been most useful to characterise the environment in which planets are formed.

Another very significant contribution of high angular resolution techniques has been the study of dust grain sizes and composition, which cannot be unambiguously determined from unresolved spectro-photometric observations. These observations have shown that micron-sized grains are ubiquitous in young circumstellar discs, so that a significant growth with respect to the original sub-micron interstellar grains is seen very early in the protoplanetary disc phase. It is not clear yet whether grain growth actually starts when the circumstellar material has settled into a disc (providing a higher particle density and therefore an enhanced collision rate), or already in the highly obscured ``Class 0'' phase where the protostar is still surrounded by an optically thick envelope. Several trends have also been noted on the spatial distribution of grains. First, the settling of larger grains towards the disc mid-plane (as predicted by theoretical models) has been observed in several cases. These large grains (up to millimetre-sized) are expected to be the seeds for further growth above the so-called ``metre-barrier''. Second, mid-infrared interferometry has suggested that dust grains are generally even larger in the innermost parts of circumstellar discs ($<2$\,AU) than in the 2--20\,AU region. Regarding grain composition and chemistry, the silicate nature of dust grains inferred from spectroscopy has been shown to be mostly consistent with spatially resolved observations. Gradients in dust composition throughout the disc have been found in a few cases, showing dust grains to be more crystalline in the inner disc ($2$\,AU) than further out. These observations provide much-needed information on the poorly constrained mixing and turbulence processes in protoplanetary discs, which are theoretically predicted to have a significant impact on planet formation models.

    \subsection{Evolution from freshly formed to mature planetary systems}

Besides characterising protoplanetary discs and revealing the first phases of planet formation through grain growth, high angular resolution optical techniques are also well suited to study the complete evolution of planetary systems, from the very moment when planetary embryos of a few Earth masses are formed in protoplanetary discs up to mature planetary systems like our own. First, direct imaging can be used to test whether the dispersal of optically thick circumstellar discs is mainly driven by the run-away growth of planetary embryos or by other physical processes such as photo-evaporation or magneto-rotational instabilities. Although observations dedicated to such goals have just started, it has already been possible to show that some discs classified as ``transitional'' by the lack of significant near-infrared excess can actually be evacuated by the presence of a short-period binary companion. In most cases, however, the physical origin of the progressive disc clearing still remains ambiguous.  High angular resolution observations have nonetheless provided some evidence for the presence of forming protoplanets in optically thick discs, either through the detection of ``hot spots'' or of evacuated cavities within the discs. More observations of this kind will eventually give a better view on the planet formation timescale, in particular with respect to the disc dissipation timescale.

As soon as the circumstellar disc becomes optically thin, direct imaging can probe the whole vertical thickness of the disc and thereby becomes more sensitive to the effects of dynamical interactions between dust, planetesimals and planetary bodies. In particular, by characterising the various structures (spirals, warps, sharp edges, etc.) seen in the outer parts of (young) debris discs with single-pupil imaging, or the hot dust content in inner disc regions with interferometry, it becomes possible to infer the dynamical history of the planetary system, including migration processes that are thought to affect young planetary systems. These observations can reveal the presence of otherwise undetectable planets. In two cases, high dynamic range observations with coronagraphic devices have even allowed the predicted planets to be directly detected, confirming the previous interpretations based on dynamical models. Precisely locating the dust populations around main sequence stars can furthermore be used to lift the well-known ambiguity of SED modelling and thereby derive significantly improved estimations of dust size distributions and optical properties.

Finally, direct imaging has produced in the last five years the first resolved images of planetary-mass objects around several young stars as well as a few main sequence stars. These pioneering observations open the way to the characterisation of a whole new family of long-period massive extrasolar planets.

    \subsection{Open questions and perspectives}

Although many important results have already been obtained by high angular resolution techniques in the visible/infrared range, most of these techniques are still in their infancy, and second generation instruments at various observatories will soon provide a significantly improved view of planet formation and evolution. In particular, extreme adaptive optics on 10-m class ground-based telescopes will bring high dynamic range single-pupil observations to a new level, in combination with new coronagraphic devices. Interferometry, on the other hand, will routinely provide reconstructed images thanks to an increased number of simultaneously recombined telescopes, and is also being extended towards new atmospheric windows. Larger apertures in space (JWST) and on the ground (30-m class telescopes) will also significantly contribute to the blossoming of high angular resolution instruments and methods.

Among the many open questions that remain in the field of extrasolar planetary systems, let us point to some topics that will strongly benefit from the new capabilities of high angular resolution imaging in the coming years. Regarding the environment for planet formation, the improved imaging capabilities of infrared interferometers will provide a new view on the disc inner regions, revealing fine structures predicted by models such as curved inner dust rims, dead zones where magneto-rotational instabilities are low, snow lines where ices start to condensate on dust grains. Such structures are expected to have a significant influence on the formation and migration of young planets. The confirmation and generalisation of ``hot spot'' discoveries within protoplanetary discs should also provide much-needed constraints on planet formation mechanisms and timescales. Such phenomena will also be accessible to some extent with single-pupil imaging, which will still mainly focus on phenomena appearing at larger scales, such as the gaps that have been predicted for long to appear as a consequence of giant planet formation and migration. Global modelling efforts including resolved observations at multiple wavelengths, which are just appearing now, will provide accurate estimations of grain growth generalised to a large number of discs, and will allow statistical studies to be performed on grain growth and settling, as well on their structure and optical properties. And most importantly, such studies will be performed on larger samples to reveal statistical trends in planet formation rates, timescales, mechanisms, etc. Another significant improvement should also concern the study of circumstellar gas, by combining spatially and spectrally resolved observations, e.g., with new generation interferometric instruments.

For more evolved systems, where planetary-mass bodies have already formed, a number of new results should also draw the attention in the coming years. One example is the origin of protoplanet disc dispersal, which can be constrained by precisely locating and comparing the remaining dust and gas populations, or by directly looking for embedded planets and stellar companions in discs. The discoveries of massive planets sculpting more evolved debris discs around main sequence stars should also flourish in the coming years. With the increased performance in high contrast imaging, dust populations will be resolved closer and closer to the star, giving a direct indication on the environment of terrestrial planets.

Finally, regarding extrasolar planet imaging, the results obtained so far are only the beginning of a brand new discipline. Imaging surveys of low-mass companions around young nearby stars will provide in the coming years the necessary inputs to constrain evolutionary models for giant planets, and in particular their cooling and contraction tracks. Planets will also be detected at older ages, opening the possibility to characterise the statistics of planetary system architectures up to very long periods, which could have once again a profound impact on the formation and migration theories. Interferometry is also expected to soon provide its first direct detection of a short period giant planet, strongly irradiated by its parent star. Such studies would be very complementary to the photometric measurements of the thermal emission of hot giant planets by providing an improved spectral resolution and giving access to non-transiting systems.

Although the future seems quite bright, high angular resolution optical techniques will not solve alone all the open questions in planet formation and evolution theories. The complementarity with future large submillimetre arrays such as ALMA will be extremely useful to constrain the physical properties of protoplanetary discs, and investigate for instance the poorly understood growth from millimetre-sized to kilometre-sized particles. Collecting photometric information at various wavelengths from the X-rays to the millimetre regime will always be needed to provide a complete view of planetary systems. And of course, indirect planet-detection techniques such as radial velocity measurements, astrometry, microlensing or photometric transits will continue to provide crucial (and complementary) information on the structure and statistics of planetary systems.


\begin{acknowledgements}
We are grateful to Prof.\ J.~Surdej for encouraging us to write this review, and for giving us a deadline (to try) to meet. We also thank Drs J.-C.~Augereau, J.~Krist, F.~Malbet, E.~Serabyn, and K.~Stapelfeldt for their constructive comments and suggestions on the manuscript. O.~Absil acknowledges the support from a Marie Curie IEF Fellowship while at LAOG, of an F.R.S.-FNRS postdoctoral fellowship while at IAGL, and from the Communaut\'e Fran{\c c}aise de Belgique (Actions de recherche concert\'ees -- Acad\'emie universitaire Wallonie-Europe). D.~Mawet\ is supported by an appointment to the National Aeronautics and Space Administration (NASA) Postdoctoral Program at the Jet Propulsion Laboratory, California Institute of Technology, administered by Oak Ridge Associated Universities through a contract with NASA.
\end{acknowledgements}


\bibliographystyle{spbasic}
\bibliography{AAreview_revised_v3}   

\begin{thebibliography}{262}
\providecommand{\natexlab}[1]{#1}
\providecommand{\url}[1]{{#1}}
\providecommand{\urlprefix}{URL }
\expandafter\ifx\csname urlstyle\endcsname\relax
  \providecommand{\doi}[1]{DOI~\discretionary{}{}{}#1}\else
  \providecommand{\doi}{DOI~\discretionary{}{}{}\begingroup
  \urlstyle{rm}\Url}\fi
\providecommand{\eprint}[2][]{\url{#2}}

\bibitem[{{{\'A}brah{\'a}m} et~al(2006){{\'A}brah{\'a}m}, {Mosoni}, {Henning},
  {K{\'o}sp{\'a}l}, {Leinert}, {Quanz}, and {Ratzka}}]{Abraham06}
{{\'A}brah{\'a}m} P, {Mosoni} L, {Henning} T, {K{\'o}sp{\'a}l} {\'A}, {Leinert}
  C, {Quanz} SP, {Ratzka} T (2006) {First AU-scale observations of V1647
  Orionis with VLTI/MIDI}. \aap 449:L13--L16, \doi{10.1051/0004-6361:20064841},
  \eprint{arXiv:astro-ph/0602334}

\bibitem[{{Absil} et~al(2006){Absil}, {Di Folco}, {M\'erand}, {Augereau},
  {Coud\'e du Foresto}, {Aufdenberg}, {Kervella}, {Ridgway}, {Berger}, {ten
  Brummelaar}, {Sturmann}, {Strumann}, {Turner}, and {McAlister}}]{Absil06}
{Absil} O, {Di Folco} E, {M\'erand} A, {Augereau} JC, {Coud\'e du Foresto} V,
  {Aufdenberg} JA, {Kervella} P, {Ridgway} ST, {Berger} DH, {ten Brummelaar}
  TA, {Sturmann} J, {Strumann} L, {Turner} NH, {McAlister} HA (2006)
  {Circumstellar material in the Vega inner system revealed by CHARA/FLUOR}.
  \aap 452:237--244

\bibitem[{{Absil} et~al(2008){Absil}, {Di Folco}, {M{\'e}rand}, {Augereau},
  {Coud{\'e} Du Foresto}, {Defr{\`e}re}, {Kervella}, {Aufdenberg}, {Desort},
  {Ehrenreich}, {Lagrange}, {Montagnier}, {Olofsson}, {Ten Brummelaar},
  {McAlister}, {Sturmann}, {Sturmann}, and {Turner}}]{Absil08}
{Absil} O, {Di Folco} E, {M{\'e}rand} A, {Augereau} JC, {Coud{\'e} Du Foresto}
  V, {Defr{\`e}re} D, {Kervella} P, {Aufdenberg} JP, {Desort} M, {Ehrenreich}
  D, {Lagrange} AM, {Montagnier} G, {Olofsson} J, {Ten Brummelaar} TA,
  {McAlister} HA, {Sturmann} J, {Sturmann} L, {Turner} NH (2008) {A
  near-infrared interferometric survey of debris disc stars. II. CHARA/FLUOR
  observations of six early-type dwarfs}. \aap 487:1041--1054,
  \doi{10.1051/0004-6361:200810008}, \eprint{arXiv:0806.4936}

\bibitem[{{Absil} et~al(2009){Absil}, {Mennesson}, {Le Bouquin}, {Di Folco},
  {Kervella}, and {Augereau}}]{Absil09}
{Absil} O, {Mennesson} B, {Le Bouquin} JB, {Di Folco} E, {Kervella} P,
  {Augereau} JC (2009) {An interferometric study of the Fomalhaut inner debris
  disk I. Near-infrared detection of hot dust with VLTI/VINCI}. \apj
  704:150--160

\bibitem[{{Acke} and {van den Ancker}(2006)}]{Acke06}
{Acke} B, {van den Ancker} ME (2006) {Resolving the disk rotation of HD 97048
  and HD 100546 in the [O I] 6300 {\AA} line: evidence for a giant planet
  orbiting HD 100546}. \aap 449:267--279, \doi{10.1051/0004-6361:20054330},
  \eprint{arXiv:astro-ph/0512562}

\bibitem[{{Adams} et~al(1987){Adams}, {Lada}, and {Shu}}]{Adams87}
{Adams} FC, {Lada} CJ, {Shu} FH (1987) {Spectral evolution of young stellar
  objects}. \apj 312:788--806, \doi{10.1086/164924}

\bibitem[{{Akeson} et~al(2000){Akeson}, {Ciardi}, {van Belle}, {Creech-Eakman},
  and {Lada}}]{Akeson00}
{Akeson} RL, {Ciardi} DR, {van Belle} GT, {Creech-Eakman} MJ, {Lada} EA (2000)
  {Infrared Interferometric Observations of Young Stellar Objects}. Astrophys\
  J 543:313--317

\bibitem[{{Akeson} et~al(2002){Akeson}, {Ciardi}, {van Belle}, and
  {Creech-Eakman}}]{Akeson02}
{Akeson} RL, {Ciardi} DR, {van Belle} GT, {Creech-Eakman} MJ (2002)
  {Constraints on Circumstellar Disk Parameters from Multiwavelength
  Observations: T Tauri and SU Aurigae}. \apj 566:1124--1131

\bibitem[{{Akeson} et~al(2005{\natexlab{a}}){Akeson}, {Boden}, {Monnier},
  {Millan-Gabet}, {Beichman}, {Beletic}, {Calvet}, {Hartmann}, {Hillenbrand},
  {Koresko}, {Sargent}, and {Tannirkulam}}]{Akeson05b}
{Akeson} RL, {Boden} AF, {Monnier} JD, {Millan-Gabet} R, {Beichman} C,
  {Beletic} J, {Calvet} N, {Hartmann} L, {Hillenbrand} L, {Koresko} C,
  {Sargent} A, {Tannirkulam} A (2005{\natexlab{a}}) {Keck Interferometer
  observations of classical and weak line T Tauri stars}. \apj 635:1173--1181

\bibitem[{{Akeson} et~al(2005{\natexlab{b}}){Akeson}, {Walker}, {Wood},
  {Eisner}, {Scire}, {Penprase}, {Ciardi}, {van Belle}, {Whitney}, and
  {Bjorkman}}]{Akeson05a}
{Akeson} RL, {Walker} CH, {Wood} K, {Eisner} JA, {Scire} E, {Penprase} B,
  {Ciardi} DR, {van Belle} GT, {Whitney} B, {Bjorkman} JE (2005{\natexlab{b}})
  {Observations and Modeling of the Inner Disk Region of T Tauri Stars}. \apj
  622:440--450

\bibitem[{{Akeson} et~al(2009){Akeson}, {Ciardi}, {Millan-Gabet}, {Merand}, {Di
  Folco}, {Monnier}, {Beichman}, {Absil}, {Aufdenberg}, {McAlister}, {ten
  Brummelaar}, {Sturmann}, {Sturmann}, and {Turner}}]{Akeson09}
{Akeson} RL, {Ciardi} DR, {Millan-Gabet} R, {Merand} A, {Di Folco} E, {Monnier}
  JD, {Beichman} CA, {Absil} O, {Aufdenberg} J, {McAlister} H, {ten Brummelaar}
  T, {Sturmann} J, {Sturmann} L, {Turner} N (2009) {Dust in the inner regions
  of debris disks around a stars}. \apj 691:1896--1908,
  \doi{10.1088/0004-637X/691/2/1896}, \eprint{0810.3701}

\bibitem[{{Alexander} et~al(2006){Alexander}, {Clarke}, and
  {Pringle}}]{Alexander06}
{Alexander} RD, {Clarke} CJ, {Pringle} JE (2006) {Photoevaporation of
  protoplanetary discs - II. Evolutionary models and observable properties}.
  \mnras 369:229--239, \doi{10.1111/j.1365-2966.2006.10294.x},
  \eprint{arXiv:astro-ph/0603254}

\bibitem[{{Alibert} et~al(2004){Alibert}, {Mordasini}, and {Benz}}]{Alibert04}
{Alibert} Y, {Mordasini} C, {Benz} W (2004) {Migration and giant planet
  formation}. \aap 417:L25--L28, \doi{10.1051/0004-6361:20040053},
  \eprint{arXiv:astro-ph/0403574}

\bibitem[{{Anglada} et~al(2007){Anglada}, {L{\'o}pez}, {Estalella}, {Masegosa},
  {Riera}, and {Raga}}]{Anglada07}
{Anglada} G, {L{\'o}pez} R, {Estalella} R, {Masegosa} J, {Riera} A, {Raga} AC
  (2007) {Proper Motions of the Jets in the Region of HH 30 and HL/XZ Tau:
  Evidence for a Binary Exciting Source of the HH 30 Jet}. \aj 133:2799--2814,
  \doi{10.1086/517493}, \eprint{arXiv:astro-ph/0703155}

\bibitem[{{Apai} et~al(2004){Apai}, {Pascucci}, {Brandner}, {Henning},
  {Lenzen}, {Potter}, {Lagrange}, and {Rousset}}]{Apai04}
{Apai} D, {Pascucci} I, {Brandner} W, {Henning} T, {Lenzen} R, {Potter} DE,
  {Lagrange} AM, {Rousset} G (2004) {NACO polarimetric differential imaging of
  TW Hya. A sharp look at the closest T Tauri disk}. \aap 415:671--676,
  \doi{10.1051/0004-6361:20034549}, \eprint{arXiv:astro-ph/0311194}

\bibitem[{{Ardila} et~al(2004){Ardila}, {Golimowski}, {Krist}, {Clampin},
  {Williams}, {Blakeslee}, {Ford}, {Hartig}, and {Illingworth}}]{Ardila04}
{Ardila} DR, {Golimowski} DA, {Krist} JE, {Clampin} M, {Williams} JP,
  {Blakeslee} JP, {Ford} HC, {Hartig} GF, {Illingworth} GD (2004) {A Resolved
  Debris Disk around the G2 V Star HD 107146}. \apjl 617:L147--L150,
  \doi{10.1086/427434}, \eprint{arXiv:astro-ph/0411422}

\bibitem[{{Ardila} et~al(2005){Ardila}, {Lubow}, {Golimowski}, {Krist},
  {Clampin}, {Ford}, {Hartig}, {Illingworth}, {Bartko}, {Ben{\'i}tez},
  {Blakeslee}, {Bouwens}, {Bradley}, {Broadhurst}, {Brown}, {Burrows}, {Cheng},
  {Cross}, {Feldman}, {Franx}, {Goto}, {Gronwall}, {Holden}, {Homeier},
  {Infante}, {Kimble}, {Lesser}, {Martel}, {Menanteau}, {Meurer}, {Miley},
  {Postman}, {Sirianni}, {Sparks}, {Tran}, {Tsvetanov}, {White}, {Zheng}, and
  {Zirm}}]{Ardila05}
{Ardila} DR, {Lubow} SH, {Golimowski} DA, {Krist} JE, {Clampin} M, {Ford} HC,
  {Hartig} GF, {Illingworth} GD, {Bartko} F, {Ben{\'i}tez} N, {Blakeslee} JP,
  {Bouwens} RJ, {Bradley} LD, {Broadhurst} TJ, {Brown} RA, {Burrows} CJ,
  {Cheng} ES, {Cross} NJG, {Feldman} PD, {Franx} M, {Goto} T, {Gronwall} C,
  {Holden} B, {Homeier} N, {Infante} L, {Kimble} RA, {Lesser} MP, {Martel} AR,
  {Menanteau} F, {Meurer} GR, {Miley} GK, {Postman} M, {Sirianni} M, {Sparks}
  WB, {Tran} HD, {Tsvetanov} ZI, {White} RL, {Zheng} W, {Zirm} AW (2005) {A
  Dynamical Simulation of the Debris Disk around HD 141569A}. \apj
  627:986--1000, \doi{10.1086/430395}, \eprint{arXiv:astro-ph/0503445}

\bibitem[{{Ardila} et~al(2007){Ardila}, {Golimowski}, {Krist}, {Clampin},
  {Ford}, and {Illingworth}}]{Ardila07}
{Ardila} DR, {Golimowski} DA, {Krist} JE, {Clampin} M, {Ford} HC, {Illingworth}
  GD (2007) {Hubble Space Telescope Advanced Camera for Surveys Coronagraphic
  Observations of the Dust Surrounding HD 100546}. \apj 665:512--534,
  \doi{10.1086/519296}

\bibitem[{{Artymowicz} and {Lubow}(1994)}]{Artymowicz94}
{Artymowicz} P, {Lubow} SH (1994) {Dynamics of binary-disk interaction. 1:
  Resonances and disk gap sizes}. \apj 421:651--667, \doi{10.1086/173679}

\bibitem[{{Augereau} and {Beust}(2006)}]{Augereau06}
{Augereau} JC, {Beust} H (2006) {On the AU Microscopii debris disk. Density
  profiles, grain properties, and dust dynamics}. \aap 455:987--999,
  \doi{10.1051/0004-6361:20054250}, \eprint{arXiv:astro-ph/0604313}

\bibitem[{{Augereau} and {Papaloizou}(2004)}]{Augereau04}
{Augereau} JC, {Papaloizou} JCB (2004) {Structuring the HD 141569 A
  circumstellar dust disk. Impact of eccentric bound stellar companions}. \aap
  414:1153--1164, \doi{10.1051/0004-6361:20031622},
  \eprint{arXiv:astro-ph/0310732}

\bibitem[{{Augereau} et~al(1999{\natexlab{a}}){Augereau}, {Lagrange},
  {Mouillet}, and {M{\'e}nard}}]{Augereau99b}
{Augereau} JC, {Lagrange} AM, {Mouillet} D, {M{\'e}nard} F (1999{\natexlab{a}})
  {HST/NICMOS2 observations of the HD 141569 A circumstellar disk}. \aap
  350:L51--L54, \eprint{arXiv:astro-ph/9909423}

\bibitem[{{Augereau} et~al(1999{\natexlab{b}}){Augereau}, {Lagrange},
  {Mouillet}, {Papaloizou}, and {Grorod}}]{Augereau99a}
{Augereau} JC, {Lagrange} AM, {Mouillet} D, {Papaloizou} JCB, {Grorod} PA
  (1999{\natexlab{b}}) {On the HR 4796 A circumstellar disk}. \aap 348:557--569

\bibitem[{{Aumann}(1985)}]{Aumann85}
{Aumann} HH (1985) {IRAS observations of matter around nearby stars}. \pasp
  97:885--891, \doi{10.1086/131620}

\bibitem[{{Aumann} et~al(1984){Aumann}, {Beichman}, {Gillett}, {de Jong},
  {Houck}, {Low}, {Neugebauer}, {Walker}, and {Wesselius}}]{Aumann84}
{Aumann} HH, {Beichman} CA, {Gillett} FC, {de Jong} T, {Houck} JR, {Low} FJ,
  {Neugebauer} G, {Walker} RG, {Wesselius} PR (1984) {Discovery of a shell
  around Alpha Lyrae}. \apjl 278:L23--L27, \doi{10.1086/184214}

\bibitem[{{Baraffe} et~al(2003){Baraffe}, {Chabrier}, {Barman}, {Allard}, and
  {Hauschildt}}]{Baraffe03}
{Baraffe} I, {Chabrier} G, {Barman} TS, {Allard} F, {Hauschildt} PH (2003)
  {Evolutionary models for cool brown dwarfs and extrasolar giant planets. The
  case of HD 209458}. \aap 402:701--712, \doi{10.1051/0004-6361:20030252},
  \eprint{arXiv:astro-ph/0302293}

\bibitem[{{Beckwith} et~al(1990){Beckwith}, {Sargent}, {Chini}, and
  {Guesten}}]{Beckwith90}
{Beckwith} SVW, {Sargent} AI, {Chini} RS, {Guesten} R (1990) {A survey for
  circumstellar disks around young stellar objects}. \aj 99:924--945,
  \doi{10.1086/115385}

\bibitem[{{Besla} and {Wu}(2007)}]{Besla07}
{Besla} G, {Wu} Y (2007) {Formation of Narrow Dust Rings in Circumstellar
  Debris Disks}. \apj 655:528--540, \doi{10.1086/509495},
  \eprint{arXiv:astro-ph/0609248}

\bibitem[{{Beust} and {Morbidelli}(2000)}]{Beust00}
{Beust} H, {Morbidelli} A (2000) {Falling Evaporating Bodies as a Clue to
  Outline the Structure of the {$\beta$} Pictoris Young Planetary System}.
  Icarus 143:170--188, \doi{10.1006/icar.1999.6238}

\bibitem[{{Biller} et~al(2006){Biller}, {Kasper}, {Close}, {Brandner}, and
  {Kellner}}]{Biller06}
{Biller} BA, {Kasper} M, {Close} LM, {Brandner} W, {Kellner} S (2006)
  {Discovery of a Brown Dwarf Very Close to the Sun: A Methane-rich Brown Dwarf
  Companion to the Low-Mass Star SCR 1845-6357}. \apjl 641:L141--L144,
  \doi{10.1086/504256}, \eprint{arXiv:astro-ph/0601440}

\bibitem[{{Biller} et~al(2007){Biller}, {Close}, {Masciadri}, {Nielsen},
  {Lenzen}, {Brandner}, {McCarthy}, {Hartung}, {Kellner}, {Mamajek}, {Henning},
  {Miller}, {Kenworthy}, and {Kulesa}}]{Biller07}
{Biller} BA, {Close} LM, {Masciadri} E, {Nielsen} E, {Lenzen} R, {Brandner} W,
  {McCarthy} D, {Hartung} M, {Kellner} S, {Mamajek} E, {Henning} T, {Miller} D,
  {Kenworthy} M, {Kulesa} C (2007) {An Imaging Survey for Extrasolar Planets
  around 45 Close, Young Stars with the Simultaneous Differential Imager at the
  Very Large Telescope and MMT}. \apjs 173:143--165, \doi{10.1086/519925}

\bibitem[{{Boccaletti} et~al(2009){Boccaletti}, {Augereau}, {Baudoz}, {Pantin},
  and {Lagrange}}]{Boccaletti09}
{Boccaletti} A, {Augereau} J, {Baudoz} P, {Pantin} E, {Lagrange} A (2009)
  {VLT/NACO coronagraphic observations of fine structures in the disk of
  {$\beta$} Pictoris}. \aap 495:523--535, \doi{10.1051/0004-6361:200811067},
  \eprint{0901.2034}

\bibitem[{{Boss}(1997)}]{Boss97}
{Boss} AP (1997) {Giant planet formation by gravitational instability.} Science
  276:1836--1839, \doi{10.1126/science.276.5320.1836}

\bibitem[{{Bouwman} et~al(2001){Bouwman}, {Meeus}, {de Koter}, {Hony},
  {Dominik}, and {Waters}}]{Bouwman01}
{Bouwman} J, {Meeus} G, {de Koter} A, {Hony} S, {Dominik} C, {Waters} LBFM
  (2001) {Processing of silicate dust grains in Herbig Ae/Be systems}. \aap
  375:950--962, \doi{10.1051/0004-6361:20010878}

\bibitem[{{Bouwman} et~al(2003){Bouwman}, {de Koter}, {Dominik}, and
  {Waters}}]{Bouwman03}
{Bouwman} J, {de Koter} A, {Dominik} C, {Waters} LBFM (2003) {The origin of
  crystalline silicates in the Herbig Be star HD 100546 and in comet
  Hale-Bopp}. \aap 401:577--592, \doi{10.1051/0004-6361:20030043},
  \eprint{arXiv:astro-ph/0301254}

\bibitem[{{Brittain} and {Rettig}(2002)}]{Brittain02}
{Brittain} SD, {Rettig} TW (2002) {CO and H3+ in the protoplanetary disk around
  the star HD141569}. \nat 418:57--59

\bibitem[{{Brown} et~al(2008){Brown}, {Blake}, {Qi}, {Dullemond}, and
  {Wilner}}]{Brown08}
{Brown} JM, {Blake} GA, {Qi} C, {Dullemond} CP, {Wilner} DJ (2008)
  {LkH{$\alpha$} 330: Evidence for Dust Clearing through Resolved Submillimeter
  Imaging}. \apjl 675:L109--L112, \doi{10.1086/533464}, \eprint{0802.0998}

\bibitem[{{Burrows} et~al(1997){Burrows}, {Marley}, {Hubbard}, {Lunine},
  {Guillot}, {Saumon}, {Freedman}, {Sudarsky}, and {Sharp}}]{Burrows97}
{Burrows} A, {Marley} M, {Hubbard} WB, {Lunine} JI, {Guillot} T, {Saumon} D,
  {Freedman} R, {Sudarsky} D, {Sharp} C (1997) {A Nongray Theory of Extrasolar
  Giant Planets and Brown Dwarfs}. \apj 491:856--875, \doi{10.1086/305002},
  \eprint{arXiv:astro-ph/9705201}

\bibitem[{{Burrows} et~al(1996){Burrows}, {Stapelfeldt}, {Watson}, {Krist},
  {Ballester}, {Clarke}, {Crisp}, {Gallagher}, {Griffiths}, {Hester},
  {Hoessel}, {Holtzman}, {Mould}, {Scowen}, {Trauger}, and
  {Westphal}}]{Burrows96}
{Burrows} CJ, {Stapelfeldt} KR, {Watson} AM, {Krist} JE, {Ballester} GE,
  {Clarke} JT, {Crisp} D, {Gallagher} JS III, {Griffiths} RE, {Hester} JJ,
  {Hoessel} JG, {Holtzman} JA, {Mould} JR, {Scowen} PA, {Trauger} JT,
  {Westphal} JA (1996) {Hubble Space Telescope Observations of the Disk and Jet
  of HH 30}. \apj 473:437--+, \doi{10.1086/178156}

\bibitem[{{Calvet} et~al(2002){Calvet}, {D'Alessio}, {Hartmann}, {Wilner},
  {Walsh}, and {Sitko}}]{Calvet02}
{Calvet} N, {D'Alessio} P, {Hartmann} L, {Wilner} D, {Walsh} A, {Sitko} M
  (2002) {Evidence for a Developing Gap in a 10 Myr Old Protoplanetary Disk}.
  \apj 568:1008--1016, \doi{10.1086/339061}, \eprint{arXiv:astro-ph/0201425}

\bibitem[{{Carson} et~al(2005){Carson}, {Eikenberry}, {Brandl}, {Wilson}, and
  {Hayward}}]{Carson05}
{Carson} JC, {Eikenberry} SS, {Brandl} BR, {Wilson} JC, {Hayward} TL (2005)
  {The Cornell High-Order Adaptive Optics Survey for Brown Dwarfs in Stellar
  Systems. I. Observations, Data Reduction, and Detection Analyses}. \aj
  130:1212--1220, \doi{10.1086/432604}, \eprint{arXiv:astro-ph/0506287}

\bibitem[{{Chauvin} et~al(2004){Chauvin}, {Lagrange}, {Dumas}, {Zuckerman},
  {Mouillet}, {Song}, {Beuzit}, and {Lowrance}}]{Chauvin04}
{Chauvin} G, {Lagrange} AM, {Dumas} C, {Zuckerman} B, {Mouillet} D, {Song} I,
  {Beuzit} JL, {Lowrance} P (2004) {A giant planet candidate near a young brown
  dwarf. Direct VLT/NACO observations using IR wavefront sensing}. \aap
  425:L29--L32, \doi{10.1051/0004-6361:200400056},
  \eprint{arXiv:astro-ph/0409323}

\bibitem[{{Chauvin} et~al(2005){Chauvin}, {Lagrange}, {Zuckerman}, {Dumas},
  {Mouillet}, {Song}, {Beuzit}, {Lowrance}, and {Bessell}}]{Chauvin05b}
{Chauvin} G, {Lagrange} AM, {Zuckerman} B, {Dumas} C, {Mouillet} D, {Song} I,
  {Beuzit} JL, {Lowrance} P, {Bessell} MS (2005) {A companion to AB Pic at the
  planet/brown dwarf boundary}. \aap 438:L29--L32

\bibitem[{{Chauvin} et~al(2006){Chauvin}, {Lagrange}, {Udry}, {Fusco},
  {Galland}, {Naef}, {Beuzit}, and {Mayor}}]{Chauvin06}
{Chauvin} G, {Lagrange} AM, {Udry} S, {Fusco} T, {Galland} F, {Naef} D,
  {Beuzit} JL, {Mayor} M (2006) {Probing long-period companions to planetary
  hosts. VLT and CFHT near infrared coronographic imaging surveys}. \aap
  456:1165--1172, \doi{10.1051/0004-6361:20054709},
  \eprint{arXiv:astro-ph/0606166}

\bibitem[{{Chen} and {Jura}(2001)}]{Chen01}
{Chen} CH, {Jura} M (2001) {A Possible Massive Asteroid Belt around {$\zeta$}
  Leporis}. \apjl 560:L171--L174, \doi{10.1086/324057},
  \eprint{arXiv:astro-ph/0109216}

\bibitem[{{Chiang} and {Murray-Clay}(2007)}]{Chiang07}
{Chiang} E, {Murray-Clay} R (2007) {Inside-out evacuation of transitional
  protoplanetary discs by the magneto-rotational instability}. Nature Physics
  3:604--608, \doi{10.1038/nphys661}, \eprint{0706.1241}

\bibitem[{{Chiang} and {Goldreich}(1997)}]{Chiang97}
{Chiang} EI, {Goldreich} P (1997) {Spectral Energy Distributions of T Tauri
  Stars with Passive Circumstellar Disks}. \apj 490:368--376

\bibitem[{{Ciardi} et~al(2001){Ciardi}, {van Belle}, {Akeson}, {Thompson},
  {Lada}, and {Howell}}]{Ciardi01}
{Ciardi} DR, {van Belle} GT, {Akeson} RL, {Thompson} RR, {Lada} EA, {Howell} SB
  (2001) {On the near-infrared size of Vega}. \apj 559:1147--1154

\bibitem[{{Cieza} et~al(2007){Cieza}, {Padgett}, {Stapelfeldt}, {Augereau},
  {Harvey}, {Evans}, {Mer{\'{\i}}n}, {Koerner}, {Sargent}, {van Dishoeck},
  {Allen}, {Blake}, {Brooke}, {Chapman}, {Huard}, {Lai}, {Mundy}, {Myers},
  {Spiesman}, and {Wahhaj}}]{Cieza07}
{Cieza} L, {Padgett} DL, {Stapelfeldt} KR, {Augereau} JC, {Harvey} P, {Evans}
  NJ II, {Mer{\'{\i}}n} B, {Koerner} D, {Sargent} A, {van Dishoeck} EF, {Allen}
  L, {Blake} G, {Brooke} T, {Chapman} N, {Huard} T, {Lai} SP, {Mundy} L,
  {Myers} PC, {Spiesman} W, {Wahhaj} Z (2007) {The Spitzer c2d Survey of
  Weak-Line T Tauri Stars. II. New Constraints on the Timescale for Planet
  Building}. \apj 667:308--328, \doi{10.1086/520698}, \eprint{0706.0563}

\bibitem[{{Clampin} et~al(2003){Clampin}, {Krist}, {Ardila}, {Golimowski},
  {Hartig}, {Ford}, {Illingworth}, {Bartko}, {Ben{\'{\i}}tez}, {Blakeslee},
  {Bouwens}, {Broadhurst}, {Brown}, {Burrows}, {Cheng}, {Cross}, {Feldman},
  {Franx}, {Gronwall}, {Infante}, {Kimble}, {Lesser}, {Martel}, {Menanteau},
  {Meurer}, {Miley}, {Postman}, {Rosati}, {Sirianni}, {Sparks}, {Tran},
  {Tsvetanov}, {White}, and {Zheng}}]{Clampin03}
{Clampin} M, {Krist} JE, {Ardila} DR, {Golimowski} DA, {Hartig} GF, {Ford} HC,
  {Illingworth} GD, {Bartko} F, {Ben{\'{\i}}tez} N, {Blakeslee} JP, {Bouwens}
  RJ, {Broadhurst} TJ, {Brown} RA, {Burrows} CJ, {Cheng} ES, {Cross} NJG,
  {Feldman} PD, {Franx} M, {Gronwall} C, {Infante} L, {Kimble} RA, {Lesser} MP,
  {Martel} AR, {Menanteau} F, {Meurer} GR, {Miley} GK, {Postman} M, {Rosati} P,
  {Sirianni} M, {Sparks} WB, {Tran} HD, {Tsvetanov} ZI, {White} RL, {Zheng} W
  (2003) {Hubble Space Telescope ACS Coronagraphic Imaging of the Circumstellar
  Disk around HD 141569A}. \aj 126:385--392, \doi{10.1086/375460},
  \eprint{arXiv:astro-ph/0303605}

\bibitem[{{Clarke} et~al(2001){Clarke}, {Gendrin}, and {Sotomayor}}]{Clarke01}
{Clarke} CJ, {Gendrin} A, {Sotomayor} M (2001) {The dispersal of circumstellar
  discs: the role of the ultraviolet switch}. \mnras 328:485--491,
  \doi{10.1046/j.1365-8711.2001.04891.x}

\bibitem[{{Colavita} et~al(2003){Colavita}, {Akeson}, {Wizinowich}, {Shao},
  {Acton}, {Beletic}, {Bell}, {Berlin}, {Boden}, {Booth}, {Boutell}, {Chaffee},
  {Chan}, {Chock}, {Cohen}, {Crawford}, {Creech-Eakman}, {Eychaner},
  {Felizardo}, {Gathright}, {Hardy}, {Henderson}, {Herstein}, {Hess},
  {Hovland}, {Hrynevych}, {Johnson}, {Kelley}, {Kendrick}, {Koresko}, {Kurpis},
  {Le Mignant}, {Lewis}, {Ligon}, {Lupton}, {McBride}, {Mennesson},
  {Millan-Gabet}, {Monnier}, {Moore}, {Nance}, {Neyman}, {Niessner}, {Palmer},
  {Reder}, {Rudeen}, {Saloga}, {Sargent}, {Serabyn}, {Smythe}, {Stomski},
  {Summers}, {Swain}, {Swanson}, {Thompson}, {Tsubota}, {Tumminello}, {van
  Belle}, {Vasisht}, {Vause}, {Walker}, {Wallace}, and {Wehmeier}}]{Colavita03}
{Colavita} M, {Akeson} R, {Wizinowich} P, {Shao} M, {Acton} S, {Beletic} J,
  {Bell} J, {Berlin} J, {Boden} A, {Booth} A, {Boutell} R, {Chaffee} F, {Chan}
  D, {Chock} J, {Cohen} R, {Crawford} S, {Creech-Eakman} M, {Eychaner} G,
  {Felizardo} C, {Gathright} J, {Hardy} G, {Henderson} H, {Herstein} J, {Hess}
  M, {Hovland} E, {Hrynevych} M, {Johnson} R, {Kelley} J, {Kendrick} R,
  {Koresko} C, {Kurpis} P, {Le Mignant} D, {Lewis} H, {Ligon} E, {Lupton} W,
  {McBride} D, {Mennesson} B, {Millan-Gabet} R, {Monnier} J, {Moore} J, {Nance}
  C, {Neyman} C, {Niessner} A, {Palmer} D, {Reder} L, {Rudeen} A, {Saloga} T,
  {Sargent} A, {Serabyn} E, {Smythe} R, {Stomski} P, {Summers} K, {Swain} M,
  {Swanson} P, {Thompson} R, {Tsubota} K, {Tumminello} A, {van Belle} G,
  {Vasisht} G, {Vause} J, {Walker} J, {Wallace} K, {Wehmeier} U (2003)
  {Observations of DG Tauri with the Keck Interferometer}. \apjl 592:L83--L86,
  \doi{10.1086/377704}, \eprint{arXiv:astro-ph/0307051}

\bibitem[{{Connelley} et~al(2007){Connelley}, {Reipurth}, and
  {Tokunaga}}]{Connelley07}
{Connelley} MS, {Reipurth} B, {Tokunaga} AT (2007) {Infrared Nebulae around
  Young Stellar Objects}. \aj 133:1528--1559, \doi{10.1086/511745},
  \eprint{arXiv:astro-ph/0611634}

\bibitem[{{Cotera} et~al(2001){Cotera}, {Whitney}, {Young}, {Wolff}, {Wood},
  {Povich}, {Schneider}, {Rieke}, and {Thompson}}]{Cotera01}
{Cotera} AS, {Whitney} BA, {Young} E, {Wolff} MJ, {Wood} K, {Povich} M,
  {Schneider} G, {Rieke} M, {Thompson} R (2001) {High-Resolution Near-Infrared
  Images and Models of the Circumstellar Disk in HH 30}. \apj 556:958--969,
  \doi{10.1086/321627}, \eprint{arXiv:astro-ph/0104066}

\bibitem[{{Cuzzi} et~al(2008){Cuzzi}, {Hogan}, and {Shariff}}]{Cuzzi08}
{Cuzzi} JN, {Hogan} RC, {Shariff} K (2008) {Toward Planetesimals: Dense
  Chondrule Clumps in the Protoplanetary Nebula}. \apj 687:1432--1447,
  \doi{10.1086/591239}, \eprint{0804.3526}

\bibitem[{{Debes} et~al(2008){Debes}, {Weinberger}, and {Schneider}}]{Debes08}
{Debes} JH, {Weinberger} AJ, {Schneider} G (2008) {Complex Organic Materials in
  the Circumstellar Disk of HR 4796A}. \apjl 673:L191--L194,
  \doi{10.1086/527546}, \eprint{arXiv:0712.3283}

\bibitem[{{Deller} and {Maddison}(2005)}]{Deller05}
{Deller} AT, {Maddison} ST (2005) {Numerical Modeling of Dusty Debris Disks}.
  \apj 625:398--413, \doi{10.1086/429365}, \eprint{arXiv:astro-ph/0502135}

\bibitem[{{Di Folco} et~al(2004){Di Folco}, {Th\'evenin}, {Kervella},
  {Domiciano de Souza}, {Coud\'e du Foresto}, {S\'egransan}, and
  {Morel}}]{DiFolco04}
{Di Folco} E, {Th\'evenin} F, {Kervella} P, {Domiciano de Souza} A, {Coud\'e du
  Foresto} V, {S\'egransan} D, {Morel} P (2004) {VLTI near-IR interferometric
  observations of Vega-like stars}. \aap 426:601--617

\bibitem[{{Di Folco} et~al(2007){Di Folco}, {Absil}, {Augereau}, {M\'erand},
  {Coud\'e du Foresto}, {Th\'evenin}, {Defr\`ere}, {Kervella}, {ten
  Brummelaar}, {McAlister}, {Ridgway}, {Sturmann}, {Strumann}, and
  {Turner}}]{DiFolco07}
{Di Folco} E, {Absil} O, {Augereau} JC, {M\'erand} A, {Coud\'e du Foresto} V,
  {Th\'evenin} F, {Defr\`ere} D, {Kervella} P, {ten Brummelaar} TA, {McAlister}
  HA, {Ridgway} ST, {Sturmann} J, {Strumann} L, {Turner} NH (2007) {A
  near-infrared interferometric survey of debris-disk stars. I. Probing the hot
  dust content around $\epsilon$~Eri and $\tau$~Cet with CHARA/FLUOR}. \aap
  475:243--250

\bibitem[{{Doucet} et~al(2006){Doucet}, {Pantin}, {Lagage}, and
  {Dullemond}}]{Doucet06}
{Doucet} C, {Pantin} E, {Lagage} PO, {Dullemond} CP (2006) {Mid-infrared
  imaging of the circumstellar dust around three Herbig Ae stars: HD~135344,
  CQ~Tau, and HD~163296}. \aap 460:117--124, \doi{10.1051/0004-6361:20054371},
  \eprint{arXiv:astro-ph/0608615}

\bibitem[{{Duch{\^e}ne}(2008)}]{Duchene08}
{Duch{\^e}ne} G (2008) {High-angular resolution imaging of disks and planets}.
  New Astronomy Review 52:117--144, \doi{10.1016/j.newar.2008.04.007}

\bibitem[{{Duch{\^e}ne} et~al(2004){Duch{\^e}ne}, {McCabe}, {Ghez}, and
  {Macintosh}}]{Duchene04}
{Duch{\^e}ne} G, {McCabe} C, {Ghez} AM, {Macintosh} BA (2004) {A
  Multiwavelength Scattered Light Analysis of the Dust Grain Population in the
  GG Tauri Circumbinary Ring}. \apj 606:969--982, \doi{10.1086/383126},
  \eprint{arXiv:astro-ph/0401560}

\bibitem[{{Dullemond} and {Dominik}(2004)}]{Dullemond04}
{Dullemond} CP, {Dominik} C (2004) {Flaring vs. self-shadowed disks: The SEDs
  of Herbig Ae/Be stars}. \aap 417:159--168, \doi{10.1051/0004-6361:20031768},
  \eprint{arXiv:astro-ph/0401495}

\bibitem[{{Dullemond} and {Dominik}(2005)}]{Dullemond05}
{Dullemond} CP, {Dominik} C (2005) {Dust coagulation in protoplanetary disks: A
  rapid depletion of small grains}. \aap 434:971--986,
  \doi{10.1051/0004-6361:20042080}, \eprint{arXiv:astro-ph/0412117}

\bibitem[{{Dullemond} et~al(2001){Dullemond}, {Dominik}, and
  {Natta}}]{Dullemond01}
{Dullemond} CP, {Dominik} C, {Natta} A (2001) {Passive Irradiated Circumstellar
  Disks with an Inner Hole}. \apj 560:957--969, \eprint{arXiv:astro-ph/0106470}

\bibitem[{{Dupuy} et~al(2009){Dupuy}, {Liu}, and {Ireland}}]{Dupuy09}
{Dupuy} TJ, {Liu} MC, {Ireland} MJ (2009) {Dynamical Mass of the Substellar
  Benchmark Binary HD 130948BC}. \apj 692:729--752,
  \doi{10.1088/0004-637X/692/1/729}, \eprint{arXiv:0807.2450}

\bibitem[{{Dutrey} et~al(2008){Dutrey}, {Guilloteau}, {Pi{\'e}tu}, {Chapillon},
  {Gueth}, {Henning}, {Launhardt}, {Pavlyuchenkov}, {Schreyer}, and
  {Semenov}}]{Dutrey08}
{Dutrey} A, {Guilloteau} S, {Pi{\'e}tu} V, {Chapillon} E, {Gueth} F, {Henning}
  T, {Launhardt} R, {Pavlyuchenkov} Y, {Schreyer} K, {Semenov} D (2008)
  {Cavities in inner disks: the GM Aurigae case}. \aap 490:L15--L18,
  \doi{10.1051/0004-6361:200810732}

\bibitem[{{Eisner}(2007)}]{Eisner07}
{Eisner} JA (2007) {Water vapour and hydrogen in the terrestrial-planet-forming
  region of a protoplanetary disk}. \nat 447:562--564

\bibitem[{{Eisner} et~al(2003){Eisner}, {Lane}, {Akeson}, {Hillenbrand}, and
  {Sargent}}]{Eisner03}
{Eisner} JA, {Lane} BF, {Akeson} RL, {Hillenbrand} LA, {Sargent} AI (2003)
  {Near-Infrared Interferometric Measurements of Herbig Ae/Be Stars}. \apj
  588:360--372

\bibitem[{{Eisner} et~al(2004){Eisner}, {Lane}, {Hillenbrand}, {Akeson}, and
  {Sargent}}]{Eisner04}
{Eisner} JA, {Lane} BF, {Hillenbrand} LA, {Akeson} RL, {Sargent} AI (2004)
  {Resolved Inner Disks around Herbig Ae/Be Stars}. \apj 613:1049--1071

\bibitem[{{Eisner} et~al(2005){Eisner}, {Hillenbrand}, {White}, {Akeson}, and
  {Sargent}}]{Eisner05}
{Eisner} JA, {Hillenbrand} LA, {White} RJ, {Akeson} RL, {Sargent} AI (2005)
  {Observations of T Tauri Disks at Sub-AU Radii: Implications for
  Magnetospheric Accretion and Planet Formation}. \apj 623:952--966

\bibitem[{{Eisner} et~al(2006){Eisner}, {Chiang}, and {Hillenbrand}}]{Eisner06}
{Eisner} JA, {Chiang} EI, {Hillenbrand} LA (2006) {Spatially Resolving the
  Inner Disk of TW Hydrae}. \apj 637:L133--L136

\bibitem[{{Eisner} et~al(2007{\natexlab{a}}){Eisner}, {Chiang}, {Lane}, and
  {Akeson}}]{Eisner07a}
{Eisner} JA, {Chiang} EI, {Lane} BF, {Akeson} RL (2007{\natexlab{a}})
  {Spectrally Dispersed K-Band Interferometric Observations of Herbig Ae/Be
  Sources: Inner Disk Temperature Profiles}. \apj 657:347--358

\bibitem[{{Eisner} et~al(2007{\natexlab{b}}){Eisner}, {Hillenbrand}, {White},
  {Bloom}, {Akeson}, and {Blake}}]{Eisner07b}
{Eisner} JA, {Hillenbrand} LA, {White} RJ, {Bloom} JS, {Akeson} RL, {Blake} CH
  (2007{\natexlab{b}}) {Near-Infrared Interferometric, Spectroscopic, and
  Photometric Monitoring of T Tauri Inner Disks}. \apj 669:1072--1084

\bibitem[{{Eisner} et~al(2009){Eisner}, {Graham}, {Akeson}, and
  {Najita}}]{Eisner09}
{Eisner} JA, {Graham} JR, {Akeson} RL, {Najita} J (2009) {Spatially Resolved
  Spectroscopy of Sub-AU-Sized Regions of T Tauri and Herbig Ae/Be Disks}. \apj
  692:309--323, \doi{10.1088/0004-637X/692/1/309}, \eprint{0809.5054}

\bibitem[{{Ferrari} et~al(2007){Ferrari}, {Soummer}, and {Aime}}]{Ferrari07}
{Ferrari} A, {Soummer} R, {Aime} C (2007) {An introduction to stellar
  coronagraphy}. Comptes Rendus Physique 8:277--287,
  \doi{10.1016/j.crhy.2007.05.008}, \eprint{arXiv:astro-ph/0703655}

\bibitem[{{Fisher} et~al(2000){Fisher}, {Telesco}, {Pi{\~n}a}, {Knacke}, and
  {Wyatt}}]{Fisher00}
{Fisher} RS, {Telesco} CM, {Pi{\~n}a} RK, {Knacke} RF, {Wyatt} MC (2000)
  {Detection of Extended Thermal Infrared Emission around the Vega-like Source
  HD 141569}. \apjl 532:L141--L144, \doi{10.1086/312575}

\bibitem[{{Fitzgerald} et~al(2007){Fitzgerald}, {Kalas}, {Duch{\^e}ne},
  {Pinte}, and {Graham}}]{Fitzgerald07}
{Fitzgerald} MP, {Kalas} PG, {Duch{\^e}ne} G, {Pinte} C, {Graham} JR (2007)
  {The AU Microscopii Debris Disk: Multiwavelength Imaging and Modeling}. \apj
  670:536--556, \doi{10.1086/521344}, \eprint{0705.4196}

\bibitem[{{Fortney} et~al(2008){Fortney}, {Marley}, {Saumon}, and
  {Lodders}}]{Fortney08}
{Fortney} JJ, {Marley} MS, {Saumon} D, {Lodders} K (2008) {Synthetic Spectra
  and Colors of Young Giant Planet Atmospheres: Effects of Initial Conditions
  and Atmospheric Metallicity}. \apj 683:1104--1116, \doi{10.1086/589942},
  \eprint{0805.1066}

\bibitem[{{Freistetter} et~al(2007){Freistetter}, {Krivov}, and
  {L{\"o}hne}}]{Freistetter07}
{Freistetter} F, {Krivov} AV, {L{\"o}hne} T (2007) {Planets of {$\beta$}
  Pictoris revisited}. \aap 466:389--393, \doi{10.1051/0004-6361:20066746},
  \eprint{arXiv:astro-ph/0701526}

\bibitem[{{Fujiwara} et~al(2006){Fujiwara}, {Honda}, {Kataza}, {Yamashita},
  {Onaka}, {Fukagawa}, {Okamoto}, {Miyata}, {Sako}, {Fujiyoshi}, and
  {Sakon}}]{Fujiwara06}
{Fujiwara} H, {Honda} M, {Kataza} H, {Yamashita} T, {Onaka} T, {Fukagawa} M,
  {Okamoto} YK, {Miyata} T, {Sako} S, {Fujiyoshi} T, {Sakon} I (2006) {The
  Asymmetric Thermal Emission of the Protoplanetary Disk Surrounding HD 142527
  Seen by Subaru/COMICS}. \apjl 644:L133--L136, \doi{10.1086/505597},
  \eprint{arXiv:astro-ph/0701808}

\bibitem[{{Fukagawa} et~al(2003){Fukagawa}, {Tamura}, {Itoh}, {Hayashi}, and
  {Oasa}}]{Fukagawa03}
{Fukagawa} M, {Tamura} M, {Itoh} Y, {Hayashi} SS, {Oasa} Y (2003)
  {Near-Infrared Imaging of the Circumstellar Disk around Herbig Ae Star HD
  150193A}. \apjl 590:L49--L52, \doi{10.1086/376685}

\bibitem[{{Fukagawa} et~al(2004){Fukagawa}, {Hayashi}, {Tamura}, {Itoh},
  {Hayashi}, {Oasa}, {Takeuchi}, {Morino}, {Murakawa}, {Oya}, {Yamashita},
  {Suto}, {Mayama}, {Naoi}, {Ishii}, {Pyo}, {Nishikawa}, {Takato}, {Usuda},
  {Ando}, {Iye}, {Miyama}, and {Kaifu}}]{Fukagawa04}
{Fukagawa} M, {Hayashi} M, {Tamura} M, {Itoh} Y, {Hayashi} SS, {Oasa} Y,
  {Takeuchi} T, {Morino} Ji, {Murakawa} K, {Oya} S, {Yamashita} T, {Suto} H,
  {Mayama} S, {Naoi} T, {Ishii} M, {Pyo} TS, {Nishikawa} T, {Takato} N, {Usuda}
  T, {Ando} H, {Iye} M, {Miyama} SM, {Kaifu} N (2004) {Spiral Structure in the
  Circumstellar Disk around AB Aurigae}. \apjl 605:L53--L56,
  \doi{10.1086/420699}

\bibitem[{{Fukagawa} et~al(2006){Fukagawa}, {Tamura}, {Itoh}, {Kudo}, {Imaeda},
  {Oasa}, {Hayashi}, and {Hayashi}}]{Fukagawa06}
{Fukagawa} M, {Tamura} M, {Itoh} Y, {Kudo} T, {Imaeda} Y, {Oasa} Y, {Hayashi}
  SS, {Hayashi} M (2006) {Near-Infrared Images of Protoplanetary Disk
  Surrounding HD 142527}. \apjl 636:L153--L156, \doi{10.1086/500128}

\bibitem[{{Gail}(2004)}]{Gail04}
{Gail} HP (2004) {Radial mixing in protoplanetary accretion disks. IV.
  Metamorphosis of the silicate dust complex}. \aap 413:571--591,
  \doi{10.1051/0004-6361:20031554}

\bibitem[{{Geers} et~al(2007){Geers}, {Pontoppidan}, {van Dishoeck},
  {Dullemond}, {Augereau}, {Mer{\'{\i}}n}, {Oliveira}, and {Pel}}]{Geers07}
{Geers} VC, {Pontoppidan} KM, {van Dishoeck} EF, {Dullemond} CP, {Augereau} JC,
  {Mer{\'{\i}}n} B, {Oliveira} I, {Pel} JW (2007) {Spatial separation of small
  and large grains in the transitional disk around the young star IRS 48}. \aap
  469:L35--L38, \doi{10.1051/0004-6361:20077524}, \eprint{arXiv:0705.2969}

\bibitem[{{Golimowski} et~al(2007){Golimowski}, {John Krist}, {Chen},
  {Stapelfeldt}, {Ardila}, {Clampin}, {Schneider}, {Silverstone}, {Ford}, and
  {Illingworth}}]{Golimowski07}
{Golimowski} D, {John Krist} J, {Chen} C, {Stapelfeldt} K, {Ardila} D,
  {Clampin} M, {Schneider} G, {Silverstone} M, {Ford} H, {Illingworth} G (2007)
  {Observations and Models of the Debris Disk around the K dwarf HD 92945}. In:
  In the Spirit of Bernard Lyot: The Direct Detection of Planets and
  Circumstellar Disks in the 21st Century

\bibitem[{{Gomes} et~al(2005){Gomes}, {Levison}, {Tsiganis}, and
  {Morbidelli}}]{Gomes05}
{Gomes} R, {Levison} HF, {Tsiganis} K, {Morbidelli} A (2005) {Origin of the
  cataclysmic Late Heavy Bombardment period of the terrestrial planets}. \nat
  435:466--469, \doi{10.1038/nature03676}

\bibitem[{{Grady} et~al(1999){Grady}, {Woodgate}, {Bruhweiler}, {Boggess},
  {Plait}, {Lindler}, {Clampin}, and {Kalas}}]{Grady99}
{Grady} CA, {Woodgate} B, {Bruhweiler} FC, {Boggess} A, {Plait} P, {Lindler}
  DJ, {Clampin} M, {Kalas} P (1999) {Hubble Space Telescope Space Telescope
  Imaging Spectrograph Coronagraphic Imaging of the Herbig AE Star AB Aurigae}.
  \apjl 523:L151--L154, \doi{10.1086/312270}

\bibitem[{{Grady} et~al(2000){Grady}, {Devine}, {Woodgate}, {Kimble},
  {Bruhweiler}, {Boggess}, {Linsky}, {Plait}, {Clampin}, and {Kalas}}]{Grady00}
{Grady} CA, {Devine} D, {Woodgate} B, {Kimble} R, {Bruhweiler} FC, {Boggess} A,
  {Linsky} JL, {Plait} P, {Clampin} M, {Kalas} P (2000) {STIS Coronagraphic
  Imaging of the Herbig AE Star: HD 163296}. \apj 544:895--902,
  \doi{10.1086/317222}

\bibitem[{{Grady} et~al(2001){Grady}, {Polomski}, {Henning}, {Stecklum},
  {Woodgate}, {Telesco}, {Pi{\~n}a}, {Gull}, {Boggess}, {Bowers}, {Bruhweiler},
  {Clampin}, {Danks}, {Green}, {Heap}, {Hutchings}, {Jenkins}, {Joseph},
  {Kaiser}, {Kimble}, {Kraemer}, {Lindler}, {Linsky}, {Maran}, {Moos}, {Plait},
  {Roesler}, {Timothy}, and {Weistrop}}]{Grady01}
{Grady} CA, {Polomski} EF, {Henning} T, {Stecklum} B, {Woodgate} BE, {Telesco}
  CM, {Pi{\~n}a} RK, {Gull} TR, {Boggess} A, {Bowers} CW, {Bruhweiler} FC,
  {Clampin} M, {Danks} AC, {Green} RF, {Heap} SR, {Hutchings} JB, {Jenkins} EB,
  {Joseph} C, {Kaiser} ME, {Kimble} RA, {Kraemer} S, {Lindler} D, {Linsky} JL,
  {Maran} SP, {Moos} HW, {Plait} P, {Roesler} F, {Timothy} JG, {Weistrop} D
  (2001) {The Disk and Environment of the Herbig Be Star HD 100546}. \aj
  122:3396--3406, \doi{10.1086/324447}

\bibitem[{{Graham} et~al(2007){Graham}, {Kalas}, and {Matthews}}]{Graham07}
{Graham} JR, {Kalas} PG, {Matthews} BC (2007) {The Signature of Primordial
  Grain Growth in the Polarized Light of the AU Microscopii Debris Disk}. \apj
  654:595--605, \doi{10.1086/509318}, \eprint{arXiv:astro-ph/0609332}

\bibitem[{{Guilloteau} et~al(1999){Guilloteau}, {Dutrey}, and
  {Simon}}]{Guilloteau99}
{Guilloteau} S, {Dutrey} A, {Simon} M (1999) {GG Tauri: the ring world}. \aap
  348:570--578

\bibitem[{{Haisch} et~al(2001){Haisch}, {Lada}, and {Lada}}]{Haisch01}
{Haisch} KE Jr, {Lada} EA, {Lada} CJ (2001) {Disk Frequencies and Lifetimes in
  Young Clusters}. \apjl 553:L153--L156, \doi{10.1086/320685},
  \eprint{arXiv:astro-ph/0104347}

\bibitem[{{Hales} et~al(2006){Hales}, {Gledhill}, {Barlow}, and
  {Lowe}}]{Hales06}
{Hales} AS, {Gledhill} TM, {Barlow} MJ, {Lowe} KTE (2006) {Near-infrared
  imaging polarimetry of dusty young stars}. \mnras 365:1348--1356,
  \doi{10.1111/j.1365-2966.2005.09820.x}, \eprint{arXiv:astro-ph/0511793}

\bibitem[{{Hartmann} and {Kenyon}(1985)}]{Hartmann85}
{Hartmann} L, {Kenyon} SJ (1985) {On the nature of FU Orionis objects}. \apj
  299:462--478

\bibitem[{{Hartmann} et~al(1998){Hartmann}, {Calvet}, {Gullbring}, and
  {D'Alessio}}]{Hartmann98}
{Hartmann} L, {Calvet} N, {Gullbring} E, {D'Alessio} P (1998) {Accretion and
  the Evolutiion of T Tauri Disks}. \apj 495:385--+, \doi{10.1086/305277}

\bibitem[{{Hartmann} et~al(1999){Hartmann}, {Calvet}, {Allen}, {Chen}, and
  {Jayawardhana}}]{Hartmann99}
{Hartmann} L, {Calvet} N, {Allen} L, {Chen} H, {Jayawardhana} R (1999) {The
  Complex Protostellar Source IRAS 04325+2402}. \aj 118:1784--1790,
  \doi{10.1086/301040}, \eprint{arXiv:astro-ph/9906280}

\bibitem[{{Heap} et~al(2000){Heap}, {Lindler}, {Lanz}, {Cornett}, {Hubeny},
  {Maran}, and {Woodgate}}]{Heap00}
{Heap} SR, {Lindler} DJ, {Lanz} TM, {Cornett} RH, {Hubeny} I, {Maran} SP,
  {Woodgate} B (2000) {Space Telescope Imaging Spectrograph Coronagraphic
  Observations of {$\beta$} Pictoris}. \apj 539:435--444, \doi{10.1086/309188}

\bibitem[{{Henning}(2008)}]{Henning08}
{Henning} T (2008) {Early phases of planet formation in protoplanetary disks}.
  Physica Scripta Volume T 130(1):014,019,
  \doi{10.1088/0031-8949/2008/T130/014019}

\bibitem[{{Hillenbrand} et~al(1992){Hillenbrand}, {Strom}, {Vrba}, and
  {Keene}}]{Hillenbrand92}
{Hillenbrand} LA, {Strom} SE, {Vrba} FJ, {Keene} J (1992) {Herbig Ae/Be stars -
  Intermediate-mass stars surrounded by massive circumstellar accretion disks}.
  \apj 397:613--643

\bibitem[{{Hines} et~al(2007){Hines}, {Schneider}, {Hollenbach}, {Mamajek},
  {Hillenbrand}, {Metchev}, {Meyer}, {Carpenter}, {Moro-Mart{\'{\i}}n},
  {Silverstone}, {Kim}, {Henning}, {Bouwman}, and {Wolf}}]{Hines07}
{Hines} DC, {Schneider} G, {Hollenbach} D, {Mamajek} EE, {Hillenbrand} LA,
  {Metchev} SA, {Meyer} MR, {Carpenter} JM, {Moro-Mart{\'{\i}}n} A,
  {Silverstone} MD, {Kim} JS, {Henning} T, {Bouwman} J, {Wolf} S (2007) {The
  Moth: An Unusual Circumstellar Structure Associated with HD 61005}. \apjl
  671:L165--L168, \doi{10.1086/525016}

\bibitem[{{Hinz} et~al(2001){Hinz}, {Hoffmann}, and {Hora}}]{Hinz01}
{Hinz} PM, {Hoffmann} WF, {Hora} JL (2001) {Constraints on Disk Sizes around
  Young Intermediate-Mass Stars: Nulling Interferometric Observations of Herbig
  Ae Objects}. \apj 561:L131--L134

\bibitem[{{Hodapp} et~al(2004){Hodapp}, {Walker}, {Reipurth}, {Wood}, {Bally},
  {Whitney}, and {Connelley}}]{Hodapp04}
{Hodapp} KW, {Walker} CH, {Reipurth} B, {Wood} K, {Bally} J, {Whitney} BA,
  {Connelley} M (2004) {A Disk Shadow around the Young Star ASR 41 in NGC
  1333}. \apjl 601:L79--L82, \doi{10.1086/381732},
  \eprint{arXiv:astro-ph/0312256}

\bibitem[{{Hughes} et~al(2007){Hughes}, {Wilner}, {Calvet}, {D'Alessio},
  {Claussen}, and {Hogerheijde}}]{Hughes07}
{Hughes} AM, {Wilner} DJ, {Calvet} N, {D'Alessio} P, {Claussen} MJ,
  {Hogerheijde} MR (2007) {An Inner Hole in the Disk around TW Hydrae Resolved
  in 7 mm Dust Emission}. \apj 664:536--542, \doi{10.1086/518885},
  \eprint{arXiv:0704.2422}

\bibitem[{{Ireland} and {Kraus}(2008)}]{Ireland08}
{Ireland} MJ, {Kraus} AL (2008) {The Disk Around CoKu Tauri/4: Circumbinary,
  Not Transitional}. \apjl 678:L59--L62, \doi{10.1086/588216},
  \eprint{arXiv:0803.2044}

\bibitem[{{Isella} and {Natta}(2005)}]{Isella05}
{Isella} A, {Natta} A (2005) {The shape of the inner rim in proto-planetary
  disks}. \aap 438:899--907

\bibitem[{{Isella} et~al(2006){Isella}, {Testi}, and {Natta}}]{Isella06}
{Isella} A, {Testi} L, {Natta} A (2006) {Large dust grains in the inner region
  of circumstellar disks}. \aap 451:951--959

\bibitem[{{Isella} et~al(2008){Isella}, {Tatulli}, {Natta}, and
  {Testi}}]{Isella08}
{Isella} A, {Tatulli} E, {Natta} A, {Testi} L (2008) {Gas and dust in the inner
  disk of the Herbig Ae star MWC 758}. \aap 483:L13--L16,
  \doi{10.1051/0004-6361:200809641}, \eprint{arXiv:0803.3606}

\bibitem[{{Jayawardhana} et~al(1998){Jayawardhana}, {Fisher}, {Hartmann},
  {Telesco}, {Pina}, and {Fazio}}]{Jayawardhana98}
{Jayawardhana} R, {Fisher} S, {Hartmann} L, {Telesco} C, {Pina} R, {Fazio} G
  (1998) {A Dust Disk Surrounding the Young A Star HR 4796A}. \apjl 503:L79+,
  \eprint{arXiv:astro-ph/9806188}

\bibitem[{{Jayawardhana} et~al(2002){Jayawardhana}, {Luhman}, {D'Alessio}, and
  {Stauffer}}]{Jayawardhana02}
{Jayawardhana} R, {Luhman} KL, {D'Alessio} P, {Stauffer} JR (2002) {Discovery
  of an Edge-On Disk in the MBM 12 Young Association}. \apjl 571:L51--L54,
  \doi{10.1086/341202}, \eprint{arXiv:astro-ph/0204272}

\bibitem[{{Johansen} et~al(2007){Johansen}, {Oishi}, {Low}, {Klahr}, {Henning},
  and {Youdin}}]{Johansen07}
{Johansen} A, {Oishi} JS, {Low} MMM, {Klahr} H, {Henning} T, {Youdin} A (2007)
  {Rapid planetesimal formation in turbulent circumstellar disks}. \nat
  448:1022--1025, \doi{10.1038/nature06086}

\bibitem[{{Johnson} et~al(2008){Johnson}, {Marcy}, {Fischer}, {Wright},
  {Reffert}, {Kregenow}, {Williams}, and {Peek}}]{Johnson08}
{Johnson} JA, {Marcy} GW, {Fischer} DA, {Wright} JT, {Reffert} S, {Kregenow}
  JM, {Williams} PKG, {Peek} KMG (2008) {Retired A Stars and Their Companions.
  II. Jovian planets orbiting {$\kappa$} CrB and HD 167042}. \apj 675:784--789,
  \doi{10.1086/526453}, \eprint{0711.4367}

\bibitem[{{Kalas} et~al(2004){Kalas}, {Liu}, and {Matthews}}]{Kalas04}
{Kalas} P, {Liu} MC, {Matthews} BC (2004) {Discovery of a Large Dust Disk
  Around the Nearby Star AU Microscopii}. Science 303:1990--1992

\bibitem[{{Kalas} et~al(2005){Kalas}, {Graham}, and {Clampin}}]{Kalas05}
{Kalas} P, {Graham} JR, {Clampin} M (2005) {A planetary system as the origin of
  structure in Fomalhaut's dust belt}. \nat 435:1067--1070

\bibitem[{{Kalas} et~al(2006){Kalas}, {Graham}, {Clampin}, and
  {Fitzgerald}}]{Kalas06}
{Kalas} P, {Graham} JR, {Clampin} MC, {Fitzgerald} MP (2006) {First Scattered
  Light Images of Debris Disks around HD 53143 and HD 139664}. \apjl
  637:L57--L60, \doi{10.1086/500305}, \eprint{arXiv:astro-ph/0601488}

\bibitem[{{Kalas} et~al(2007{\natexlab{a}}){Kalas}, {Duchene}, {Fitzgerald},
  and {Graham}}]{Kalas07b}
{Kalas} P, {Duchene} G, {Fitzgerald} MP, {Graham} JR (2007{\natexlab{a}})
  {Discovery of an Extended Debris Disk around the F2 V Star HD 15745}. \apjl
  671:L161--L164, \doi{10.1086/525252}, \eprint{0712.0378}

\bibitem[{{Kalas} et~al(2007{\natexlab{b}}){Kalas}, {Fitzgerald}, and
  {Graham}}]{Kalas07a}
{Kalas} P, {Fitzgerald} MP, {Graham} JR (2007{\natexlab{b}}) {Discovery of
  Extreme Asymmetry in the Debris Disk Surrounding HD 15115}. \apjl
  661:L85--L88, \doi{10.1086/518652}, \eprint{0704.0645}

\bibitem[{{Kalas} et~al(2008){Kalas}, {Graham}, {Chiang}, {Fitzgerald},
  {Clampin}, {Kite}, {Stapelfeldt}, {Marois}, and {Krist}}]{Kalas08}
{Kalas} P, {Graham} JR, {Chiang} E, {Fitzgerald} MP, {Clampin} M, {Kite} ES,
  {Stapelfeldt} K, {Marois} C, {Krist} J (2008) {Optical Images of an Exosolar
  Planet 25 Light-Years from Earth}. Science 322:1345--,
  \doi{10.1126/science.1166609}, \eprint{0811.1994}

\bibitem[{{Kashyap} et~al(2008){Kashyap}, {Drake}, and {Saar}}]{Kashyap08}
{Kashyap} VL, {Drake} JJ, {Saar} SH (2008) {Extrasolar Giant Planets and X-Ray
  Activity}. \apj 687:1339--1354, \doi{10.1086/591922}, \eprint{0807.1308}

\bibitem[{{Kelsall} et~al(1998){Kelsall}, {Weiland}, {Franz}, {Reach},
  {Arendt}, {Dwek}, {Freudenreich}, {Hauser}, {Moseley}, {Odegard},
  {Silverberg}, and {Wright}}]{Kelsall98}
{Kelsall} T, {Weiland} JL, {Franz} BA, {Reach} WT, {Arendt} RG, {Dwek} E,
  {Freudenreich} HT, {Hauser} MG, {Moseley} SH, {Odegard} NP, {Silverberg} RF,
  {Wright} EL (1998) {The COBE Diffuse Infrared Background Experiment Search
  for the Cosmic Infrared Background. II. Model of the Interplanetary Dust
  Cloud}. \apj 508:44--73, \doi{10.1086/306380},
  \eprint{arXiv:astro-ph/9806250}

\bibitem[{{Kenyon} and {Hartmann}(1991)}]{Kenyon91}
{Kenyon} SJ, {Hartmann} LW (1991) {The dusty envelopes of FU Orionis
  variables}. \apj 383:664--673, \doi{10.1086/170823}

\bibitem[{{Kessler-Silacci} et~al(2006){Kessler-Silacci}, {Augereau},
  {Dullemond}, {Geers}, {Lahuis}, {Evans}, {van Dishoeck}, {Blake}, {Boogert},
  {Brown}, {J{\o}rgensen}, {Knez}, and {Pontoppidan}}]{KesslerSilacci06}
{Kessler-Silacci} J, {Augereau} JC, {Dullemond} CP, {Geers} V, {Lahuis} F,
  {Evans} NJ II, {van Dishoeck} EF, {Blake} GA, {Boogert} ACA, {Brown} J,
  {J{\o}rgensen} JK, {Knez} C, {Pontoppidan} KM (2006) {c2d Spitzer IRS Spectra
  of Disks around T Tauri Stars. I. Silicate Emission and Grain Growth}. \apj
  639:275--291, \doi{10.1086/499330}, \eprint{arXiv:astro-ph/0511092}

\bibitem[{{Kirkpatrick} et~al(1999){Kirkpatrick}, {Reid}, {Liebert}, {Cutri},
  {Nelson}, {Beichman}, {Dahn}, {Monet}, {Gizis}, and
  {Skrutskie}}]{Kirkpatrick99}
{Kirkpatrick} JD, {Reid} IN, {Liebert} J, {Cutri} RM, {Nelson} B, {Beichman}
  CA, {Dahn} CC, {Monet} DG, {Gizis} JE, {Skrutskie} MF (1999) {Dwarfs Cooler
  than ``M'': The Definition of Spectral Type ``L'' Using Discoveries from the
  2 Micron All-Sky Survey (2MASS)}. \apj 519:802--833, \doi{10.1086/307414}

\bibitem[{{Koerner} et~al(1998){Koerner}, {Ressler}, {Werner}, and
  {Backman}}]{Koerner98}
{Koerner} DW, {Ressler} ME, {Werner} MW, {Backman} DE (1998) {Mid-Infrared
  Imaging of a Circumstellar Disk around HR 4796: Mapping the Debris of
  Planetary Formation}. \apjl 503:L83+, \doi{10.1086/311525},
  \eprint{arXiv:astro-ph/9806268}

\bibitem[{{K{\"o}hler} et~al(2008){K{\"o}hler}, {Mann}, and {Li}}]{Kohler08}
{K{\"o}hler} M, {Mann} I, {Li} A (2008) {Complex Organic Materials in the HR
  4796A Disk?} \apjl 686:L95--L98, \doi{10.1086/592961}, \eprint{0808.4113}

\bibitem[{{Kraus} et~al(2008{\natexlab{a}}){Kraus}, {Hofmann}, {Benisty},
  {Berger}, {Chesneau}, {Isella}, {Malbet}, {Meilland}, {Nardetto}, {Natta},
  {Preibisch}, {Schertl}, {Smith}, {Stee}, {Tatulli}, {Testi}, and
  {Weigelt}}]{Kraus08b}
{Kraus} S, {Hofmann} KH, {Benisty} M, {Berger} JP, {Chesneau} O, {Isella} A,
  {Malbet} F, {Meilland} A, {Nardetto} N, {Natta} A, {Preibisch} T, {Schertl}
  D, {Smith} M, {Stee} P, {Tatulli} E, {Testi} L, {Weigelt} G
  (2008{\natexlab{a}}) {The origin of hydrogen line emission for five Herbig
  Ae/Be stars spatially resolved by VLTI/AMBER spectro-interferometry}. \aap
  489:1157--1173, \doi{10.1051/0004-6361:200809946}, \eprint{0807.1119}

\bibitem[{{Kraus} et~al(2008{\natexlab{b}}){Kraus}, {Preibisch}, and
  {Ohnaka}}]{Kraus08a}
{Kraus} S, {Preibisch} T, {Ohnaka} K (2008{\natexlab{b}}) {Detection of an
  Inner Gaseous Component in a Herbig Be Star Accretion Disk: Near- and
  Mid-Infrared Spectrointerferometry and Radiative Transfer modeling of MWC
  147}. \apj 676:490--508, \eprint{arXiv:0711.4988}

\bibitem[{{Krist} et~al(2000){Krist}, {Stapelfeldt}, {M{\'e}nard}, {Padgett},
  and {Burrows}}]{Krist00}
{Krist} JE, {Stapelfeldt} KR, {M{\'e}nard} F, {Padgett} DL, {Burrows} CJ (2000)
  {WFPC2 Images of a Face-on Disk Surrounding TW Hydrae}. \apj 538:793--800,
  \doi{10.1086/309170}

\bibitem[{{Krist} et~al(2005{\natexlab{a}}){Krist}, {Ardila}, {Golimowski},
  {Clampin}, {Ford}, {Illingworth}, {Hartig}, {Bartko}, {Ben{\'{\i}}tez},
  {Blakeslee}, {Bouwens}, {Bradley}, {Broadhurst}, {Brown}, {Burrows}, {Cheng},
  {Cross}, {Demarco}, {Feldman}, {Franx}, {Goto}, {Gronwall}, {Holden},
  {Homeier}, {Infante}, {Kimble}, {Lesser}, {Martel}, {Mei}, {Menanteau},
  {Meurer}, {Miley}, {Motta}, {Postman}, {Rosati}, {Sirianni}, {Sparks},
  {Tran}, {Tsvetanov}, {White}, and {Zheng}}]{Krist05a}
{Krist} JE, {Ardila} DR, {Golimowski} DA, {Clampin} M, {Ford} HC, {Illingworth}
  GD, {Hartig} GF, {Bartko} F, {Ben{\'{\i}}tez} N, {Blakeslee} JP, {Bouwens}
  RJ, {Bradley} LD, {Broadhurst} TJ, {Brown} RA, {Burrows} CJ, {Cheng} ES,
  {Cross} NJG, {Demarco} R, {Feldman} PD, {Franx} M, {Goto} T, {Gronwall} C,
  {Holden} B, {Homeier} N, {Infante} L, {Kimble} RA, {Lesser} MP, {Martel} AR,
  {Mei} S, {Menanteau} F, {Meurer} GR, {Miley} GK, {Motta} V, {Postman} M,
  {Rosati} P, {Sirianni} M, {Sparks} WB, {Tran} HD, {Tsvetanov} ZI, {White} RL,
  {Zheng} W (2005{\natexlab{a}}) Hubble space telescope advanced camera for
  surveys coronagraphic imaging of the au microscopii debris disk. \aj
  129:1008--1017

\bibitem[{{Krist} et~al(2005{\natexlab{b}}){Krist}, {Stapelfeldt},
  {Golimowski}, {Ardila}, {Clampin}, {Martel}, {Ford}, {Illingworth}, and
  {Hartig}}]{Krist05b}
{Krist} JE, {Stapelfeldt} KR, {Golimowski} DA, {Ardila} DR, {Clampin} M,
  {Martel} AR, {Ford} HC, {Illingworth} GD, {Hartig} GF (2005{\natexlab{b}})
  {Hubble Space Telescope ACS Images of the GG Tauri Circumbinary Disk}. \aj
  130:2778--2787, \doi{10.1086/497069}, \eprint{arXiv:astro-ph/0508222}

\bibitem[{{Kuchner} and {Holman}(2003)}]{Kuchner03}
{Kuchner} MJ, {Holman} MJ (2003) {The Geometry of Resonant Signatures in Debris
  Disks with Planets}. \apj 588:1110--1120, \doi{10.1086/374213},
  \eprint{arXiv:astro-ph/0209261}

\bibitem[{{Kuhn} et~al(2001){Kuhn}, {Potter}, and {Parise}}]{Kuhn01}
{Kuhn} JR, {Potter} D, {Parise} B (2001) {Imaging Polarimetric Observations of
  a New Circumstellar Disk System}. \apjl 553:L189--L191, \doi{10.1086/320686},
  \eprint{arXiv:astro-ph/0105239}

\bibitem[{{Lafreni{\`e}re} et~al(2007){Lafreni{\`e}re}, {Doyon}, {Marois},
  {Nadeau}, {Oppenheimer}, {Roche}, {Rigaut}, {Graham}, {Jayawardhana},
  {Johnstone}, {Kalas}, {Macintosh}, and {Racine}}]{Lafreniere07}
{Lafreni{\`e}re} D, {Doyon} R, {Marois} C, {Nadeau} D, {Oppenheimer} BR,
  {Roche} PF, {Rigaut} F, {Graham} JR, {Jayawardhana} R, {Johnstone} D, {Kalas}
  PG, {Macintosh} B, {Racine} R (2007) {The Gemini Deep Planet Survey}. \apj
  670:1367--1390, \doi{10.1086/522826}, \eprint{0705.4290}

\bibitem[{{Lafreni{\`e}re} et~al(2008){Lafreni{\`e}re}, {Jayawardhana}, and
  {van Kerkwijk}}]{Lafreniere08}
{Lafreni{\`e}re} D, {Jayawardhana} R, {van Kerkwijk} MH (2008) {Direct Imaging
  and Spectroscopy of a Planetary-Mass Candidate Companion to a Young Solar
  Analog}. \apjl 689:L153--L156, \doi{10.1086/595870}, \eprint{0809.1424}

\bibitem[{{Lagage} et~al(2006){Lagage}, {Doucet}, {Pantin}, {Habart},
  {Duch{\^e}ne}, {M{\'e}nard}, {Pinte}, {Charnoz}, and {Pel}}]{Lagage06}
{Lagage} PO, {Doucet} C, {Pantin} E, {Habart} E, {Duch{\^e}ne} G, {M{\'e}nard}
  F, {Pinte} C, {Charnoz} S, {Pel} JW (2006) {Anatomy of a Flaring
  Proto-Planetary Disk Around a Young Intermediate-Mass Star}. Science
  314:621--623, \doi{10.1126/science.1131436}

\bibitem[{{Lagrange} et~al(2009){Lagrange}, {Gratadour}, {Chauvin}, {Fusco},
  {Ehrenreich}, {Mouillet}, {Rousset}, {Rouan}, {Allard}, {Gendron}, {Charton},
  {Mugnier}, {Rabou}, {Montri}, and {Lacombe}}]{Lagrange09}
{Lagrange} AM, {Gratadour} D, {Chauvin} G, {Fusco} T, {Ehrenreich} D,
  {Mouillet} D, {Rousset} G, {Rouan} D, {Allard} F, {Gendron} {\'E}, {Charton}
  J, {Mugnier} L, {Rabou} P, {Montri} J, {Lacombe} F (2009) {A probable giant
  planet imaged in the {$\beta$} Pictoris disk. VLT/NaCo deep L'-band imaging}.
  \aap 493:L21--L25, \doi{10.1051/0004-6361:200811325}

\bibitem[{{Lawson}(2000)}]{Lawson00}
{Lawson} PR (ed)  (2000) {Principles of Long Baseline Stellar Interferometry}.
  NASA/JPL, California Institute of Technology, Pasadena, CA

\bibitem[{{Leinert} and {Moster}(2007)}]{Leinert07}
{Leinert} C, {Moster} B (2007) {Evidence for dust accumulation just outside the
  orbit of Venus}. \aap 472:335--340, \doi{10.1051/0004-6361:20077682},
  \eprint{0708.0912}

\bibitem[{{Leinert} et~al(2004){Leinert}, {van Boekel}, {Waters}, {Chesneau},
  {Malbet}, {K{\"o}hler}, {Jaffe}, {Ratzka}, {Dutrey}, {Preibisch}, {Graser},
  {Bakker}, {Chagnon}, {Cotton}, {Dominik}, {Dullemond}, {Glazenborg-Kluttig},
  {Glindemann}, {Henning}, {Hofmann}, {de Jong}, {Lenzen}, {Ligori}, {Lopez},
  {Meisner}, {Morel}, {Paresce}, {Pel}, {Percheron}, {Perrin}, {Przygodda},
  {Richichi}, {Sch{\"o}ller}, {Schuller}, {Stecklum}, {van den Ancker}, {von
  der L{\"u}he}, and {Weigelt}}]{Leinert04}
{Leinert} C, {van Boekel} R, {Waters} LBFM, {Chesneau} O, {Malbet} F,
  {K{\"o}hler} R, {Jaffe} W, {Ratzka} T, {Dutrey} A, {Preibisch} T, {Graser} U,
  {Bakker} E, {Chagnon} G, {Cotton} WD, {Dominik} C, {Dullemond} CP,
  {Glazenborg-Kluttig} AW, {Glindemann} A, {Henning} T, {Hofmann} KH, {de Jong}
  J, {Lenzen} R, {Ligori} S, {Lopez} B, {Meisner} J, {Morel} S, {Paresce} F,
  {Pel} JW, {Percheron} I, {Perrin} G, {Przygodda} F, {Richichi} A,
  {Sch{\"o}ller} M, {Schuller} P, {Stecklum} B, {van den Ancker} ME, {von der
  L{\"u}he} O, {Weigelt} G (2004) {Mid-infrared sizes of circumstellar disks
  around Herbig Ae/Be stars measured with MIDI on the VLTI}. \aap 423:537--548

\bibitem[{{Lin} and {Papaloizou}(1979)}]{Lin79}
{Lin} DNC, {Papaloizou} J (1979) {Tidal torques on accretion discs in binary
  systems with extreme mass ratios}. \mnras 186:799--812

\bibitem[{{Lisse} et~al(2008){Lisse}, {Chen}, {Wyatt}, and {Morlok}}]{Lisse08}
{Lisse} CM, {Chen} CH, {Wyatt} MC, {Morlok} A (2008) {Circumstellar Dust
  Created by Terrestrial Planet Formation in HD 113766}. \apj 673:1106--1122,
  \doi{10.1086/523626}, \eprint{0710.0839}

\bibitem[{{Liu} et~al(2003){Liu}, {Hinz}, {Meyer}, {Mamajek}, {Hoffmann}, and
  {Hora}}]{Liu03}
{Liu} WM, {Hinz} PM, {Meyer} MR, {Mamajek} EE, {Hoffmann} WF, {Hora} JL (2003)
  {A Resolved Circumstellar Disk around the Herbig Ae Star HD 100546 in the
  Thermal Infrared}. \apjl 598:L111--L114, \doi{10.1086/380827},
  \eprint{arXiv:astro-ph/0310564}

\bibitem[{{Liu} et~al(2004){Liu}, {Hinz}, {Hoffmann}, {Brusa}, {Wildi},
  {Miller}, {Lloyd-Hart}, {Kenworthy}, {McGuire}, and {Angel}}]{Liu04}
{Liu} WM, {Hinz} PM, {Hoffmann} WF, {Brusa} G, {Wildi} F, {Miller} D,
  {Lloyd-Hart} M, {Kenworthy} MA, {McGuire} PC, {Angel} JRP (2004) {Adaptive
  Optics Nulling Interferometric Constraints on the Mid-Infrared Exozodiacal
  Dust Emission around Vega}. \apj 610:L125--L128

\bibitem[{{L{\"o}hne} et~al(2008){L{\"o}hne}, {Krivov}, and
  {Rodmann}}]{Lohne08}
{L{\"o}hne} T, {Krivov} AV, {Rodmann} J (2008) {Long-Term Collisional Evolution
  of Debris Disks}. \apj 673:1123--1137, \doi{10.1086/524840},
  \eprint{0710.4294}

\bibitem[{{Lowrance} et~al(2005){Lowrance}, {Becklin}, {Schneider},
  {Kirkpatrick}, {Weinberger}, {Zuckerman}, {Dumas}, {Beuzit}, {Plait},
  {Malumuth}, {Heap}, {Terrile}, and {Hines}}]{Lowrance05}
{Lowrance} PJ, {Becklin} EE, {Schneider} G, {Kirkpatrick} JD, {Weinberger} AJ,
  {Zuckerman} B, {Dumas} C, {Beuzit} JL, {Plait} P, {Malumuth} E, {Heap} S,
  {Terrile} RJ, {Hines} DC (2005) {An Infrared Coronagraphic Survey for
  Substellar Companions}. \aj 130:1845--1861, \doi{10.1086/432839},
  \eprint{arXiv:astro-ph/0506358}

\bibitem[{{Lubow} et~al(1999){Lubow}, {Seibert}, and {Artymowicz}}]{Lubow99}
{Lubow} SH, {Seibert} M, {Artymowicz} P (1999) {Disk Accretion onto High-Mass
  Planets}. \apj 526:1001--1012, \doi{10.1086/308045},
  \eprint{arXiv:astro-ph/9910404}

\bibitem[{{Malbet} et~al(1998){Malbet}, {Berger}, {Colavita}, {Koresko},
  {Beichman}, {Boden}, {Kulkarni}, {Lane}, {Mobley}, {Pan}, {Shao}, {van
  Belle}, and {Wallace}}]{Malbet98}
{Malbet} F, {Berger} JP, {Colavita} MM, {Koresko} CD, {Beichman} C, {Boden} AF,
  {Kulkarni} SR, {Lane} BF, {Mobley} DW, {Pan} XP, {Shao} M, {van Belle} GT,
  {Wallace} JK (1998) {FU Orionis Resolved by Infrared Long-Baseline
  Interferometry at a 2 AU Scale}. \apj 507:L149--L152

\bibitem[{{Malbet} et~al(2005){Malbet}, {Lachaume}, {Berger}, {Colavita}, {di
  Folco}, {Eisner}, {Lane}, {Millan-Gabet}, {S{\'e}gransan}, and
  {Traub}}]{Malbet05}
{Malbet} F, {Lachaume} R, {Berger} JP, {Colavita} MM, {di Folco} E, {Eisner}
  JA, {Lane} BF, {Millan-Gabet} R, {S{\'e}gransan} D, {Traub} WA (2005) {New
  insights on the AU-scale circumstellar structure of FU Orionis}. \aap
  437:627--636

\bibitem[{{Malbet} et~al(2007){Malbet}, {Benisty}, {de Wit}, {Kraus},
  {Meilland}, {Millour}, {Tatulli}, {Berger}, {Chesneau}, {Hofmann}, {Isella},
  {Natta}, {Petrov}, {Preibisch}, {Stee}, {Testi}, {Weigelt}, {Antonelli},
  {Beckmann}, {Bresson}, {Chelli}, {Dugu{\'e}}, {Duvert}, {Gennari},
  {Gl{\"u}ck}, {Kern}, {Lagarde}, {Le Coarer}, {Lisi}, {Perraut}, {Puget},
  {Rantakyr{\"o}}, {Robbe-Dubois}, {Roussel}, {Zins}, {Accardo}, {Acke},
  {Agabi}, {Altariba}, {Arezki}, {Aristidi}, {Baffa}, {Behrend}, {Bl{\"o}cker},
  {Bonhomme}, {Busoni}, {Cassaing}, {Clausse}, {Colin}, {Connot},
  {Delboulb{\'e}}, {Domiciano de Souza}, {Driebe}, {Feautrier}, {Ferruzzi},
  {Forveille}, {Fossat}, {Foy}, {Fraix-Burnet}, {Gallardo}, {Giani}, {Gil},
  {Glentzlin}, {Heiden}, {Heininger}, {Hernandez Utrera}, {Kamm}, {Kiekebusch},
  {Le Contel}, {Le Contel}, {Lesourd}, {Lopez}, {Lopez}, {Magnard}, {Marconi},
  {Mars}, {Martinot-Lagarde}, {Mathias}, {M{\`e}ge}, {Monin}, {Mouillet},
  {Mourard}, {Nussbaum}, {Ohnaka}, {Pacheco}, {Perrier}, {Rabbia}, {Rebattu},
  {Reynaud}, {Richichi}, {Robini}, {Sacchettini}, {Schertl}, {Sch{\"o}ller},
  {Solscheid}, {Spang}, {Stefanini}, {Tallon}, {Tallon-Bosc}, {Tasso},
  {Vakili}, {von der L{\"u}he}, {Valtier}, {Vannier}, and {Ventura}}]{Malbet07}
{Malbet} F, {Benisty} M, {de Wit} WJ, {Kraus} S, {Meilland} A, {Millour} F,
  {Tatulli} E, {Berger} JP, {Chesneau} O, {Hofmann} KH, {Isella} A, {Natta} A,
  {Petrov} RG, {Preibisch} T, {Stee} P, {Testi} L, {Weigelt} G, {Antonelli} P,
  {Beckmann} U, {Bresson} Y, {Chelli} A, {Dugu{\'e}} M, {Duvert} G, {Gennari}
  S, {Gl{\"u}ck} L, {Kern} P, {Lagarde} S, {Le Coarer} E, {Lisi} F, {Perraut}
  K, {Puget} P, {Rantakyr{\"o}} F, {Robbe-Dubois} S, {Roussel} A, {Zins} G,
  {Accardo} M, {Acke} B, {Agabi} K, {Altariba} E, {Arezki} B, {Aristidi} E,
  {Baffa} C, {Behrend} J, {Bl{\"o}cker} T, {Bonhomme} S, {Busoni} S, {Cassaing}
  F, {Clausse} JM, {Colin} J, {Connot} C, {Delboulb{\'e}} A, {Domiciano de
  Souza} A, {Driebe} T, {Feautrier} P, {Ferruzzi} D, {Forveille} T, {Fossat} E,
  {Foy} R, {Fraix-Burnet} D, {Gallardo} A, {Giani} E, {Gil} C, {Glentzlin} A,
  {Heiden} M, {Heininger} M, {Hernandez Utrera} O, {Kamm} D, {Kiekebusch} M,
  {Le Contel} D, {Le Contel} JM, {Lesourd} T, {Lopez} B, {Lopez} M, {Magnard}
  Y, {Marconi} A, {Mars} G, {Martinot-Lagarde} G, {Mathias} P, {M{\`e}ge} P,
  {Monin} JL, {Mouillet} D, {Mourard} D, {Nussbaum} E, {Ohnaka} K, {Pacheco} J,
  {Perrier} C, {Rabbia} Y, {Rebattu} S, {Reynaud} F, {Richichi} A, {Robini} A,
  {Sacchettini} M, {Schertl} D, {Sch{\"o}ller} M, {Solscheid} W, {Spang} A,
  {Stefanini} P, {Tallon} M, {Tallon-Bosc} I, {Tasso} D, {Vakili} F, {von der
  L{\"u}he} O, {Valtier} JC, {Vannier} M, {Ventura} N (2007) {Disk and wind
  interaction in the young stellar object MWC~297 spatially resolved with
  AMBER/VLTI}. \aap 464:43--53

\bibitem[{{Mann} et~al(2006){Mann}, {K{\"o}hler}, {Kimura}, {Cechowski}, and
  {Minato}}]{Mann06}
{Mann} I, {K{\"o}hler} M, {Kimura} H, {Cechowski} A, {Minato} T (2006) {Dust in
  the solar system and in extra-solar planetary systems}. \aapr 13:159--228,
  \doi{10.1007/s00159-006-0028-0}

\bibitem[{{Marley} et~al(2007){Marley}, {Fortney}, {Hubickyj}, {Bodenheimer},
  and {Lissauer}}]{Marley07}
{Marley} MS, {Fortney} JJ, {Hubickyj} O, {Bodenheimer} P, {Lissauer} JJ (2007)
  {On the Luminosity of Young Jupiters}. \apj 655:541--549,
  \doi{10.1086/509759}, \eprint{arXiv:astro-ph/0609739}

\bibitem[{{Marois} et~al(2007){Marois}, {Macintosh}, and {Barman}}]{Marois07}
{Marois} C, {Macintosh} B, {Barman} T (2007) {GQ Lup B Visible and
  Near-Infrared Photometric Analysis}. \apjl 654:L151--L154,
  \doi{10.1086/511071}, \eprint{arXiv:astro-ph/0610058}

\bibitem[{{Marois} et~al(2008){Marois}, {Macintosh}, {Barman}, {Zuckerman},
  {Song}, {Patience}, {Lafreni{\`e}re}, and {Doyon}}]{Marois08}
{Marois} C, {Macintosh} B, {Barman} T, {Zuckerman} B, {Song} I, {Patience} J,
  {Lafreni{\`e}re} D, {Doyon} R (2008) {Direct Imaging of Multiple Planets
  Orbiting the Star HR 8799}. Science 322:1348--,
  \doi{10.1126/science.1166585}, \eprint{0811.2606}

\bibitem[{{Marsh} et~al(2002){Marsh}, {Silverstone}, {Becklin}, {Koerner},
  {Werner}, {Weinberger}, and {Ressler}}]{Marsh02}
{Marsh} KA, {Silverstone} MD, {Becklin} EE, {Koerner} DW, {Werner} MW,
  {Weinberger} AJ, {Ressler} ME (2002) {Mid-Infrared Images of the Debris Disk
  around HD 141569}. \apj 573:425--430, \doi{10.1086/340488}

\bibitem[{{Masciadri} et~al(2005){Masciadri}, {Mundt}, {Henning}, {Alvarez},
  and {Barrado y Navascu{\'e}s}}]{Masciadri05}
{Masciadri} E, {Mundt} R, {Henning} T, {Alvarez} C, {Barrado y Navascu{\'e}s} D
  (2005) {A Search for Hot Massive Extrasolar Planets around Nearby Young Stars
  with the Adaptive Optics System NACO}. \apj 625:1004--1018,
  \doi{10.1086/429687}, \eprint{arXiv:astro-ph/0502376}

\bibitem[{{Mawet} et~al(2009){Mawet}, {Serabyn}, {Stapelfeldt}, and
  {Crepp}}]{Mawet09}
{Mawet} D, {Serabyn} E, {Stapelfeldt} K, {Crepp} J (2009) {Imaging the Debris
  Disk of HD 32297 with a Phase-Mask Coronagraph at High Strehl Ratio}. \apjl
  702:L47--L50, \doi{10.1088/0004-637X/702/1/L47}

\bibitem[{{Mayor} and {Queloz}(1995)}]{Mayor95}
{Mayor} M, {Queloz} D (1995) {A Jupiter-Mass Companion to a Solar-Type Star}.
  \nat 378:355--359, \doi{10.1038/378355a0}

\bibitem[{{McCabe} et~al(2002){McCabe}, {Duch{\^e}ne}, and {Ghez}}]{McCabe02}
{McCabe} C, {Duch{\^e}ne} G, {Ghez} AM (2002) {NICMOS Images of the GG Tauri
  Circumbinary Disk}. \apj 575:974--988, \doi{10.1086/341479},
  \eprint{arXiv:astro-ph/0204465}

\bibitem[{{McCabe} et~al(2003){McCabe}, {Duch{\^e}ne}, and {Ghez}}]{McCabe03}
{McCabe} C, {Duch{\^e}ne} G, {Ghez} AM (2003) {The First Detection of Spatially
  Resolved Mid-Infrared Scattered Light from a Protoplanetary Disk}. \apjl
  588:L113--L116, \doi{10.1086/375632}, \eprint{arXiv:astro-ph/0304083}

\bibitem[{{McCarthy} and {Zuckerman}(2004)}]{McCarthy04}
{McCarthy} C, {Zuckerman} B (2004) {The Brown Dwarf Desert at 75-1200 AU}. \aj
  127:2871--2884, \doi{10.1086/383559}

\bibitem[{{McCaughrean} and {O'dell}(1996)}]{McCaughrean96}
{McCaughrean} MJ, {O'dell} CR (1996) {Direct Imaging of Circumstellar Disks in
  the Orion Nebula}. \aj 111:1977--1986, \doi{10.1086/117934}

\bibitem[{{McElwain} et~al(2007){McElwain}, {Metchev}, {Larkin}, {Barczys},
  {Iserlohe}, {Krabbe}, {Quirrenbach}, {Weiss}, and {Wright}}]{McElwain07}
{McElwain} MW, {Metchev} SA, {Larkin} JE, {Barczys} M, {Iserlohe} C, {Krabbe}
  A, {Quirrenbach} A, {Weiss} J, {Wright} SA (2007) {First High-Contrast
  Science with an Integral Field Spectrograph: The Substellar Companion to GQ
  Lupi}. \apj 656:505--514, \doi{10.1086/510063},
  \eprint{arXiv:astro-ph/0610265}

\bibitem[{{McKee} and {Ostriker}(2007)}]{McKee07}
{McKee} CF, {Ostriker} EC (2007) {Theory of Star Formation}. \araa 45:565--687,
  \doi{10.1146/annurev.astro.45.051806.110602}, \eprint{0707.3514}

\bibitem[{{Meeus} et~al(2001){Meeus}, {Waters}, {Bouwman}, {van den Ancker},
  {Waelkens}, and {Malfait}}]{Meeus01}
{Meeus} G, {Waters} LBFM, {Bouwman} J, {van den Ancker} ME, {Waelkens} C,
  {Malfait} K (2001) {ISO spectroscopy of circumstellar dust in 14 Herbig Ae/Be
  systems: Towards an understanding of dust processing}. \aap 365:476--490,
  \doi{10.1051/0004-6361:20000144}, \eprint{arXiv:astro-ph/0012295}

\bibitem[{{Mer{\'{\i}}n} et~al(2004){Mer{\'{\i}}n}, {Montesinos}, {Eiroa},
  {Solano}, {Mora}, {D'Alessio}, {Calvet}, {Oudmaijer}, {de Winter}, {Davies},
  {Harris}, {Cameron}, {Deeg}, {Ferlet}, {Garz{\'o}n}, {Grady}, {Horne},
  {Miranda}, {Palacios}, {Penny}, {Quirrenbach}, {Rauer}, {Schneider}, and
  {Wesselius}}]{Merin04}
{Mer{\'{\i}}n} B, {Montesinos} B, {Eiroa} C, {Solano} E, {Mora} A, {D'Alessio}
  P, {Calvet} N, {Oudmaijer} RD, {de Winter} D, {Davies} JK, {Harris} AW,
  {Cameron} A, {Deeg} HJ, {Ferlet} R, {Garz{\'o}n} F, {Grady} CA, {Horne} K,
  {Miranda} LF, {Palacios} J, {Penny} A, {Quirrenbach} A, {Rauer} H,
  {Schneider} J, {Wesselius} PR (2004) {Study of the properties and spectral
  energy distributions of the Herbig AeBe stars HD 34282 and HD 141569}. \aap
  419:301--318, \doi{10.1051/0004-6361:20034561},
  \eprint{arXiv:astro-ph/0402599}

\bibitem[{{Metchev} et~al(2005){Metchev}, {Eisner}, {Hillenbrand}, and
  {Wolf}}]{Metchev05}
{Metchev} SA, {Eisner} JA, {Hillenbrand} LA, {Wolf} S (2005) {Adaptive Optics
  Imaging of the AU Microscopii Circumstellar Disk: Evidence for Dynamical
  Evolution}. \apj 622:451--462, \doi{10.1086/427869},
  \eprint{arXiv:astro-ph/0412143}

\bibitem[{{Millan-Gabet} et~al(1999){Millan-Gabet}, {Schloerb}, {Traub},
  {Malbet}, {Berger}, and {Bregman}}]{MillanGabet99}
{Millan-Gabet} R, {Schloerb} FP, {Traub} WA, {Malbet} F, {Berger} JP, {Bregman}
  JD (1999) {Sub-Astronomical Unit Structure of the Near-Infrared Emission from
  AB Aurigae}. \apj 513:L131--L134

\bibitem[{{Millan-Gabet} et~al(2001){Millan-Gabet}, {Schloerb}, and
  {Traub}}]{MillanGabet01}
{Millan-Gabet} R, {Schloerb} FP, {Traub} WA (2001) {Spatially Resolved
  Circumstellar Structure of Herbig Ae/Be Stars in the Near-Infrared}. \apj
  546:358--381

\bibitem[{{Millan-Gabet} et~al(2006{\natexlab{a}}){Millan-Gabet}, {Monnier},
  {Akeson}, {Hartmann}, {Berger}, {Tannirkulam}, {Melnikov}, {Billmeier},
  {Calvet}, {D'Alessio}, {Hillenbrand}, {Kuchner}, {Traub}, {Tuthill},
  {Beichman}, {Boden}, {Booth}, {Colavita}, {Creech-Eakman}, {Gathright},
  {Hrynevych}, {Koresko}, {Le Mignant}, {Ligon}, {Mennesson}, {Neyman},
  {Sargent}, {Shao}, {Swain}, {Thompson}, {Unwin}, {van Belle}, {Vasisht}, and
  {Wizinowich}}]{MillanGabet06a}
{Millan-Gabet} R, {Monnier} JD, {Akeson} RL, {Hartmann} L, {Berger} JP,
  {Tannirkulam} A, {Melnikov} S, {Billmeier} R, {Calvet} N, {D'Alessio} P,
  {Hillenbrand} LA, {Kuchner} M, {Traub} WA, {Tuthill} PG, {Beichman} C,
  {Boden} A, {Booth} A, {Colavita} M, {Creech-Eakman} M, {Gathright} J,
  {Hrynevych} M, {Koresko} C, {Le Mignant} D, {Ligon} R, {Mennesson} B,
  {Neyman} C, {Sargent} A, {Shao} M, {Swain} M, {Thompson} R, {Unwin} S, {van
  Belle} G, {Vasisht} G, {Wizinowich} P (2006{\natexlab{a}}) {Keck
  Interferometer Observations of FU Orionis Objects}. \apj 641:547--555

\bibitem[{{Millan-Gabet} et~al(2006{\natexlab{b}}){Millan-Gabet}, {Monnier},
  {Berger}, {Traub}, {Schloerb}, {Pedretti}, {Benisty}, {Carleton},
  {Haguenauer}, {Kern}, {Labeye}, {Lacasse}, {Malbet}, {Perraut}, {Pearlman},
  and {Thureau}}]{MillanGabet06b}
{Millan-Gabet} R, {Monnier} JD, {Berger} JP, {Traub} WA, {Schloerb} FP,
  {Pedretti} E, {Benisty} M, {Carleton} NP, {Haguenauer} P, {Kern} P, {Labeye}
  P, {Lacasse} MG, {Malbet} F, {Perraut} K, {Pearlman} M, {Thureau} N
  (2006{\natexlab{b}}) {Bright Localized Near-Infrared Emission at 1-4 AU in
  the AB Aurigae Disk Revealed by IOTA Closure Phases}. \apj 645:L77--L80

\bibitem[{{Moerchen} et~al(2007){Moerchen}, {Telesco}, {Packham}, and
  {Kehoe}}]{Moerchen07}
{Moerchen} MM, {Telesco} CM, {Packham} C, {Kehoe} TJJ (2007) {Mid-Infrared
  Resolution of a 3 AU Radius Debris Disk around {$\zeta$} Leporis}. \apjl
  655:L109--L112, \doi{10.1086/511955}, \eprint{arXiv:astro-ph/0612550}

\bibitem[{{Monnier}(2003)}]{Monnier03}
{Monnier} JD (2003) {Optical interferometry in astronomy}. Reports on Progress
  in Physics 66:789--857, \doi{10.1088/0034-4885/66/5/203},
  \eprint{arXiv:astro-ph/0307036}

\bibitem[{{Monnier} and {Millan-Gabet}(2002)}]{Monnier02}
{Monnier} JD, {Millan-Gabet} R (2002) {On the Interferometric Sizes of Young
  Stellar Objects}. \apj 579:694--698

\bibitem[{{Monnier} et~al(2005){Monnier}, {Millan-Gabet}, {Billmeier},
  {Akeson}, {Wallace}, {Berger}, {Calvet}, {D'Alessio}, {Danchi}, {Hartmann},
  {Hillenbrand}, {Kuchner}, {Rajagopal}, {Traub}, {Tuthill}, {Boden}, {Booth},
  {Colavita}, {Gathright}, {Hrynevych}, {Le Mignant}, {Ligon}, {Neyman},
  {Swain}, {Thompson}, {Vasisht}, {Wizinowich}, {Beichman}, {Beletic},
  {Creech-Eakman}, {Koresko}, {Sargent}, {Shao}, and {van Belle}}]{Monnier05}
{Monnier} JD, {Millan-Gabet} R, {Billmeier} R, {Akeson} RL, {Wallace} D,
  {Berger} JP, {Calvet} N, {D'Alessio} P, {Danchi} WC, {Hartmann} L,
  {Hillenbrand} LA, {Kuchner} M, {Rajagopal} J, {Traub} WA, {Tuthill} PG,
  {Boden} A, {Booth} A, {Colavita} M, {Gathright} J, {Hrynevych} M, {Le
  Mignant} D, {Ligon} R, {Neyman} C, {Swain} M, {Thompson} R, {Vasisht} G,
  {Wizinowich} P, {Beichman} C, {Beletic} J, {Creech-Eakman} M, {Koresko} C,
  {Sargent} A, {Shao} M, {van Belle} G (2005) {The Near-Infrared
  Size-Luminosity Relations for Herbig Ae/Be Disks}. \apj 624:832--840

\bibitem[{{Monnier} et~al(2006){Monnier}, {Berger}, {Millan-Gabet}, {Traub},
  {Schloerb}, {Pedretti}, {Benisty}, {Carleton}, {Haguenauer}, {Kern},
  {Labeye}, {Lacasse}, {Malbet}, {Perraut}, {Pearlman}, and {Zhao}}]{Monnier06}
{Monnier} JD, {Berger} JP, {Millan-Gabet} R, {Traub} WA, {Schloerb} FP,
  {Pedretti} E, {Benisty} M, {Carleton} NP, {Haguenauer} P, {Kern} P, {Labeye}
  P, {Lacasse} MG, {Malbet} F, {Perraut} K, {Pearlman} M, {Zhao} M (2006) {Few
  Skewed Disks Found in First Closure-Phase Survey of Herbig Ae/Be Stars}. \apj
  647:444--463

\bibitem[{{Morbidelli} et~al(2008){Morbidelli}, {Levison}, and
  {Gomes}}]{Morbidelli08}
{Morbidelli} A, {Levison} HF, {Gomes} R (2008) {The Dynamical Structure of the
  Kuiper Belt and Its Primordial Origin}, The Solar System Beyond Neptune, pp
  275--292

\bibitem[{{Mouillet} et~al(2001){Mouillet}, {Lagrange}, {Augereau}, and
  {M{\'e}nard}}]{Mouillet01}
{Mouillet} D, {Lagrange} AM, {Augereau} JC, {M{\'e}nard} F (2001) {Asymmetries
  in the HD 141569 circumstellar disk}. \aap 372:L61--L64,
  \doi{10.1051/0004-6361:20010660}

\bibitem[{{Muzerolle} et~al(2003){Muzerolle}, {Calvet}, {Hartmann}, and
  {D'Alessio}}]{Muzerolle03}
{Muzerolle} J, {Calvet} N, {Hartmann} L, {D'Alessio} P (2003) {Unveiling the
  Inner Disk Structure of T Tauri Stars}. \apjl 597:L149--L152,
  \doi{10.1086/379921}, \eprint{arXiv:astro-ph/0310067}

\bibitem[{{Muzerolle} et~al(2004){Muzerolle}, {D'Alessio}, {Calvet}, and
  {Hartmann}}]{Muzerolle04}
{Muzerolle} J, {D'Alessio} P, {Calvet} N, {Hartmann} L (2004) {Magnetospheres
  and Disk Accretion in Herbig Ae/Be Stars}. \apj 617:406--417,
  \eprint{arXiv:astro-ph/0409008}

\bibitem[{{Najita} et~al(2009){Najita}, {Doppmann}, {Carr}, {Graham}, and
  {Eisner}}]{Najita09}
{Najita} JR, {Doppmann} GW, {Carr} JS, {Graham} JR, {Eisner} JA (2009)
  {High-Resolution K-Band Spectroscopy of MWC 480 and V1331 Cyg}. \apj
  691:738--748, \doi{10.1088/0004-637X/691/1/738}, \eprint{0809.4267}

\bibitem[{{Nakajima} et~al(1995){Nakajima}, {Oppenheimer}, {Kulkarni},
  {Golimowski}, {Matthews}, and {Durrance}}]{Nakajima95}
{Nakajima} T, {Oppenheimer} BR, {Kulkarni} SR, {Golimowski} DA, {Matthews} K,
  {Durrance} ST (1995) {Discovery of a Cool Brown Dwarf}. \nat 378:463--+,
  \doi{10.1038/378463a0}

\bibitem[{{Natta} et~al(2001){Natta}, {Prusti}, {Neri}, {Wooden}, {Grinin}, and
  {Mannings}}]{Natta01}
{Natta} A, {Prusti} T, {Neri} R, {Wooden} D, {Grinin} VP, {Mannings} V (2001)
  {A reconsideration of disk properties in Herbig Ae stars}. \aap 371:186--197

\bibitem[{{Neuh{\"a}user} et~al(2005){Neuh{\"a}user}, {Guenther}, {Wuchterl},
  {Mugrauer}, {Bedalov}, and {Hauschildt}}]{Neuhauser05}
{Neuh{\"a}user} R, {Guenther} EW, {Wuchterl} G, {Mugrauer} M, {Bedalov} A,
  {Hauschildt} PH (2005) {Evidence for a co-moving sub-stellar companion of GQ
  Lup}. \aap 435:L13--L16

\bibitem[{{Nielsen} et~al(2008){Nielsen}, {Close}, {Biller}, {Masciadri}, and
  {Lenzen}}]{Nielsen08}
{Nielsen} EL, {Close} LM, {Biller} BA, {Masciadri} E, {Lenzen} R (2008)
  {Constraints on Extrasolar Planet Populations from VLT NACO/SDI and MMT SDI
  and Direct Adaptive Optics Imaging Surveys: Giant Planets are Rare at Large
  Separations}. \apj 674:466--481, \doi{10.1086/524344}, \eprint{0706.4331}

\bibitem[{{Okamoto} et~al(2004){Okamoto}, {Kataza}, {Honda}, {Yamashita},
  {Onaka}, {Watanabe}, {Miyata}, {Sako}, {Fujiyoshi}, and {Sakon}}]{Okamoto04}
{Okamoto} YK, {Kataza} H, {Honda} M, {Yamashita} T, {Onaka} T, {Watanabe} Ji,
  {Miyata} T, {Sako} S, {Fujiyoshi} T, {Sakon} I (2004) {An early extrasolar
  planetary system revealed by planetesimal belts in {$\beta$} Pictoris}. \nat
  431:660--663, \doi{10.1038/nature02948}

\bibitem[{{Oppenheimer} et~al(2008){Oppenheimer}, {Brenner}, {Hinkley},
  {Zimmerman}, {Sivaramakrishnan}, {Soummer}, {Kuhn}, {Graham}, {Perrin},
  {Lloyd}, {Roberts}, and {Harrington}}]{Oppenheimer08}
{Oppenheimer} BR, {Brenner} D, {Hinkley} S, {Zimmerman} N, {Sivaramakrishnan}
  A, {Soummer} R, {Kuhn} J, {Graham} JR, {Perrin} M, {Lloyd} JP, {Roberts} LC
  Jr, {Harrington} DM (2008) {The Solar-System-Scale Disk around AB Aurigae}.
  \apj 679:1574--1581, \doi{10.1086/587778}, \eprint{0803.3629}

\bibitem[{{Padgett} et~al(1999){Padgett}, {Brandner}, {Stapelfeldt}, {Strom},
  {Terebey}, and {Koerner}}]{Padgett99}
{Padgett} DL, {Brandner} W, {Stapelfeldt} KR, {Strom} SE, {Terebey} S,
  {Koerner} D (1999) {HUBBLE SPACE TELESCOPE/NICMOS Imaging of Disks and
  Envelopes around Very Young Stars}. \aj 117:1490--1504, \doi{10.1086/300781},
  \eprint{arXiv:astro-ph/9902101}

\bibitem[{{Padgett} et~al(2006){Padgett}, {Cieza}, {Stapelfeldt}, {Evans},
  {Koerner}, {Sargent}, {Fukagawa}, {van Dishoeck}, {Augereau}, {Allen},
  {Blake}, {Brooke}, {Chapman}, {Harvey}, {Porras}, {Lai}, {Mundy}, {Myers},
  {Spiesman}, and {Wahhaj}}]{Padgett06}
{Padgett} DL, {Cieza} L, {Stapelfeldt} KR, {Evans} NJ II, {Koerner} D,
  {Sargent} A, {Fukagawa} M, {van Dishoeck} EF, {Augereau} JC, {Allen} L,
  {Blake} G, {Brooke} T, {Chapman} N, {Harvey} P, {Porras} A, {Lai} SP, {Mundy}
  L, {Myers} PC, {Spiesman} W, {Wahhaj} Z (2006) {The SPITZER c2d Survey of
  Weak-Line T Tauri Stars. I. Initial Results}. \apj 645:1283--1296,
  \doi{10.1086/504374}

\bibitem[{{Perrin} et~al(2004){Perrin}, {Graham}, {Kalas}, {Lloyd}, {Max},
  {Gavel}, {Pennington}, and {Gates}}]{Perrin2004}
{Perrin} MD, {Graham} JR, {Kalas} P, {Lloyd} JP, {Max} CE, {Gavel} DT,
  {Pennington} DM, {Gates} EL (2004) {Laser Guide Star Adaptive Optics Imaging
  Polarimetry of Herbig Ae/Be Stars}. Science 303:1345--1348

\bibitem[{{Perrin} et~al(2006){Perrin}, {Duch{\^e}ne}, {Kalas}, and
  {Graham}}]{Perrin2006}
{Perrin} MD, {Duch{\^e}ne} G, {Kalas} P, {Graham} JR (2006) {Discovery of an
  Optically Thick, Edge-on Disk around the Herbig Ae Star PDS 144N}. \apj
  645:1272--1282, \doi{10.1086/504510}, \eprint{arXiv:astro-ph/0603667}

\bibitem[{{Pi{\'e}tu} et~al(2006){Pi{\'e}tu}, {Dutrey}, {Guilloteau},
  {Chapillon}, and {Pety}}]{Pietu06}
{Pi{\'e}tu} V, {Dutrey} A, {Guilloteau} S, {Chapillon} E, {Pety} J (2006)
  {Resolving the inner dust disks surrounding LkCa 15 and MWC 480 at mm
  wavelengths}. \aap 460:L43--L47, \doi{10.1051/0004-6361:20065968},
  \eprint{arXiv:astro-ph/0610200}

\bibitem[{{Pinte} et~al(2007){Pinte}, {Fouchet}, {M{\'e}nard}, {Gonzalez}, and
  {Duch{\^e}ne}}]{Pinte07}
{Pinte} C, {Fouchet} L, {M{\'e}nard} F, {Gonzalez} JF, {Duch{\^e}ne} G (2007)
  {On the stratified dust distribution of the GG Tauri circumbinary ring}. \aap
  469:963--971, \doi{10.1051/0004-6361:20077137}, \eprint{arXiv:0704.2747}

\bibitem[{{Pinte} et~al(2008{\natexlab{a}}){Pinte}, {M{\'e}nard}, {Berger},
  {Benisty}, and {Malbet}}]{Pinte08a}
{Pinte} C, {M{\'e}nard} F, {Berger} JP, {Benisty} M, {Malbet} F
  (2008{\natexlab{a}}) {The Inner Radius of T Tauri Disks Estimated from
  Near-Infrared Interferometry: The Importance of Scattered Light}. \apjl
  673:L63--L66

\bibitem[{{Pinte} et~al(2008{\natexlab{b}}){Pinte}, {Padgett}, {M{\'e}nard},
  {Stapelfeldt}, {Schneider}, {Olofsson}, {Pani{\'c}}, {Augereau},
  {Duch{\^e}ne}, {Krist}, {Pontoppidan}, {Perrin}, {Grady}, {Kessler-Silacci},
  {van Dishoeck}, {Lommen}, {Silverstone}, {Hines}, {Wolf}, {Blake}, {Henning},
  and {Stecklum}}]{Pinte08b}
{Pinte} C, {Padgett} DL, {M{\'e}nard} F, {Stapelfeldt} KR, {Schneider} G,
  {Olofsson} J, {Pani{\'c}} O, {Augereau} JC, {Duch{\^e}ne} G, {Krist} J,
  {Pontoppidan} K, {Perrin} MD, {Grady} CA, {Kessler-Silacci} J, {van Dishoeck}
  EF, {Lommen} D, {Silverstone} M, {Hines} DC, {Wolf} S, {Blake} GA, {Henning}
  T, {Stecklum} B (2008{\natexlab{b}}) {Probing dust grain evolution in IM
  Lupi's circumstellar disc. Multi-wavelength observations and modelling of the
  dust disc}. \aap 489:633--650, \doi{10.1051/0004-6361:200810121},
  \eprint{0808.0619}

\bibitem[{{Pollack}(1984)}]{Pollack84}
{Pollack} JB (1984) {Origin and History of the Outer Planets: Theoretical
  Models and Observational Contraints}. \araa 22:389--424,
  \doi{10.1146/annurev.aa.22.090184.002133}

\bibitem[{{Pollack} et~al(1996){Pollack}, {Hubickyj}, {Bodenheimer},
  {Lissauer}, {Podolak}, and {Greenzweig}}]{Pollack96}
{Pollack} JB, {Hubickyj} O, {Bodenheimer} P, {Lissauer} JJ, {Podolak} M,
  {Greenzweig} Y (1996) {Formation of the Giant Planets by Concurrent Accretion
  of Solids and Gas}. Icarus 124:62--85, \doi{10.1006/icar.1996.0190}

\bibitem[{{Pontoppidan} et~al(2008){Pontoppidan}, {Blake}, {van Dishoeck},
  {Smette}, {Ireland}, and {Brown}}]{Pontoppidan08}
{Pontoppidan} KM, {Blake} GA, {van Dishoeck} EF, {Smette} A, {Ireland} MJ,
  {Brown} J (2008) {Spectroastrometric Imaging of Molecular Gas within
  Protoplanetary Disk Gaps}. \apj 684:1323--1329, \doi{10.1086/590400},
  \eprint{0805.3314}

\bibitem[{{Preibisch} et~al(2006){Preibisch}, {Kraus}, {Driebe}, {van Boekel},
  and {Weigelt}}]{Preibisch06}
{Preibisch} T, {Kraus} S, {Driebe} T, {van Boekel} R, {Weigelt} G (2006) {A
  compact dusty disk around the Herbig Ae star HR 5999 resolved with VLTI /
  MIDI}. \aap 458:235--243

\bibitem[{{Quanz} et~al(2006){Quanz}, {Henning}, {Bouwman}, {Ratzka}, and
  {Leinert}}]{Quanz06}
{Quanz} SP, {Henning} T, {Bouwman} J, {Ratzka} T, {Leinert} C (2006) {FU
  Orionis: The MIDI VLTI Perspective}. \apj 648:472--483

\bibitem[{{Quillen}(2006{\natexlab{a}})}]{Quillen06b}
{Quillen} AC (2006{\natexlab{a}}) {Predictions for a planet just inside
  Fomalhaut's eccentric ring}. \mnras 372:L14--L18,
  \doi{10.1111/j.1745-3933.2006.00216.x}, \eprint{arXiv:astro-ph/0605372}

\bibitem[{{Quillen}(2006{\natexlab{b}})}]{Quillen06a}
{Quillen} AC (2006{\natexlab{b}}) {The Warped Circumstellar Disk of HD 100546}.
  \apj 640:1078--1085, \doi{10.1086/500165}, \eprint{arXiv:astro-ph/0505088}

\bibitem[{{Quillen} et~al(2005){Quillen}, {Varni{\`e}re}, {Minchev}, and
  {Frank}}]{Quillen05}
{Quillen} AC, {Varni{\`e}re} P, {Minchev} I, {Frank} A (2005) {Driving Spiral
  Arms in the Circumstellar Disks of HD 100546 and HD 141569A}. \aj
  129:2481--2495, \doi{10.1086/428954}

\bibitem[{{Quirrenbach}(2001)}]{Quirrenbach01}
{Quirrenbach} A (2001) {Optical Interferometry}. \araa 39:353--401,
  \doi{10.1146/annurev.astro.39.1.353}

\bibitem[{{Ratzka} et~al(2007){Ratzka}, {Leinert}, {Henning}, {Bouwman},
  {Dullemond}, and {Jaffe}}]{Ratzka07}
{Ratzka} T, {Leinert} C, {Henning} T, {Bouwman} J, {Dullemond} CP, {Jaffe} W
  (2007) {High spatial resolution mid-infrared observations of the low-mass
  young star TW Hydrae}. \aap 471:173--185

\bibitem[{{Reche} et~al(2008){Reche}, {Beust}, {Augereau}, and
  {Absil}}]{Reche08}
{Reche} R, {Beust} H, {Augereau} JC, {Absil} O (2008) {On the observability of
  resonant structures in planetesimal disks due to planetary migration}. \aap
  480:551--561, \doi{10.1051/0004-6361:20077934}, \eprint{0801.2691}

\bibitem[{{Reche} et~al(2009){Reche}, {Beust}, and {Augereau}}]{Reche09}
{Reche} R, {Beust} H, {Augereau} JC (2009) {Investigating the flyby scenario
  for the HD 141569 system}. \aap 493:661--669,
  \doi{10.1051/0004-6361:200810419}, \eprint{0809.4421}

\bibitem[{{Riaud} et~al(2006){Riaud}, {Mawet}, {Absil}, {Boccaletti}, {Baudoz},
  {Herwats}, and {Surdej}}]{Riaud06}
{Riaud} P, {Mawet} D, {Absil} O, {Boccaletti} A, {Baudoz} P, {Herwats} E,
  {Surdej} J (2006) {Coronagraphic imaging of three weak-line T Tauri stars:
  evidence of planetary formation around PDS 70}. \aap 458:317--325

\bibitem[{{Roberge} et~al(2005){Roberge}, {Weinberger}, and
  {Malumuth}}]{Roberge05}
{Roberge} A, {Weinberger} AJ, {Malumuth} EM (2005) {Spatially Resolved
  Spectroscopy and Coronagraphic Imaging of the TW Hydrae Circumstellar Disk}.
  \apj 622:1171--1181, \doi{10.1086/427974}, \eprint{arXiv:astro-ph/0410251}

\bibitem[{{Safronov}(1969)}]{Safronov69}
{Safronov} VS (1969) {Evolution of the Protoplanetary Cloud and Formation of
  the Earth and Planets}. Nauka, Moscow, Russia, translated for NASA and NSF by
  Israel Program for Scientific Translation, 1972; NASA TT F-677

\bibitem[{{Schegerer} et~al(2008){Schegerer}, {Wolf}, {Ratzka}, and
  {Leinert}}]{Schegerer08}
{Schegerer} AA, {Wolf} S, {Ratzka} T, {Leinert} C (2008) {The T Tauri star RY
  Tauri as a case study of the inner regions of circumstellar dust disks}. \aap
  478:779--793

\bibitem[{{Schmidt} et~al(2008){Schmidt}, {Neuh{\"a}user}, {Seifahrt}, {Vogt},
  {Bedalov}, {Helling}, {Witte}, and {Hauschildt}}]{Schmidt08}
{Schmidt} TOB, {Neuh{\"a}user} R, {Seifahrt} A, {Vogt} N, {Bedalov} A,
  {Helling} C, {Witte} S, {Hauschildt} PH (2008) {Direct evidence of a
  sub-stellar companion around CT Chamaeleontis}. \aap 491:311--320,
  \doi{10.1051/0004-6361:20078840}, \eprint{0809.2812}

\bibitem[{{Schneider} et~al(1999){Schneider}, {Smith}, {Becklin}, {Koerner},
  {Meier}, {Hines}, {Lowrance}, {Terrile}, {Thompson}, and
  {Rieke}}]{Schneider99}
{Schneider} G, {Smith} BA, {Becklin} EE, {Koerner} DW, {Meier} R, {Hines} DC,
  {Lowrance} PJ, {Terrile} RJ, {Thompson} RI, {Rieke} M (1999) {NICMOS Imaging
  of the HR 4796A Circumstellar Disk}. \apjl 513:L127--L130,
  \doi{10.1086/311921}, \eprint{arXiv:astro-ph/9901218}

\bibitem[{{Schneider} et~al(2003){Schneider}, {Wood}, {Silverstone}, {Hines},
  {Koerner}, {Whitney}, {Bjorkman}, and {Lowrance}}]{Schneider03}
{Schneider} G, {Wood} K, {Silverstone} MD, {Hines} DC, {Koerner} DW, {Whitney}
  BA, {Bjorkman} JE, {Lowrance} PJ (2003) {NICMOS Coronagraphic Observations of
  the GM Aurigae Circumstellar Disk}. \aj 125:1467--1479, \doi{10.1086/367596}

\bibitem[{{Schneider} et~al(2005){Schneider}, {Silverstone}, and
  {Hines}}]{Schneider05}
{Schneider} G, {Silverstone} MD, {Hines} DC (2005) {Discovery of a Nearly
  Edge-on Disk around HD 32297}. \apjl 629:L117--L120, \doi{10.1086/452631},
  \eprint{arXiv:astro-ph/0507355}

\bibitem[{{Schneider} et~al(2006){Schneider}, {Silverstone}, {Hines},
  {Augereau}, {Pinte}, {M{\'e}nard}, {Krist}, {Clampin}, {Grady}, {Golimowski},
  {Ardila}, {Henning}, {Wolf}, and {Rodmann}}]{Schneider06}
{Schneider} G, {Silverstone} MD, {Hines} DC, {Augereau} J, {Pinte} C,
  {M{\'e}nard} F, {Krist} J, {Clampin} M, {Grady} C, {Golimowski} D, {Ardila}
  D, {Henning} T, {Wolf} S, {Rodmann} J (2006) {Discovery of an 86 AU Radius
  Debris Ring around HD 181327}. \apj 650:414--431, \doi{10.1086/506507},
  \eprint{arXiv:astro-ph/0606213}

\bibitem[{{Schneider} et~al(2009){Schneider}, {Weinberger}, {Becklin}, {Debes},
  and {Smith}}]{Schneider09}
{Schneider} G, {Weinberger} AJ, {Becklin} EE, {Debes} JH, {Smith} BA (2009)
  {STIS Imaging of the HR 4796A Circumstellar Debris Ring}. \aj 137:53--61,
  \doi{10.1088/0004-6256/137/1/53}, \eprint{0810.0286}

\bibitem[{{Serabyn} et~al(2006){Serabyn}, {Booth}, {Colavita}, {Crawford},
  {Garcia}, {Gathright}, {Hrynevych}, {Koresko}, {Ligon}, {Mennesson},
  {Panteleeva}, {Ragland}, {Summers}, {Traub}, {Tsubota}, {Wetherell},
  {Wizinowich}, and {Woillez}}]{Serabyn06}
{Serabyn} E, {Booth} A, {Colavita} MM, {Crawford} S, {Garcia} J, {Gathright} J,
  {Hrynevych} M, {Koresko} C, {Ligon} R, {Mennesson} B, {Panteleeva} T,
  {Ragland} S, {Summers} K, {Traub} W, {Tsubota} K, {Wetherell} E, {Wizinowich}
  P, {Woillez} J (2006) {Science observations with the Keck Interferometer
  Nuller}. In: {Monnier} JD, {Sch\"oller} M, {Danchi} WC (eds) Advances in
  Stellar Interferometry, Proc.\ SPIE, vol 6268, p~15, \doi{10.1117/12.672621}

\bibitem[{{Shuping} et~al(2003){Shuping}, {Bally}, {Morris}, and
  {Throop}}]{Shuping03}
{Shuping} RY, {Bally} J, {Morris} M, {Throop} H (2003) {Evidence for Grain
  Growth in the Protostellar Disks of Orion}. \apjl 587:L109--L112,
  \doi{10.1086/375334}

\bibitem[{{Sicilia-Aguilar} et~al(2007){Sicilia-Aguilar}, {Hartmann}, {Watson},
  {Bohac}, {Henning}, and {Bouwman}}]{SiciliaAguilar07}
{Sicilia-Aguilar} A, {Hartmann} LW, {Watson} D, {Bohac} C, {Henning} T,
  {Bouwman} J (2007) {Silicate Dust in Evolved Protoplanetary Disks: Growth,
  Sedimentation, and Accretion}. \apj 659:1637--1660, \doi{10.1086/512121},
  \eprint{arXiv:astro-ph/0701321}

\bibitem[{{Silber} et~al(2000){Silber}, {Gledhill}, {Duch{\^e}ne}, and
  {M{\'e}nard}}]{Silber00}
{Silber} J, {Gledhill} T, {Duch{\^e}ne} G, {M{\'e}nard} F (2000) {Near-Infrared
  Imaging Polarimetry of the GG Tauri Circumbinary Ring}. \apjl 536:L89--L92,
  \doi{10.1086/312731}, \eprint{arXiv:astro-ph/0005303}

\bibitem[{{Sivaramakrishnan} et~al(2007){Sivaramakrishnan}, {Oppenheimer},
  {Hinkley}, {Brenner}, {Soummer}, {Mey}, {Lloyd}, {Perrin}, {Graham},
  {Makidon}, {Roberts}, and {Kuhn}}]{Sivaramakrishnan07}
{Sivaramakrishnan} A, {Oppenheimer} BR, {Hinkley} S, {Brenner} D, {Soummer} R,
  {Mey} JL, {Lloyd} JP, {Perrin} MD, {Graham} JR, {Makidon} RB, {Roberts} LC,
  {Kuhn} JR (2007) {The Lyot Project: status and results}. Comptes Rendus
  Physique 8:355--364, \doi{10.1016/j.crhy.2007.04.001}

\bibitem[{{Skrutskie} et~al(1990){Skrutskie}, {Dutkevitch}, {Strom}, {Edwards},
  {Strom}, and {Shure}}]{Skrutskie90}
{Skrutskie} MF, {Dutkevitch} D, {Strom} SE, {Edwards} S, {Strom} KM, {Shure} MA
  (1990) {A sensitive 10-micron search for emission arising from circumstellar
  dust associated with solar-type pre-main-sequence stars}. \aj 99:1187--1195,
  \doi{10.1086/115407}

\bibitem[{{Smith} and {Terrile}(1984)}]{Smith84}
{Smith} BA, {Terrile} RJ (1984) {A circumstellar disk around Beta Pictoris}.
  Science 226:1421--1424, \doi{10.1126/science.226.4681.1421}

\bibitem[{{Smith} et~al(2009){Smith}, {Churcher}, {Wyatt}, {Moerchen}, and
  {Telesco}}]{Smith09}
{Smith} R, {Churcher} LJ, {Wyatt} MC, {Moerchen} MM, {Telesco} CM (2009)
  {Resolved debris disc emission around {$\eta$} Telescopii: a young solar
  system or ongoing planet formation?} \aap 493:299--308,
  \doi{10.1051/0004-6361:200810706}, \eprint{0810.5087}

\bibitem[{{Stapelfeldt} et~al(2007){Stapelfeldt}, {Krist}, {Bryden}, and
  {Chen}}]{Stapelfeldt07}
{Stapelfeldt} K, {Krist} J, {Bryden} G, {Chen} C (2007) {An HST/Spitzer Study
  of the HD 10647 Debris Disk}. In: In the Spirit of Bernard Lyot: The Direct
  Detection of Planets and Circumstellar Disks in the 21st Century, pp 47--+

\bibitem[{{Stapelfeldt} et~al(1998){Stapelfeldt}, {Krist}, {Menard}, {Bouvier},
  {Padgett}, and {Burrows}}]{Stapelfeldt98}
{Stapelfeldt} KR, {Krist} JE, {Menard} F, {Bouvier} J, {Padgett} DL, {Burrows}
  CJ (1998) {An Edge-On Circumstellar Disk in the Young Binary System HK
  Tauri}. \apjl 502:L65+, \doi{10.1086/311479}

\bibitem[{{Stapelfeldt} et~al(2003){Stapelfeldt}, {M{\'e}nard}, {Watson},
  {Krist}, {Dougados}, {Padgett}, and {Brandner}}]{Stapelfeldt03}
{Stapelfeldt} KR, {M{\'e}nard} F, {Watson} AM, {Krist} JE, {Dougados} C,
  {Padgett} DL, {Brandner} W (2003) {Hubble Space Telescope WFPC2 Imaging of
  the Disk and Jet of HV Tauri C}. \apj 589:410--418, \doi{10.1086/374374}

\bibitem[{{Stark} et~al(2009){Stark}, {Kuchner}, {Traub}, {Monnier}, {Serabyn},
  {Colavita}, {Koresko}, {Mennesson}, and {Keller}}]{Stark09}
{Stark} CC, {Kuchner} MJ, {Traub} WA, {Monnier} JD, {Serabyn} E, {Colavita} M,
  {Koresko} C, {Mennesson} B, {Keller} LD (2009) {51 Ophiuchus: A Possible Beta
  Pictoris Analog Measured with the Keck Interferometer Nuller}. \apj
  703:1188--1197, \doi{10.1088/0004-637X/703/2/1188}, \eprint{0909.1821}

\bibitem[{{Stecklum} et~al(2004){Stecklum}, {Launhardt}, {Fischer}, {Henden},
  {Leinert}, and {Meusinger}}]{Stecklum04}
{Stecklum} B, {Launhardt} R, {Fischer} O, {Henden} A, {Leinert} C, {Meusinger}
  H (2004) {High-Resolution Near-Infrared Observations of the Circumstellar
  Disk System in the Bok Globule CB 26}. \apj 617:418--424,
  \doi{10.1086/425291}, \eprint{arXiv:astro-ph/0408482}

\bibitem[{{Stern} and {Colwell}(1997)}]{Stern97}
{Stern} SA, {Colwell} JE (1997) {Accretion in the Edgeworth-Kuiper Belt:
  Forming 100-1000 KM Radius Bodies at 30 AU and Beyond.} \aj 114:841--+,
  \doi{10.1086/118518}

\bibitem[{{Strom} et~al(1989){Strom}, {Strom}, {Edwards}, {Cabrit}, and
  {Skrutskie}}]{Strom89}
{Strom} KM, {Strom} SE, {Edwards} S, {Cabrit} S, {Skrutskie} MF (1989)
  {Circumstellar material associated with solar-type pre-main-sequence stars -
  A possible constraint on the timescale for planet building}. \aj
  97:1451--1470, \doi{10.1086/115085}

\bibitem[{{Strubbe} and {Chiang}(2006)}]{Strubbe06}
{Strubbe} LE, {Chiang} EI (2006) {Dust Dynamics, Surface Brightness Profiles,
  and Thermal Spectra of Debris Disks: The Case of AU Microscopii}. \apj
  648:652--665, \doi{10.1086/505736}, \eprint{arXiv:astro-ph/0510527}

\bibitem[{{Tamura} et~al(1991){Tamura}, {Gatley}, {Waller}, and
  {Werner}}]{Tamura91}
{Tamura} M, {Gatley} I, {Waller} W, {Werner} MW (1991) {Two micron morphology
  of candidate protostars}. \apjl 374:L25--L28, \doi{10.1086/186063}

\bibitem[{{Tannirkulam} et~al(2007){Tannirkulam}, {Harries}, and
  {Monnier}}]{Tannirkulam07}
{Tannirkulam} A, {Harries} TJ, {Monnier} JD (2007) {The Inner Rim of YSO Disks:
  Effects of Dust Grain Evolution}. \apj 661:374--384, \doi{10.1086/513265},
  \eprint{arXiv:astro-ph/0702044}

\bibitem[{{Tannirkulam} et~al(2008{\natexlab{a}}){Tannirkulam}, {Monnier},
  {Harries}, {Millan-Gabet}, {Zhu}, {Pedretti}, {Ireland}, {Tuthill}, {ten
  Brummelaar}, {McAlister}, {Farrington}, {Goldfinger}, {Sturmann}, {Sturmann},
  and {Turner}}]{Tannirkulam08b}
{Tannirkulam} A, {Monnier} JD, {Harries} TJ, {Millan-Gabet} R, {Zhu} Z,
  {Pedretti} E, {Ireland} M, {Tuthill} P, {ten Brummelaar} T, {McAlister} H,
  {Farrington} C, {Goldfinger} PJ, {Sturmann} J, {Sturmann} L, {Turner} N
  (2008{\natexlab{a}}) {A Tale of Two Herbig Ae stars -MWC275 and AB Aurigae:
  Comprehensive Models for SED and Interferometry}. ArXiv e-prints 808,
  \eprint{0808.1728}

\bibitem[{{Tannirkulam} et~al(2008{\natexlab{b}}){Tannirkulam}, {Monnier},
  {Millan-Gabet}, {Harries}, {Pedretti}, {ten Brummelaar}, {McAlister},
  {Turner}, {Sturmann}, and {Sturmann}}]{Tannirkulam08a}
{Tannirkulam} A, {Monnier} JD, {Millan-Gabet} R, {Harries} TJ, {Pedretti} E,
  {ten Brummelaar} TA, {McAlister} H, {Turner} N, {Sturmann} J, {Sturmann} L
  (2008{\natexlab{b}}) {Strong Near-Infrared Emission Interior to the Dust
  Sublimation Radius of Young Stellar Objects MWC 275 and AB Aurigae}. \apjl
  677:L51--L54

\bibitem[{{Tatulli} et~al(2007){Tatulli}, {Isella}, {Natta}, {Testi},
  {Marconi}, {Malbet}, {Stee}, {Petrov}, {Millour}, {Chelli}, {Duvert},
  {Antonelli}, {Beckmann}, {Bresson}, {Dugu{\'e}}, {Gennari}, {Gl{\"u}ck},
  {Kern}, {Lagarde}, {Le Coarer}, {Lisi}, {Perraut}, {Puget}, {Rantakyr{\"o}},
  {Robbe-Dubois}, {Roussel}, {Weigelt}, {Zins}, {Accardo}, {Acke}, {Agabi},
  {Altariba}, {Arezki}, {Aristidi}, {Baffa}, {Behrend}, {Bl{\"o}cker},
  {Bonhomme}, {Busoni}, {Cassaing}, {Clausse}, {Colin}, {Connot},
  {Delboulb{\'e}}, {Domiciano de Souza}, {Driebe}, {Feautrier}, {Ferruzzi},
  {Forveille}, {Fossat}, {Foy}, {Fraix-Burnet}, {Gallardo}, {Giani}, {Gil},
  {Glentzlin}, {Heiden}, {Heininger}, {Hernandez Utrera}, {Hofmann}, {Kamm},
  {Kiekebusch}, {Kraus}, {Le Contel}, {Le Contel}, {Lesourd}, {Lopez}, {Lopez},
  {Magnard}, {Mars}, {Martinot-Lagarde}, {Mathias}, {M{\`e}ge}, {Monin},
  {Mouillet}, {Mourard}, {Nussbaum}, {Ohnaka}, {Pacheco}, {Perrier}, {Rabbia},
  {Rebattu}, {Reynaud}, {Richichi}, {Robini}, {Sacchettini}, {Schertl},
  {Sch{\"o}ller}, {Solscheid}, {Spang}, {Stefanini}, {Tallon}, {Tallon-Bosc},
  {Tasso}, {Vakili}, {von der L{\"u}he}, {Valtier}, {Vannier}, and
  {Ventura}}]{Tatulli07}
{Tatulli} E, {Isella} A, {Natta} A, {Testi} L, {Marconi} A, {Malbet} F, {Stee}
  P, {Petrov} RG, {Millour} F, {Chelli} A, {Duvert} G, {Antonelli} P,
  {Beckmann} U, {Bresson} Y, {Dugu{\'e}} M, {Gennari} S, {Gl{\"u}ck} L, {Kern}
  P, {Lagarde} S, {Le Coarer} E, {Lisi} F, {Perraut} K, {Puget} P,
  {Rantakyr{\"o}} F, {Robbe-Dubois} S, {Roussel} A, {Weigelt} G, {Zins} G,
  {Accardo} M, {Acke} B, {Agabi} K, {Altariba} E, {Arezki} B, {Aristidi} E,
  {Baffa} C, {Behrend} J, {Bl{\"o}cker} T, {Bonhomme} S, {Busoni} S, {Cassaing}
  F, {Clausse} JM, {Colin} J, {Connot} C, {Delboulb{\'e}} A, {Domiciano de
  Souza} A, {Driebe} T, {Feautrier} P, {Ferruzzi} D, {Forveille} T, {Fossat} E,
  {Foy} R, {Fraix-Burnet} D, {Gallardo} A, {Giani} E, {Gil} C, {Glentzlin} A,
  {Heiden} M, {Heininger} M, {Hernandez Utrera} O, {Hofmann} KH, {Kamm} D,
  {Kiekebusch} M, {Kraus} S, {Le Contel} D, {Le Contel} JM, {Lesourd} T,
  {Lopez} B, {Lopez} M, {Magnard} Y, {Mars} G, {Martinot-Lagarde} G, {Mathias}
  P, {M{\`e}ge} P, {Monin} JL, {Mouillet} D, {Mourard} D, {Nussbaum} E,
  {Ohnaka} K, {Pacheco} J, {Perrier} C, {Rabbia} Y, {Rebattu} S, {Reynaud} F,
  {Richichi} A, {Robini} A, {Sacchettini} M, {Schertl} D, {Sch{\"o}ller} M,
  {Solscheid} W, {Spang} A, {Stefanini} P, {Tallon} M, {Tallon-Bosc} I, {Tasso}
  D, {Vakili} F, {von der L{\"u}he} O, {Valtier} JC, {Vannier} M, {Ventura} N
  (2007) {Constraining the wind launching region in Herbig Ae stars: AMBER/VLTI
  spectroscopy of HD~104237}. \aap 464:55--58

\bibitem[{{Tatulli} et~al(2008){Tatulli}, {Malbet}, {M{\'e}nard}, {Gil},
  {Testi}, {Natta}, {Kraus}, {Stee}, and {Robbe-Dubois}}]{Tatulli08}
{Tatulli} E, {Malbet} F, {M{\'e}nard} F, {Gil} C, {Testi} L, {Natta} A, {Kraus}
  S, {Stee} P, {Robbe-Dubois} S (2008) {Spatially resolving the hot CO around
  the young Be star 51 Ophiuchi}. \aap 489:1151--1155,
  \doi{10.1051/0004-6361:200809627}, \eprint{0806.4937}

\bibitem[{{Telesco} et~al(2000){Telesco}, {Fisher}, {Pi{\~n}a}, {Knacke},
  {Dermott}, {Wyatt}, {Grogan}, {Holmes}, {Ghez}, {Prato}, {Hartmann}, and
  {Jayawardhana}}]{Telesco00}
{Telesco} CM, {Fisher} RS, {Pi{\~n}a} RK, {Knacke} RF, {Dermott} SF, {Wyatt}
  MC, {Grogan} K, {Holmes} EK, {Ghez} AM, {Prato} L, {Hartmann} LW,
  {Jayawardhana} R (2000) {Deep 10 and 18 Micron Imaging of the HR 4796A
  Circumstellar Disk: Transient Dust Particles and Tentative Evidence for a
  Brightness Asymmetry}. \apj 530:329--341, \doi{10.1086/308332},
  \eprint{arXiv:astro-ph/9909363}

\bibitem[{{Telesco} et~al(2005){Telesco}, {Fisher}, {Wyatt}, {Dermott},
  {Kehoe}, {Novotny}, {Mari{\~n}as}, {Radomski}, {Packham}, {De Buizer}, and
  {Hayward}}]{Telesco05}
{Telesco} CM, {Fisher} RS, {Wyatt} MC, {Dermott} SF, {Kehoe} TJJ, {Novotny} S,
  {Mari{\~n}as} N, {Radomski} JT, {Packham} C, {De Buizer} J, {Hayward} TL
  (2005) {Mid-infrared images of {$\beta$} Pictoris and the possible role of
  planetesimal collisions in the central disk}. \nat 433:133--136,
  \doi{10.1038/nature03255}

\bibitem[{{Throop} et~al(2001){Throop}, {Bally}, {Esposito}, and
  {McCaughrean}}]{Throop01}
{Throop} HB, {Bally} J, {Esposito} LW, {McCaughrean} MJ (2001) {Evidence for
  Dust Grain Growth in Young Circumstellar Disks}. Science 292:1686--1698,
  \doi{10.1126/science.1059093}, \eprint{arXiv:astro-ph/0104445}

\bibitem[{{Trilling} et~al(2001){Trilling}, {Koerner}, {Barnes}, {Ftaclas}, and
  {Brown}}]{Trilling01}
{Trilling} DE, {Koerner} DW, {Barnes} JW, {Ftaclas} C, {Brown} RH (2001)
  {Near-Infrared Coronagraphic Imaging of the Circumstellar Disk around TW
  Hydrae}. \apjl 552:L151--L154, \doi{10.1086/320332},
  \eprint{arXiv:astro-ph/0103458}

\bibitem[{{Tuthill} et~al(2000){Tuthill}, {Monnier}, {Danchi}, {Wishnow}, and
  {Haniff}}]{Tuthill00}
{Tuthill} PG, {Monnier} JD, {Danchi} WC, {Wishnow} EH, {Haniff} CA (2000)
  {Michelson Interferometry with the Keck I Telescope}. \pasp 112:555--565,
  \eprint{arXiv:astro-ph/0003146}

\bibitem[{{Tuthill} et~al(2001){Tuthill}, {Monnier}, and {Danchi}}]{Tuthill01}
{Tuthill} PG, {Monnier} JD, {Danchi} WC (2001) {A dusty torus around the
  luminous young star LkH{$\alpha$}101}. \nat 409:1012--1014,
  \eprint{arXiv:astro-ph/0102240}

\bibitem[{{Tuthill} et~al(2002){Tuthill}, {Monnier}, {Danchi}, {Hale}, and
  {Townes}}]{Tuthill02}
{Tuthill} PG, {Monnier} JD, {Danchi} WC, {Hale} DDS, {Townes} CH (2002)
  {Imaging the Disk around the Luminous Young Star LkH{$\alpha$} 101 with
  Infrared Interferometry}. \apj 577:826--838, \doi{10.1086/342235},
  \eprint{arXiv:astro-ph/0206105}

\bibitem[{{van Boekel} et~al(2004){van Boekel}, {Min}, {Leinert}, {Waters},
  {Richichi}, {Chesneau}, {Dominik}, {Jaffe}, {Dutrey}, {Graser}, {Henning},
  {de Jong}, {K{\"o}hler}, {de Koter}, {Lopez}, {Malbet}, {Morel}, {Paresce},
  {Perrin}, {Preibisch}, {Przygodda}, {Sch{\"o}ller}, and
  {Wittkowski}}]{vanBoekel04}
{van Boekel} R, {Min} M, {Leinert} C, {Waters} LBFM, {Richichi} A, {Chesneau}
  O, {Dominik} C, {Jaffe} W, {Dutrey} A, {Graser} U, {Henning} T, {de Jong} J,
  {K{\"o}hler} R, {de Koter} A, {Lopez} B, {Malbet} F, {Morel} S, {Paresce} F,
  {Perrin} G, {Preibisch} T, {Przygodda} F, {Sch{\"o}ller} M, {Wittkowski} M
  (2004) {The building blocks of planets within the `terrestrial' region of
  protoplanetary disks}. \nat 432:479--482

\bibitem[{{van Boekel} et~al(2005){van Boekel}, {Dullemond}, and
  {Dominik}}]{vanBoekel05}
{van Boekel} R, {Dullemond} CP, {Dominik} C (2005) {Flaring and self-shadowed
  disks around Herbig Ae stars: simulations for 10 {$\mu$}m interferometers}.
  \aap 441:563--571, \doi{10.1051/0004-6361:20042252},
  \eprint{arXiv:astro-ph/0506759}

\bibitem[{{Vinkovi{\'c}} and {Jurki{\'c}}(2007)}]{Vinkovic07}
{Vinkovi{\'c}} D, {Jurki{\'c}} T (2007) {Relation between the Luminosity of
  Young Stellar Objects and Their Circumstellar Environment}. \apj
  658:462--479, \eprint{arXiv:astro-ph/0612039}

\bibitem[{{Vinkovi\'c} et~al(2006){Vinkovi\'c}, {Ivezi\'c}, {Juki\'c}, and
  {Elitzur}}]{Vinkovic06}
{Vinkovi\'c} D, {Ivezi\'c} {\v Z}, {Juki\'c} T, {Elitzur} M (2006)
  {Near-Infrared and the Inner Regions of Protoplanetary Disks}. \apj
  636:348--361

\bibitem[{{Wahhaj} et~al(2005){Wahhaj}, {Koerner}, {Backman}, {Werner},
  {Serabyn}, {Ressler}, and {Lis}}]{Wahhaj05}
{Wahhaj} Z, {Koerner} DW, {Backman} DE, {Werner} MW, {Serabyn} E, {Ressler} ME,
  {Lis} DC (2005) {Radial Distribution of Dust Grains around HR 4796A}. \apj
  618:385--396, \doi{10.1086/425858}, \eprint{arXiv:astro-ph/0409283}

\bibitem[{{Wahhaj} et~al(2007){Wahhaj}, {Koerner}, and {Sargent}}]{Wahhaj07}
{Wahhaj} Z, {Koerner} DW, {Sargent} AI (2007) {High-Resolution Imaging of the
  Dust Disk around 49 Ceti}. \apj 661:368--373, \doi{10.1086/512716},
  \eprint{arXiv:astro-ph/0701352}

\bibitem[{{Watson} and {Stapelfeldt}(2004)}]{Watson04}
{Watson} AM, {Stapelfeldt} KR (2004) {The Visible and Near-Infrared Dust
  Opacity Law in the HH 30 Circumstellar Disk}. \apj 602:860--874,
  \doi{10.1086/381142}

\bibitem[{{Watson} and {Stapelfeldt}(2007)}]{Watson07}
{Watson} AM, {Stapelfeldt} KR (2007) {Asymmetry and Variability in the HH 30
  Circumstellar Disk}. \aj 133:845--861, \doi{10.1086/510455}

\bibitem[{{Weidenschilling}(1984)}]{Weidenschilling84}
{Weidenschilling} SJ (1984) {Evolution of grains in a turbulent solar nebula}.
  Icarus 60:553--567, \doi{10.1016/0019-1035(84)90164-7}

\bibitem[{{Weinberger} et~al(1999){Weinberger}, {Becklin}, {Schneider},
  {Smith}, {Lowrance}, {Silverstone}, {Zuckerman}, and
  {Terrile}}]{Weinberger99}
{Weinberger} AJ, {Becklin} EE, {Schneider} G, {Smith} BA, {Lowrance} PJ,
  {Silverstone} MD, {Zuckerman} B, {Terrile} RJ (1999) {The Circumstellar Disk
  of HD 141569 Imaged with NICMOS}. \apjl 525:L53--L56, \doi{10.1086/312334},
  \eprint{arXiv:astro-ph/9909097}

\bibitem[{{Weinberger} et~al(2002){Weinberger}, {Becklin}, {Schneider},
  {Chiang}, {Lowrance}, {Silverstone}, {Zuckerman}, {Hines}, and
  {Smith}}]{Weinberger02}
{Weinberger} AJ, {Becklin} EE, {Schneider} G, {Chiang} EI, {Lowrance} PJ,
  {Silverstone} M, {Zuckerman} B, {Hines} DC, {Smith} BA (2002) {Infrared Views
  of the TW Hydra Disk}. \apj 566:409--418, \doi{10.1086/338076},
  \eprint{arXiv:astro-ph/0110342}

\bibitem[{{Wetherill}(1980)}]{Wetherill80}
{Wetherill} GW (1980) {Formation of the terrestrial planets}. \araa 18:77--113,
  \doi{10.1146/annurev.aa.18.090180.000453}

\bibitem[{{Wyatt}(2003)}]{Wyatt03}
{Wyatt} MC (2003) {Resonant Trapping of Planetesimals by Planet Migration:
  Debris Disk Clumps and Vega's Similarity to the Solar System}. \apj
  598:1321--1340, \doi{10.1086/379064}, \eprint{arXiv:astro-ph/0308253}

\bibitem[{{Wyatt}(2005)}]{Wyatt05}
{Wyatt} MC (2005) {Spiral structure when setting up pericentre glow: possible
  giant planets at hundreds of AU in the HD 141569 disk}. \aap 440:937--948,
  \doi{10.1051/0004-6361:20053391}, \eprint{arXiv:astro-ph/0506208}

\bibitem[{{Wyatt} et~al(1999){Wyatt}, {Dermott}, {Telesco}, {Fisher}, {Grogan},
  {Holmes}, and {Pi{\~n}a}}]{Wyatt99}
{Wyatt} MC, {Dermott} SF, {Telesco} CM, {Fisher} RS, {Grogan} K, {Holmes} EK,
  {Pi{\~n}a} RK (1999) {How Observations of Circumstellar Disk Asymmetries Can
  Reveal Hidden Planets: Pericenter Glow and Its Application to the HR 4796
  Disk}. \apj 527:918--944, \doi{10.1086/308093},
  \eprint{arXiv:astro-ph/9908267}

\bibitem[{{Wyatt} et~al(2007){Wyatt}, {Smith}, {Su}, {Rieke}, {Greaves},
  {Beichman}, and {Bryden}}]{Wyatt07}
{Wyatt} MC, {Smith} R, {Su} KYL, {Rieke} GH, {Greaves} JS, {Beichman} CA,
  {Bryden} G (2007) {Steady State Evolution of Debris Disks around A Stars}.
  \apj 663:365--382, \doi{10.1086/518404}, \eprint{arXiv:astro-ph/0703608}

\end{thebibliography}

\end{document}